\documentclass[aps,prc,preprint,superscriptaddress,showpacs,nofootinbib,floatfix]{revtex4}

\usepackage{bm}
\usepackage{epsfig}
\usepackage{longtable}

\def\bea {\begin{eqnarray}}
\def\eea {\end{eqnarray}}
\def\be {\begin{equation}}
\def\ee {\end{equation}}
\def\ben{\begin{enumerate}}
\def\een{\end{enumerate}}
\def\bi{\begin{itemize}}
\def\ei{\end{itemize}}

\def\viz{{\it viz.}\ }

\def\etal{{\it et al.}\ }

\def\F{{\cal F}}
\def\prl {Phys. Rev. Lett.\ }
\def\pl {Phys. Lett.\ }
\def\pr {Phys. Rev.\ }
\def\np {Nucl. Phys.\ }
\def\gV{g_{\mbox{\tiny V}}}
\def\gA{g_{\mbox{\tiny A}}}
\def\fM{f_{\mbox{\tiny M}}}
\def\fT{f_{\mbox{\tiny T}}}
\def\fS{f_{\mbox{\tiny S}}}
\def\fP{f_{\mbox{\tiny P}}}
\def\MN{M_{\mbox{\tiny N}}}
\def\MW{M_{\mbox{\tiny W}}}
\def\GV{G_{\mbox{\tiny V}}}
\def\GF{G_{\mbox{\tiny F}}}
\def\DRV{\Delta_{\mbox{\tiny R}}^{\mbox{\tiny V}}}
\def\mA{m_{\mbox{\tiny A}}}
\def\mZ{m_{\mbox{\tiny Z}}}
\def\dbar{\mid \! \mid }

\def\hhphen{\, {\mbox{--}}}
\newcommand{\sfrac}[2]{\mbox{\small{$\frac{#1}{#2}$}}}

\begin{document} 

\title{Superallowed $0^+\rightarrow 0^+$ nuclear $\beta$-decays: \\
   A critical survey with tests of CVC and the standard model}

\author{J.C. Hardy} 
\email{hardy@comp.tamu.edu}

\affiliation{Cyclotron Institute, Texas A\&M University, College Station, Texas 77843}
\author{I.S. Towner}
\affiliation{Cyclotron Institute, Texas A\&M University, College Station, Texas 77843}
\affiliation{Physics Department, Queen's University, Kingston, Ontario K7L 3N6, Canada}

\date{\today}

\begin{abstract}
A complete and critical survey is presented of all half-life, decay-energy
and branching-ratio measurements related to 20 superallowed $0^+ \rightarrow
0^+$ decays; no measurements are ignored, though some are rejected
for cause and others updated.  
A new calculation of the statistical rate function $f$ is described and 
experimental $ft$ values determined.  The associated theoretical corrections
needed to convert these results into $\F t$ values are discussed, and
careful attention is paid to the origin and magnitude of their
uncertainties.  As an exacting confirmation of the conserved
vector current hypothesis, the $\F t$ values are seen to be constant 
to 3 parts in $10^4$.  These data are also used to set a new limit
on any possible scalar interaction: $C_S/C_V = - ( 0.00005 \pm 0.00130 )$.  The
average $\F t$ value obtained from the survey, when combined with the
muon liftime, yields the up-down quark-mixing element of the
Cabibbo-Kobayashi-Maskawa matrix, $V_{ud} = 0.9738 \pm 0.0004$;
and the unitarity test on the top row of the matrix becomes 
$|V_{ud}|^2 + |V_{us}|^2 + |V_{ub}|^2 = 0.9966 \pm 0.0014$
using the Particle Data Group's currently recommended values for
$V_{us}$ and $V_{ub}$.  We also express this result in terms of the possible
existence of right-hand currents.  Finally, we discuss the priorities
for future theoretical and experimental work with the goal of making the CKM
unitarity test more definitive.

\end{abstract}

\maketitle

\section{\label{s:intro} Introduction}
 
Precise measurements of the beta decay between nuclear analog states of spin,
$J^{\pi} = 0^+$, and isospin, $T = 1$, provide demanding and
fundamental tests of the properties of the electroweak interaction.
Collectively, these transitions can sensitively probe the conservation of
the vector weak current, set tight limits on the presence of scalar or right-hand currents
and contribute to the most demanding available test of the unitarity of the
Cabibbo-Kobayashi-Maskawa (CKM) matrix, a fundamental tenet of the electroweak
standard model. 

Eight transitions, $^{14}$O, $^{26}$Al$^m$, $^{34}$Cl, $^{38}$K$^m$,
$^{42}$Sc, $^{46}$V, $^{50}$Mn and $^{54}$Co are particularly amenable
to experiment and, because of their significance to physics, have 
consequently received a good deal of attention over the past few
decades.  In each of these cases, the experimental $ft$-value is known to
better than $0.1 \%$.  In the 1990s, $^{10}$C was added to this list; its
$ft$ value is known to a precision of $0.15 \%$.  More recently,
three more cases have been added: $^{22}$Mg, $^{34}$Ar and $^{74}$Rb,
with $ft$-value standard deviations ranging from from $0.24 \%$ to $0.40 \%$.  In
the near future these uncertainties will undoubtedly be reduced
and an additional eight cases could well be added to the list.
Though improvements are still possible, with current data we can test
the conserved vector current hypothesis at the level of 3 parts in $10^4$
and the three-generation Standard Model at the level of its quantum corrections.

Over the past decade, it has become increasingly clear that the CKM
unitarity test made possible by these measurements does not, in fact, quite
agree with standard-model expectations.  The test involves the top row of the
CKM matrix and requires that the sum of squares of the three
experimentally-determined elements, $|V_{ud}|^2 + |V_{us}|^2 + |V_{ub}|^2$,
should equal 1.  With results from superallowed $\beta$-decay providing
the input for $V_{ud}$, and values for $V_{us}$ and $V_{ub}$ taken
from the Particle Data Group reviews, the sum falls short of unity by 0.3 \%,
more than twice the quoted standard deviation \cite{TH03} -- a provocative but
hardly definitive disagreement.  Nevertheless, it has stimulated experimental
activity not only on the nuclear decays used to determine $V_{ud}$ but also on
the $K_{e3}$ branching ratio used for $V_{us}$.  Strikingly, a new measurement
of the $K^+_{e3}$ branching ratio \cite{Sh03} has thrown the accepted value
of $V_{us}$ into doubt.  Although the new branching-ratio result disagrees significantly
with previous measurements, it
would, if taken by itself, lead to a larger value for $V_{us}$ and thus
bring the CKM top-row sum into agreement with unity.  At this time, the value
of $V_{us}$ remains controversial and there are a number of kaon-decay experiments
currently underway, which should lead to a settled outcome within a very few years.

With all this activity in progress, and the likelihood that a new and reliable value
of $V_{us}$ will soon be forthcoming, this is an opportune time to produce a complete
new survey of the nuclear data used to establish $V_{ud}$.  This way, we will be able
to view the value of $V_{ud}$ with renewed confidence in anticipation of a revised result
for $V_{us}$.  ($V_{ub}$ is very small and contributes a negligible .001\% to the unitarity
sum.) We have published four previous surveys, refs.~\cite{TH73,HT75,Ko84,HT90}: the most
recent appeared fifteen years ago and included only eight superallowed transitions.
In addition to bringing the results for these cases up to date, we are now incorporating
data on twelve more transitions and have continued the practice we began in 1984 \cite{Ko84}
of updating all original data to take account of the most modern calibration standards.
We have also made completely new calculations of the statistical rate function, $f$, and
employed the most complete radiative and isospin-symmetry-breaking corrections in dealing
with the $ft$-values in the context of fundamental weak-interaction tests.

Superallowed Fermi beta decay between $0^+$ states depends uniquely
on the vector part of the hadronic weak interaction.  When it occurs
between isospin $T=1$ analog states, the conserved vector current (CVC)
hypothesis indicates that the $ft$ values should be the same
irrespective of the nucleus, {\it viz.}
\be
ft = \frac{K}{\GV^2 | M_F |^2} = {\rm ~constant},
\label{ftconst}
\ee
where $K/(\hbar c )^6 = 2 \pi^3 \hbar \ln 2 / (m_e c^2)^5 =
( 8120.271 \pm 0.012 ) \times 10^{-10}$ GeV$^{-4}$s, $\GV $ is
the vector coupling constant for semi-leptonic weak interactions,
and $M_F$ is the Fermi matrix element, which for $T = 1$ states
has the value $M_F = \sqrt{2}$.  The CVC hypothesis
asserts that the vector coupling constant, $\GV$, is a true constant
and not renormalised to another value in the nuclear medium.
A demonstration with the data assembled here that
the $ft$ values are indeed constant would provide a
stringent test of the CVC hypothesis.

Unfortunately, Eq.~(\ref{ftconst}) has to be amended slightly.  Firstly, 
there are radiative corrections because, for example, the emitted
electron may emit a bremsstrahlung photon, which goes undetected
in the experiment.  Secondly, isospin is not an exact symmetry in nuclei
so the nuclear matrix element, $M_F$ is slightly reduced from its ideal
value, leading us to write: $|M_F|^2 = 2 ( 1 - \delta_C )$.  Thus,
we define a ``corrected" $ft$ value as
\be
\F t \equiv ft (1 + \delta_R )(1 - \delta_C ) = \frac{K}{2 \GV^2 
(1 + \DRV )} = {\rm ~constant},
\label{Ftconst}
\ee
where $\delta_C$ is the isospin-symmetry-breaking correction,
$\delta_R$ is the transition-dependent part of the radiative
correction, and $\DRV$ is the transition-independent part. 
Fortunately these corrections are all of order 1\% but, even so, to
maintain an accuracy criterion of 0.1\% they must be calculated
with an accuracy of 10\% of their central value.  This is a demanding
request, especially for the nuclear-structure-dependent corrections.

To separate out those terms that are dependent on nuclear
structure from those that are not, we split the transition-dependent radiative
correction into two terms,
\be
\delta_R = \delta_R^{\prime} + \delta_{NS},
\label{dr}
\ee
of which the first, $\delta_R^{\prime}$, is a function only of the electron's
energy and the charge of the daughter nucleus $Z$; it therefore
depends on the particular nuclear decay, but is {\em independent} of
nuclear structure.  The second term, $\delta_{NS}$, like $\delta_C$, 
depends in its evaluation on the details of nuclear structure.
To emphasize the different sensitivities of the correction terms,
we rewrite the expression for $\F t$ as
\be
\F t \equiv ft (1 + \delta_R^{\prime}) (1 + \delta_{NS} - \delta_C ),
\label{Ftfactor}
\ee
where the first correction in brackets is independent of nuclear
structure, while the second incorporates the structure-dependent terms.

Our procedure in this paper will be to examine all experimental
data related to 20 superallowed transitions, comprising those that have
been well studied, together with others that have only recently become
accessible to precision measurement. The methods used and the data
accepted are presented in Sect.~\ref{Exp data}.  The calculations and
corrections required to extract final $\F t$-values from these data are
described and applied in Sect.~\ref{s:Ft}; in the same section, we use
the resulting $\F t$-values to test CVC.  Finally, in Sect.~\ref{s:iwip}
we explore the impact of these results on a number of weak-interaction issues: CKM
unitarity as well as the possible existence of scalar or right-hand currents.

\section{\label{Exp data} Experimental data}

The $ft$-value that characterizes any $\beta$-transition depends on three measured quantities: the
total transition energy, $Q_{EC}$, the half-life, $t_{1/2}$, of the parent state and the branching ratio,
$R$, for the particular transition of interest.  The $Q_{EC}$-value is required to determine the statistical
rate function, $f$, while the half-life and branching ratio combine to yield the partial half-life, $t$.
In tables~\ref{QEC}-\ref{reject} we present the measured values of these three quantities and supporting
information for a total of twenty superallowed transitions, incorporating the eight cases we dealt with
in our last complete survey \cite{HT90}, but now including four more cases that have been measured more
recently with comparable precision, and a further eight that are likely to become accessible to precision
measurements within the next few years.

\subsection{\label{eval} Evaluation principles}

In our treatment of the data, we considered all measurements formally published before November 2004 and
those we knew to be in an advanced state of preparation for publication by that date.  We scrutinized all
the original experimental reports in detail.  Where necessary and possible, we used the information provided
there to correct the results for calibration data that have improved since the measurement was made.  If corrections
were evidently required but insufficient information was provided to make them, the results were rejected.  Of
the surviving results, only those with (updated) uncertainties that are within a factor of ten of the most precise
measurement for each quantity were retained for averaging in the tables.  Each datum appearing in the tables is
attributed to its original journal reference {\it via} an alphanumeric code comprising the initial two letters
of the first author's name and the two last digits of the publication date.  These alphanumeric codes are correlated
with the actual reference numbers in Table~\ref{ref}.

The statistical procedures we have followed in analyzing the tabulated data are based on those used by the Particle
Data Group in their periodic reviews of particle properties, e.g. ref \cite{PDG04}, and adopted by us in earlier
surveys \cite{HT75,HT90} of superallowed $0^+\rightarrow 0^+$ beta decay.  In the tables and throughout this work,
``error bars" and ``uncertainties" always refer to plus-and-minus one standard deviation (68\% confidence level).  For
a set of $N$ uncoupled measurements, $x_i \pm \delta x_i$, of a particular quantity, a gaussian distribution is
assumed, the weighted average being calculated according to:
\be
  \overline{x} \pm \delta\overline{x} = \frac{\sum_i w_i x_i}{\sum_i w_i} \pm \left ({\textstyle \sum_i} w_i \right )^{-1/2} ,
\label{ave}
\ee
where
\vspace{-0.4cm}
\begin{displaymath}
 w_i = 1/(\delta x_i)^2
\nonumber  
\end{displaymath}
and the sums extend over all $N$ measurements.  For each average, the $\chi^2$ is also calculated
and a scale factor, $S$, determined:
\be
   S = \left [\chi^2/(N-1) \right ]^{1/2}.
\label{scale}
\ee
This factor is then used to establish the quoted uncertainty.  If $S \le 1$, the value of $\delta \overline{x}$
from Eq.\,(\ref{ave}) is left unchanged.  If $S > 1$ and the input $\delta x_i$ are all about the same size, then we increase
$\delta \overline{x}$ by the factor $S$, which is equivalent to assuming that all the experimental errors were underestimated
by the same factor.  Finally, if $S > 1$ but the $\delta x_i$ are of widely varying magnitudes, $S$ is recalculated with
only those results for which $\delta x_i \le 3 N^{1/2} \delta\overline{x}$ being retained; the recalculated scale factor is
then applied in the usual way.  In all three cases, no change is made to the original average $\overline{x}$ calculated with
Eq.\,(\ref{ave}).

The data for $Q_{EC}$ include measurements of both individual $Q_{EC}$-values and the differences between pairs of
$Q_{EC}$-values.  This required a two-step analysis procedure.  We first treated the individual $Q_{EC}$-value measurements for
each particular transition in the manner already described, obtaining an average result with uncertainty
in each case, $\tilde{x}_j \pm \delta \tilde{x}_j$, where the subscript $j$ now designates a particular transition.  For
transitions unconnected by difference measurements, these uncertainties were scaled if necessary and then the values
were quoted as final results.  For those transitions involved in
one or more difference measurements we combined their average $Q_{EC}$ values, $\tilde{x}_j \pm \delta \tilde{x}_j$, with the
difference measurements, $d_k \pm \delta d_k$, in a single fitting procedure.  If $M_1$ is the number of transitions that are
connected by difference measurements, and $M_2$ is the number of those difference measurements, then we have a total of
$M_1 + M_2$ input data values from which we need to extract a final set of $M_1$ average $Q_{EC}$-values,
$\overline{x}_j \pm \delta \overline{x}_j$.  We accomplish this by minimizing $\chi^2$, where 
\be
   \chi^2 = \sum_{j=1}^{M_1} \left(\frac{\tilde{x}_j - \overline{x}_j}{\delta \tilde{x}_j}\right)^2 + 
            \sum_{k=1}^{M_2} \left(\frac{d_k - \overline{d}_k}{\delta d_k}\right)^2
\label{chi2}
\ee
and
\vspace{-0.4cm}
\begin{displaymath}
 \overline{d}_k = \overline{x}_{j_1} - \overline{x}_{j_2},
\nonumber  
\end{displaymath}
with $j_1$ and $j_2$ designating the two transitions whose $Q_{EC}$-value difference is determined in a particular $d_k$
measurement.  For each of these individual $Q_{EC}$-values, we obtained its scale factor from Eq.\,(\ref{scale}), where
the $\chi^2$ used in that equation is now given by
\be
   \chi^2 = \sum_{i} \left(\frac{x_i - \overline{x}_j}{\delta x_i}\right)^2 + 
            \sum_{l} \left(\frac{d_l - \overline{d}_l}{\delta d_l}\right)^2 ,
\label{doubletchi2}
\ee
where $j$ is the particular transition of interest.  The sum in $i$ extends over all individual $Q_{EC}$-value measurements
of transition $j$, and the sum in $l$ extends over all doublet measurements that include transition $j$ as one
component.  The resultant value of $S$ was applied to the uncertainty, $\delta \overline{x}_j$, with the same conventions
as were described previously.

\subsection{Data tables}

The $Q_{EC}$-value data appear in Tables~\ref{QEC} and \ref{Qdiff}.  For the best-known nine superallowed decays -- those of $^{10}$C,
$^{14}$O, $^{26}$Al$^m$, $^{34}$Cl, $^{38}$K$^m$, $^{42}$Sc, $^{46}$V, $^{50}$Mn and $^{54}$Co -- the daughter nuclei are stable, and
the most precise determinations of their $Q_{EC}$-values have come from direct measurements of that property {\it via}, for example,
(p,n) or ($^3$He,t) reactions.  Such measurements are identified in column 3 of Table~\ref{QEC} by ``$Q_{EC}(sa)$" and each individual
result is itemized with its appropriate reference in the next three columns.  The weighted average (see Eq.\,(\ref{ave})) of all
measurements for a particular decay appears in column 7, with the corresponding scale factor (see Eq.\,(\ref{scale})) in column 8.
A few of these cases, like $^{34}$Cl and $^{46}$V, have no further complications.  There are other cases, however, in which $Q_{EC}$-value
differences have been measured in addition to the individual $Q_{EC}$-values.  These measurements are presented in Table~\ref{Qdiff}.
They have been dealt with in combination with the direct $Q_{EC}$-value measurements, as described in section~\ref{eval} (see, in
particular, Eq.\,(\ref{chi2})), with the final average $Q_{EC}$-value appearing in column 7 of Table~\ref{QEC} and the average
difference in column 4 of Table~\ref{Qdiff}.  Both are flagged with footnotes to indicate the interconnection.

\begingroup
\squeezetable
\setcounter{LTchunksize}{59}
\setlength{\LTcapwidth}{6.5in}
\begin{center}
\begin{longtable*}{lllllllll}
\caption {Decay energies, $Q_{EC}$, for superallowed $\beta$-decay branches.  (See Table~\ref{ref} for the correlation between the
  alphanumeric reference code used in this table and the actual reference numbers.)
\label{QEC}} \\

\hline\hline
&&&&&&&& \\

\multicolumn{2}{c}{Parent/Daughter}
 & Property\footnotemark[1] 
 & \multicolumn{3}{c}{Measured Energy, $Q_{EC}$ (keV)}
 & \multicolumn{1}{c}{}
 & \multicolumn{2}{c}{Average value} \\[1mm]
\cline{4-6} 
\cline{8-9} 
& & & & & & & &  \\[-2mm]
   \multicolumn{2}{c}{nuclei} & 
 & \multicolumn{1}{c}{1}
 & \multicolumn{1}{c}{2} 
 & \multicolumn{1}{c}{3}
 & \multicolumn{1}{c}{} 
 & \multicolumn{1}{l}{~~~Energy (keV)} 
 & \multicolumn{1}{c}{scale} \\[1mm]

\hline\hline &&&&&&&& \\

\endfirsthead

\multicolumn{9}{l}
{\tablename\ \thetable{} (continued)} \\[2mm]
\hline\hline
\\

\multicolumn{2}{c}{Parent/Daughter}
 & Property\footnotemark[1] 
 & \multicolumn{3}{c}{Measured Energy (keV)}
 & \multicolumn{1}{c}{}
 & \multicolumn{2}{c}{Average value} \\[1mm]
\cline{4-6} 
\cline{8-9} 
& & & & & & & &  \\[-2mm]
   \multicolumn{2}{c}{nuclei} & 
 & \multicolumn{1}{c}{1}
 & \multicolumn{1}{c}{2} 
 & \multicolumn{1}{c}{3}
 & \multicolumn{1}{c}{}  
 & \multicolumn{1}{l}{~~~~Energy (keV)} 
 & \multicolumn{1}{c}{scale} \\[1mm]
\hline
& & & & & & & & \\[-1mm]
\endhead

\hline
\endfoot

\hline\hline
\vspace{-15mm}
\endlastfoot

& & & & & & & \\[-5mm]
$T_z = -1$: & & & & & & & \\
~~ $^{10}$C & $^{10}$B & $Q_{EC}(gs)$ & ~~3647.84 $\pm$ 0.34 ~$\,$[Ba84] & ~~3647.95 $\pm$ 0.12 [Ba98] & & & ~$\:$3647.94 $\pm$ 0.11
  & 1.0   \\
 & & $E_x(d0^+)$ & ~$\;$1740.15 $\pm$ 0.17 ~$\,$[Aj88] & ~~1740.07 $\pm$ 0.02 \footnotemark[2] & & & ~$\:$1740.07 $\pm$ 0.02 & 1.0  \\
 & & $\bm{Q_{EC}(sa)}$ & & & & & {\bf 1907.87 $\pm$ 0.11} & \\[2mm]
~~ $^{14}$O & $^{14}$N & $Q_{EC}(gs)$ & ~$\;$5143.35 $\pm$ 0.60 ~$\,$[Bu61] & ~~5145.09 $\pm$ 0.46 [Ba62] & $\,$5145.57 $\pm$ 0.48 [Ro70]
 & & & \\
 & & & ~$\;$5142.71 $\pm$ 0.80 ~$\,$[Vo77] & ~~5143.43 $\pm$ 0.37 [Wh77] & $\,$5144.34 $\pm$ 0.17 [To03] & & ~$\:$5144.29 $\pm$ 0.28 & 2.1  \\
 & & $E_x(d0^+)$ & ~$\!$2312.798 $\pm$ 0.011$\:$[Aj91] & & & & 2312.798 $\pm$ 0.011 & \\
 & & $\bm{Q_{EC}(sa)}$ & & & & & {\bf 2831.18 $\pm$ 0.24 \footnotemark[3]} & {\bf 2.5} \\[2mm]
~~ $^{18}$Ne & $^{18}$F & $ME(p)$ & $\:$~~~5316.8 $\pm$ 1.5 ~~$\,$[Ma94] & ~~5317.63 $\pm$ 0.36 [Bl04b] & & & ~$\:$5317.58 $\pm$ 0.35 & 1.0 \\
 & & $ME(d)$ & $\:$~~~873.31 $\pm$ 0.94 ~[Bo64] & ~~~~~875.5 $\pm$ 2.2 ~$\,$[Ho64] & ~~~~876.5 $\pm$ 2.8 ~[Pr67] & & & \\
 & & & ~~~~~877.2 $\pm$ 3.0 ~~$\,$[Se73] & ~~~$\,$873.96 $\pm$ 0.61 [Ro75] & & & ~~~874.02 $\pm$ 0.48 & 1.0 \\
 & & $Q_{EC}(gs)$ & ~~~~~$\;$4438 $\pm$ 9 ~~~~$\:$[Fr63] & & & & ~$\:$4443.54 $\pm$ 0.60 & 1.0 \\
 & & $E_x(d0^+)$ & ~$\;$1041.55 $\pm$ 0.08 ~[Ti95] & & & & ~$\:$1041.55 $\pm$ 0.08 & \\
 & & $\bm{Q_{EC}(sa)}$ & & & & & {\bf 3401.99 $\pm$ 0.60} & \\[2mm]
~~ $^{22}$Mg & $^{22}$Na & $ME(p)$ & ~~~$\;$-401.3 $\pm$ 3.0 ~~$\,$[Ha74c] & ~~~~-400.5 $\pm$ 1.3 ~$\,$\footnotemark[4] & & & ~~~~-400.6 $\pm$ 1.2 & 1.0 \\
  & & $ME(d)$ & ~$\:\,$-5184.3 $\pm$ 1.5 ~~$\,$[We68] & ~~$\,$-5182.5 $\pm$ 0.5 ~$\,$[Be68] & ~$\,$-5181.3 $\pm$ 1.7 ~[An70] & & & \\
 & & & ~$\:\,$-5183.2 $\pm$ 1.0 ~~$\,$[Gi72] & ~-5181.56 $\pm$ 0.16 [Mu04] & -5181.08 $\pm$ 0.30
 [Sa04] & & ~-5181.58 $\pm$ 0.27 & 2.4 \\
 & & $Q_{EC}(gs)$ & ~$\;$4781.64 $\pm$ 0.28 ~[Mu04] & ~~4781.40 $\pm$ 0.67 [Sa04] & & & ~$\;$4781.58 $\pm$ 0.25 & 1.0 \\
 & & $E_x(d0^+)$ & ~~~$\,$657.00 $\pm$ 0.14 ~[En98] & & & & ~~~657.00 $\pm$ 0.14 & \\
 & & $\bm{Q_{EC}(sa)}$ & & & & & {\bf 4124.58 $\pm$ 0.29} & \\[2mm] 
~~ $^{26}$Si & $^{26}$Al & $ME(p)$ & ~~~~~-7159 $\pm$ 18 ~~~[Mi67] & ~~~~~-7149 $\pm$ 30 ~~$\,$[Mc67] & ~~~~-7139 $\pm$ 30 ~~[Ha68] & & & \\
 & & & ~~$\;$-7145.5 $\pm$ 3.0 ~~[Ha74c] & & & & ~~~-7145.8 $\pm$ 2.9 & 1.0 \\
 & & $ME(d0^+)$ & -11982.08 $\pm$ 0.26~$\,$\footnotemark[5] & & & & -11982.08 $\pm$ 0.26 & \\
 & & $\bm{Q_{EC}(sa)}$ & ~~~~~~4850 $\pm$ 13 ~~~[Fr63] & & & & ~~$\,${\bf 4836.9 $\pm$ 3.0} & {\bf 1.0} \\[2mm]
~~ $^{30}$S & $^{30}$P & $ME(p)$ & $\:$~~$\;$-14060 $\pm$ 15 ~~~[Mi67] & ~~~$\,$-14054 $\pm$ 25 ~~$\,$[Mc67] & ~~~-14068 $\pm$ 30 ~~[Ha68] & & & \\ 
 & & & ~-14063.4 $\pm$ 3.0 ~~$\,$[Ha74c] & & & & ~$\:$-14063.2 $\pm$ 2.9& 1.0 \\
 & & $ME(d)$ & ~~~$\,$-20203 $\pm$ 3 ~~~~~[Ha67] &  -20200.58 $\pm$ 0.40 [Re85]& & & -20200.62 $\pm$ 0.40 & 1.0 \\
 & & $Q_{EC}(gs)$ & & & & & ~~~~6137.4 $\pm$ 2.9 & \\
 & & $E_x(d0^+)$ & $\:$~~~677.29 $\pm$ 0.07 $\:$[En98] & & & & ~~~~677.29 $\pm$ 0.07 & \\
 & & $\bm{Q_{EC}(sa)}$ & ~~~~~~5437 $\pm$ 17 ~~~[Fr63] & & & & ~~$\,${\bf 5459.5 $\pm$ 3.9} & {\bf 1.3} \\[2mm]
~~ $^{34}$Ar & $^{34}$Cl & $ME(p)$ & ~-18380.2 $\pm$ 3.0 ~~$\,$[Ha74c] & -18377.10 $\pm$ 0.41$\,$[He02] & & & -18377.17 $\pm$ 0.40 & 1.0 \\
 & & $ME(d)$ & $\!$-24440.01 $\pm$ 0.23~$\,$\footnotemark[5] & & & & -24440.01 $\pm$ 0.23 & \\
 & & $\bm{Q_{EC}(sa)}$ & & & & & {\bf ~6062.83 $\pm$ 0.46} & \\[2mm] 
~~ $^{38}$Ca & $^{38}$K & $ME(p)$ & ~$\,$-22056.0 $\pm$ 5.0 ~$\,$[Se74] &  & & & ~$\:$-22056.0 $\pm$ 5.0 & \\
 & & $ME(d0^+)$ & -28670.20 $\pm$ 0.32~$\,$\footnotemark[5] & & & & -28670.20 $\pm$ 0.32 & \\
 & & $\bm{Q_{EC}(sa)}$ & & & & & ~~~{\bf 6614.2 $\pm$ 5.0} & \\[2mm]
~~ $^{42}$Ti & $^{42}$Sc & $ME(p)$ & $\:$~~~-25121 $\pm$ 6 ~~~~$\,$[Mi67] & ~~~-25086 $\pm$ 30 ~~~[Ha68] & ~~~-25124 $\pm$ 13 ~~[Zi72] & & ~$\:$-25120.7 $\pm$ 5.3 & 1.0 \\
 & & $ME(d)$ & -32121.55 $\pm$ 0.80~$\,$\footnotemark[5] & & & & -32121.55 $\pm$ 0.80 & \\
 & & $\bm{Q_{EC}(sa)}$ & & & & & ~~~{\bf 7000.9 $\pm$ 5.4} & \\[2mm]
$T_z = 0$: & & & & & & & & \\
~~ $^{26}$Al$^m$ & $^{26}$Mg & $Q_{EC}(gs)$ & ~$\;$4004.79 $\pm$ 0.55 ~[De69] & ~~4004.41 $\pm$ 0.10 \footnotemark[6] & & & ~~~4004.42 $\pm$ 0.10 & 1.0 \\
 & & $E_x(p0^+)$ & ~$\;$228.305 $\pm$ 0.013$\,$[En98] & & & & ~~~228.305 $\pm$ 0.013 & \\
 & & $\bm{Q_{EC}(sa)}$ & ~$\;$4232.71 $\pm$ 0.60 ~[Vo77] & ~~4232.19 $\pm$ 0.12 [Br94] & & & ~$\:${\bf 4232.55 $\pm$ 0.17 \footnotemark[3]} & {\bf 2.7} \\[2mm]
~~ $^{34}$Cl & $^{34}$S & $\bm{Q_{EC}(sa)}$ & ~~~$\,$5490.3 $\pm$ 1.9 ~~[Ry73a] & ~~~$\,$5491.6 $\pm$ 2.3 ~$\,$[Ha74d] & ~~5491.71 $\pm$ 0.54$\:$[Ba77c]
 & & &  \\
 & & & ~~~$\,$5492.2 $\pm$ 0.4 ~~[Vo77] & ~~5491.65 $\pm$ 0.26 \footnotemark[7] & & & ~$\:${\bf 5491.78 $\pm$ 0.20} & {\bf 1.0} \\[2mm]
~~ $^{38}$K$^m$ & $^{38}$Ar & $Q_{EC}(gs)$ & ~~5914.76 $\pm$ 0.60 $\,$[Ja78] & & & & ~~~5914.76 $\pm$ 0.60 &  \\
 & & $E_x(p0^+)$ & ~~~~~130.4 $\pm$ 0.3 ~~[En98] & & & & ~~~~~~130.4 $\pm$ 0.3 & \\
 & & $\bm{Q_{EC}(sa)}$ & ~~~$\,$6044.6 $\pm$ 1.5 ~~[Bu79] & ~~6044.38 $\pm$ 0.12 [Ha98] & & & ~$\:${\bf 6044.40 $\pm$ 0.11} & {\bf 1.0} \\[2mm]
~~ $^{42}$Sc & $^{42}$Ca & $\bm{Q_{EC}(sa)}$ & ~~6423.71 $\pm$ 0.40 $\:$[Vo77] & ~~6425.84 $\pm$ 0.17 \footnotemark[8] & & & ~$\:${\bf 6425.63 $\pm$ 0.38 \footnotemark[3]} &     {\bf 3.2}  \\[2mm]
~~ $^{46}$V  & $^{46}$Ti & $\bm{Q_{EC}(sa)}$ & ~~~$\,$7053.3 $\pm$ 1.8 ~~[Sq76] & ~~7050.41 $\pm$ 0.60 [Vo77] & & & ~$\:${\bf 7050.71 $\pm$ 0.89} & {\bf 1.6}  \\[2mm]
~~ $^{50}$Mn & $^{50}$Cr & $\bm{Q_{EC}(sa)}$ & ~~~$\,$7632.8 $\pm$ 2.8 ~~[Ha74d] & ~~7631.91 $\pm$ 0.40 [Vo77] & & & ~$\:${\bf 7632.43 $\pm$ 0.23 \footnotemark[3]} &  {\bf 1.0} \\[2mm]
~~ $^{54}$Co & $^{54}$Fe & $\bm{Q_{EC}(sa)}$ & ~~~$\,$8241.2 $\pm$ 1.8 ~~[Ho74] & ~~~$\,$8245.6 $\pm$ 3.0 ~$\,$[Ha74d] & ~~8241.61 $\pm$ 0.60 [Vo77] & &   ~$\:${\bf 8242.60 $\pm$   0.29 \footnotemark[3]} & {\bf   1.5} \\[2mm]
~~ $^{62}$Ga & $^{62}$Zn & $\bm{Q_{EC}(sa)}$ & ~~~~~~9171 $\pm$ 26 $\:$~~[Da79] & & & & ~~~~~~{\bf 9171 $\pm$ 26} &  \\[2mm]
~~ $^{66}$As & $^{66}$Ge & $\bm{Q_{EC}(sa)}$ & ~~~~~~9550 $\pm$ 50 ~~~[Da80] & & & & {\bf ~~~~~9550 $\pm$ 50} &  \\[2mm]
~~ $^{70}$Br & $^{70}$Se & $\bm{Q_{EC}(sa)}$ & ~~~~~~9970 $\pm$ 170$\,$~~[Da80] & & & & {\bf ~~~~~9970 $\pm$ 170} &  \\[2mm]
~~ $^{74}$Rb & $^{74}$Kr & $ME(p)$ & $\,$~~~-51905 $\pm$ 18 ~~~[He02] & ~-51915.2 $\pm$ 4.0 ~~[Ke04] & & & ~$\:$-51914.7 $\pm$ 3.9 & 1.0 \\
 & & $ME(d)$ & ~-62330.3 $\pm$ 2.4 $\,$~~[He02] & ~-62332.0 $\pm$ 2.1 ~~[Ke04] & & & ~$\:$-62331.3 $\pm$ 1.6 & 1.0  \\
 & & $\bm{Q_{EC}(sa)}$ & & & & & ~$\,${\bf 10416.5 $\pm$ 4.2} & \\
\vspace{-12mm}
\footnotetext[1]{Abbreviations used in this column are as follows: ``$gs$", transition between ground states; ``$sa$", superallowed
transition; ``$p$", parent; ``$d$", daughter; ``$ME$", mass excess; ``$E_x(0^+)$", excitation energy of the $0^+$ (analog) state.
Thus, for example,    ``$Q_{EC}(sa)$" signifies the $Q_{EC}$-value for the superallowed transition, ``$ME(d)$", the mass excess
of the daughter nucleus; and ``$ME(d0^+)$, the mass excess of the daughter's $0^+$ state.}
\footnotetext[2]{Result based on references [Ba88] and [Ba89].}
\footnotetext[3]{Average result includes the results of $Q_{EC}$ pairs; see Table~\ref{Qdiff}.}
\footnotetext[4]{Result based on references [Bi03] and [Se04].}
\footnotetext[5]{Result obtained from data elsewhere in this table.}
\footnotetext[6]{Result based on references [Is80], [Al82], [Hu82], [Be85], [Pr90], [Ki91] and [Wa92].}
\footnotetext[7]{Result based on references [Wa83], [Ra83] and [Li94].}
\footnotetext[8]{Result based on references [Zi87] and [Ki89].}
\end{longtable*}
\end{center}
\endgroup

\newpage

\begin{table*}
\caption{$Q_{EC}$-value differences for superallowed $\beta$-decay branches.  These data are also used as input to determine
some of the average $Q_{EC}$-values listed in Table~\ref{QEC}.   (See Table~\ref{ref} for the correlation between the
alphabetical reference code used in this table and the actual reference numbers.)
\label{Qdiff}}
\begin{ruledtabular}
\begin{tabular}{llll}
Parent   
 & \multicolumn{1}{l}{Parent}
 & \multicolumn{2}{c}{$Q_{EC2} - Q_{EC1}$ (keV)} \\[1mm]
\cline{3-4} 
nucleus 1 
 & \multicolumn{1}{l}{nucleus 2}
 & \multicolumn{1}{c}{measurement} 
 & \multicolumn{1}{c}{average\footnotemark[1]} \\
\hline
$^{14}$O & $^{26}$Al$^m$ & 1401.68 $\pm$ 0.13 [Ko87] & 1401.37 $\pm$ 0.29 \\
$^{26}$Al$^m$ & $^{42}$Sc & ~$\,$2193.5 $\pm$ 0.2 ~$\,$[Ko87] & 2193.09 $\pm$ 0.42 \\
$^{42}$Sc & $^{50}$Mn & ~$\,$1207.6 $\pm$ 2.3 ~$\,$[Ha74d] & 1206.79 $\pm$ 0.44 \\
$^{42}$Sc & $^{54}$Co & ~$\,$1817.2 $\pm$ 0.2 ~$\,$[Ko87] & 1816.97 $\pm$ 0.48 \\
$^{50}$Mn & $^{54}$Co & ~$\,$610.09 $\pm$ 0.17 $^{\rm[Ko87]}_{\rm[Ko97b]}$ & ~$\,$610.18 $\pm$ 0.37 \\[-4mm]
\footnotetext[1]{Average values include the results of direct $Q_{EC}$-value measurements: see Table~\ref{QEC}.}
\end{tabular}
\end{ruledtabular}
\end{table*}

\begingroup
\squeezetable
\begin{table*}
\caption{Half-lives, $t_{1/2}$, of superallowed $\beta$-emitters.  (See Table~\ref{ref} for the correlation between the
  alphabetical reference code used in this table and the actual reference numbers.)
\label{t1/2}}
\begin{ruledtabular}
\begin{tabular}{llllllll}
Parent   
 & \multicolumn{4}{c}{Measured half-lives, $t_{1/2}$ (ms)}
 & \multicolumn{1}{c}{}
 & \multicolumn{2}{c}{Average value} \\[1mm]
\cline{2-5} 
\cline{7-8} 
nucleus 
 & \multicolumn{1}{c}{1}
 & \multicolumn{1}{c}{2} 
 & \multicolumn{1}{c}{3} 
 & \multicolumn{1}{c}{4} 
 & \multicolumn{1}{c}{}
 & \multicolumn{1}{c}{$t_{1/2}$ (ms)} 
 & \multicolumn{1}{c}{scale} \\
\hline
 & & & & & & & \\[-4mm]
$T_z = -1$: & & & & & & & \\
~~ $^{10}$C  & ~19280 $\pm$ 20 $\,$~[Az74] & $\:$19295 $\pm$ 15 ~[Ba90] & &  & & $\,$~~19290 $\pm$ 12  & 1.0   \\
~~ $^{14}$O  & ~70480 $\pm$ 150 [Al72] & $\:$70588 $\pm$ 28 ~[Cl73] &~70430 $\pm$ 180 [Az74] &~70684 $\pm$ 77 $\,$~[Be78] & & &  \\
             & ~70613 $\pm$ 25 $\,$~[Wi78] & $\:$70560 $\pm$ 49 ~[Ga01] &~70641 $\pm$ 20 $\,$~[Ba04] & & &$\,$~~70616 $\pm$ 14 & 1.1  \\
~~ $^{18}$Ne & $\,$~~1690 $\pm$ 40 $\,$~[As70] & ~~1670 $\pm$ 20 ~[Al70] & $\,$~~1669 $\pm$ 4 ~~~[Al75]  &$\,$~~1687 $\pm$ 9 ~~~[Ha75]
             & &$\,$~1672.1 $\pm$ 4.6 & 1.3            \\
~~ $^{22}$Mg & $\,$~~3857 $\pm$ 9 ~~~[Ha75] & 3875.5 $\pm$ 1.2 [Ha03]   & & & & $\,$~3875.2 $\pm$ 2.4 & 2.0 \\
~~ $^{26}$Si & $\,$~~2210 $\pm$ 21 $\,$~[Ha75] & ~~2240 $\pm$ 10 ~[Wi80] & & & & ~~~~2234 $\pm$ 12 & 1.3 \\
~~ $^{30}$S  & $\,$~~1180 $\pm$ 40 $\,$~[Ba67] & ~~1220 $\pm$ 30 ~[Mo71] &1178.3 $\pm$ 4.8 ~[Wi80] &   & & $\,$~1179.4 $\pm$ 4.7 & 1.0 \\
~~ $^{34}$Ar & $\,$~844.5 $\pm$ 3.4 ~[Ha74a] & ~847.0 $\pm$ 3.7 $\,$[Ia03]  & & & & ~~~845.6 $\pm$ 2.5 & 1.0 \\
~~ $^{38}$Ca & ~~~~470 $\pm$ 20 $\:$~[Ka68] & ~~~\,439 $\pm$ 12 $\,$~[Ga69] & ~~~~450 $\pm$ 70 ~~[Zi72] &~~~~430 $\pm$ 12 $\,$~[Wi80]
             & & ~~~440.0 $\pm$ 7.8  & 1.2          \\
~~ $^{42}$Ti & ~~~~200 $\pm$ 20 $\:$~[Ni69] & ~~~\,202 $\pm$ 5 ~~~[Ga69] & ~~~~173 $\pm$ 14 ~~[Al69] & & & ~~~198.8 $\pm$ 6.3 & 1.4 \\[1mm]
$T_z = 0$: & & & & & & &  \\
~~ $^{26}$Al$^m$ &  $\,$~~6346 $\pm$ 5 $\,$~~~[Fr69a] & $\,$~~6346 $\pm$ 5 $\,$~~~[Az75] & 6339.5 $\pm$ 4.5 $\,$~[Al77] & 6346.2 $\pm$ 2.6 ~[Ko83] & &    $\,$~6345.0 $\pm$ 1.9    & 1.0      \\
             & $\,$~~6345 $\pm$ 14 ~~[Sc05] & & & & & & \\
~~ $^{34}$Cl & $\,$~~1526 $\pm$ 2 $\,$~~~[Ry73a] & 1525.2 $\pm$ 1.1 $\,$~[Wi76] & 1527.7 $\pm$ 2.2 $\,$~[Ko83] & 1527.1 $\pm$ 0.5 ~[Ia03]
             & & 1526.77 $\pm$ 0.44   &  1.0   \\
~~ $^{38}$K$^m$ & $\,$~925.6 $\pm$ 0.7 $\,$~[Sq75] & $\,$~922.3 $\pm$ 1.1 $\,$~[Wi76] & 921.71 $\pm$ 0.65 [Wi78]  & 924.15 $\pm$ 0.31$\,$[Ko83]
            & &  &  \\
 & $\,$~924.4 $\pm$ 0.6 $\,$~[Ba00] & 924.46 $\pm$ 0.14 [Ba05] &  & & & $\,$~924.33 $\pm$ 0.27 & 2.3  \\
~~ $^{42}$Sc &  680.98 $\pm$ 0.62 [Wi76] & 680.67 $\pm$ 0.28 [Ko97a]  &    &  &  & $\,$~680.72 $\pm$ 0.26  &  1.0      \\
~~ $^{46}$V &  422.47 $\pm$ 0.39 [Al77] & 422.28 $\pm$ 0.23 [Ba77a] & 422.57 $\pm$ 0.13 [Ko97a]  &   &  & $\,$~422.50 $\pm$ 0.11 & 1.0    \\
~~ $^{50}$Mn &  $\,$~284.0 $\pm$ 0.4 $\,$~[Ha74b] & $\,$~282.8 $\pm$ 0.3 $\,$~[Fr75] & 282.72 $\pm$ 0.26 [Wi76] & 283.29 $\pm$ 0.08 [Ko97a]
             & & $\,$~283.24 $\pm$ 0.13 & 1.8       \\
~~ $^{54}$Co & $\,$~193.4 $\pm$ 0.4 $\,$~[Ha74b] & $\,$~193.0 $\pm$ 0.3 $\,$~[Ho74] & 193.28 $\pm$ 0.18 [Al77] & 193.28 $\pm$ 0.07 [Ko97a] 
             & & 193.271 $\pm$ 0.063 & 1.0        \\
~~ $^{62}$Ga & 115.95 $\pm$ 0.30 [Al78] & 116.34 $\pm$ 0.35 [Da79] & 115.84 $\pm$ 0.25 [Hy03] & 116.19 $\pm$ 0.04 [Bl04a] & & 116.175 $\pm$ 0.038 & 1.0        \\
             & 116.09 $\pm$ 0.17 [Ca05] & & & & & & \\
~~ $^{66}$As &  $\,$~95.78 $\pm$ 0.39 [Al78] & $\,$~95.77 $\pm$ 0.28 [Bu88] &   &  &  & ~~~95.77 $\pm$ 0.23    & 1.0         \\
~~ $^{70}$Br &  ~~~80.2 $\pm$ 0.8 $\,$~[Al78] & $\,$~78.54 $\pm$ 0.59 [Bu88] &    &  &  & ~~~79.12 $\pm$ 0.79 & 1.7       \\
~~ $^{74}$Rb &  $\,$~64.90 $\pm$ 0.09 [Oi01] & 64.761 $\pm$ 0.031$\,$[Ba01] &    &   &  & $\,$~64.776 $\pm$ 0.043 &  1.5   \\

\end{tabular}
\end{ruledtabular}
\end{table*}
\endgroup

\begingroup
\squeezetable
\begin{table*}

\vskip -1in

\caption{Branching ratios, R, for superallowed $\beta$-transitions. ( See Table~\ref{ref} for the correlation between the
  alphabetical reference code used in this table and the actual reference numbers.)
\label{R}}
\begin{ruledtabular}
\begin{tabular}{llllllll}
\multicolumn{2}{c}{Parent/Daughter}
 & Daughter state
 & \multicolumn{2}{c}{Measured Branching Ratio, R (\%)}
 & \multicolumn{1}{c}{}
 & \multicolumn{2}{c}{Average value} \\[1mm]
\cline{4-6} 
\cline{7-8} 
   \multicolumn{2}{c}{nuclei}
 & \multicolumn{1}{c}{$E_x$ (MeV)}
 & \multicolumn{1}{c}{1}
 & \multicolumn{1}{c}{2} 
 & \multicolumn{1}{c}{} 
 & \multicolumn{1}{c}{R (\%)} 
 & \multicolumn{1}{c}{scale} \\

\hline
& & & & & & & \\[-4mm]
 $T_z = -1$: & & & & & & & \\
~~ $^{10}$C & $^{10}$B & 2.16 & 0$^{+0.0008}_{-0}$ [Go72] & & & 0$^{+0.0008}_{-0}$ &   \\
 & & {\bf 1.74} & 1.468 $\pm$ 0.014 [Ro72]& 1.473 $\pm$ 0.007 [Na91] & & & \\
 & & & 1.465 $\pm$ 0.009 [Kr91] & 1.4625 $\pm$ 0.0025 [Sa95] & & & \\
 & & & 1.4665 $\pm$ 0.0038 [Fu99] & & & {\bf 1.4646 $\pm$ 0.0019} & {\bf 1.0} \\
~~ $^{14}$O & $^{14}$N & gs & 0.60 $\pm$ 0.10 [Sh55] & 0.65 $\pm$ 0.05 [Fr63] & & & \\
 & & & 0.61 $\pm$ 0.01 [Si66] & & & 0.611 $\pm$ 0.010 & 1.0 \\
 & & 3.95 & 0.062 $\pm$ 0.007 [Ka69] & 0.058 $\pm$ 0.004 [Wi80] & & & \\
 & & & 0.053 $\pm$ 0.002 [He81] & & & 0.0545 $\pm$ 0.0019 & 1.1 \\
 & & {\bf 2.31} & & & & {\bf 99.334 $\pm$ 0.010} & \\
~~ $^{18}$Ne & $^{18}$F & {\bf 1.04} & 9 $\pm$ 3 [Fr63] & 7.70 $\pm$ 0.21\footnotemark[1] [Ha75] & & {\bf 7.70 $\pm$ 0.21} & {\bf 1.0} \\
~~ $^{22}$Mg & $^{22}$Na & {\bf 0.66} & 54.0 $\pm$ 1.1 [Ha75] & 53.15 $\pm$ 0.12 [Ha03] & & {\bf 53.16 $\pm$ 0.12} & {\bf 1.0} \\
~~ $^{26}$Si & $^{26}$Al & 1.06 & 21.8 $\pm$ 0.8 [Ha75] & & & 21.8 $\pm$ 0.8 & \\
 & & {\bf 0.23} & & & & {\bf 75.09 $\pm$ 0.92\footnotemark[1]} & \\
~~ $^{30}$S & $^{30}$P & gs & 20 $\pm$ 1 [Fr63] & & & 20 $\pm$ 1 &  \\
 & & {\bf 0.68} & & & & {\bf 77.4 $\pm$ 1.0\footnotemark[1]} & \\
~~ $^{34}$Ar & $^{34}$Cl & 0.67 & 2.49 $\pm$ 0.10 [Ha74a] & & & 2.49 $\pm$ 0.10 & \\
 & & {\bf gs} & & & & {\bf 94.45 $\pm$ 0.25\footnotemark[1]} & \\
~~ $^{42}$Ti & $^{42}$Sc & 0.61 & 56 $\pm$ 14 [Al69] & & & 56 $\pm$ 14 & \\
 & & {\bf gs} & & & & {\bf 43 $\pm$ 14\footnotemark[1]} & \\[1mm]
 $T_z = 0$: & & & & & & &  \\
~~ $^{26}$Al$^m$ & $^{26}$Mg & {\bf gs} & $>$99.997 [Ki91] & & & 100.000$^{+0}_{-0.003}$ & \\
~~ $^{34}$Cl & $^{34}$S & {\bf gs} & $>$99.988 [Dr75] & & & 100.000$^{+0}_{-0.012}$ & \\
~~ $^{38}$K$^m$ & $^{38}$Ar & 3.38 & $<$0.0019 [Ha94] & & & 0$^{+0.002}_{-0}$ & \\
 & & {\bf gs} & $>$99.998 & & & {\bf 100.000}$\bm{^{+0}_{-0.002}}$ & \\
~~ $^{42}$Sc & $^{42}$Ca & 1.84 & 0.0063 $\pm$ 0.0026 [In77] & 0.0022 $\pm$ 0.0017 [De78] & & & \\
 & & & 0.0103 $\pm$ 0.0031 [Sa80] & 0.0070 $\pm$ 0.0012 [Da85] & & 0.0059 $\pm$ 0.0014 & 1.6 \\
  & & {\bf gs} & & & & {\bf 99.9941 $\pm$ 0.0014} & \\
~~$^{46}$V & $^{46}$Ti & 2.61 & 0.0039 $\pm$ 0.0004 [Ha94] & & & 0.0039 $\pm$ 0.0004 & \\
 & & 4.32 & 0.0113 $\pm$ 0.0012 [Ha94] & & & 0.0113 $\pm$ 0.0012 & \\
 & & $\Sigma$GT\footnotemark[2] & $<$0.004 & & & 0$^{+0.004}_{-0}$ & \\
 & & {\bf gs} & & & & {\bf 99.9848}$\bm{^{+0.0013}_{-0.0042}}$ & \\
~~$^{50}$Mn & $^{50}$Cr & 3.63 & 0.057 $\pm$ 0.003 [Ha94] & & & 0.057 $\pm$ 0.003 & \\
 & & 3.85 & $<$0.0003 [Ha94] & & & $0^{+0.0003}_{-0}$ & \\
 & & 5.00 & 0.0007 $\pm$ 0.0001 [Ha94] & & & 0.0007 $\pm$ 0.0001 & \\
 & & {\bf gs} & & & & {\bf 99.9423 $\pm$ 0.0030} & \\
~~$^{54}$Co & $^{54}$Fe & 2.56 & 0.0045 $\pm$ 0.0006 [Ha94] & & & 0.0045 $\pm$ 0.0006 & \\
 & & $\Sigma$GT\footnotemark[2] & $<$0.03 & & & 0$^{+0.03}_{-0}$ & \\
 & & {\bf gs} & & & & {\bf 99.9955}$\bm{^{+0.0006}_{-0.0300}}$ & \\
~~$^{62}$Ga & $^{62}$Zn & $\Sigma$GT\footnotemark[2] & 0.15$^{+0.15}_{-0.05}$ [Hy03],[Bl02] & & & 0.15$^{+0.15}_{-0.05}$ & \\
 & & {\bf gs} & & & & {\bf 99.85}$\bm{^{+0.05}_{-0.15}}$ & \\
~~$^{74}$Rb & $^{74}$Kr & $\Sigma$GT\footnotemark[2] & 0.50 $\pm$ 0.10 [Pi03] & & & 0.50 $\pm$ 0.10 & \\
 & & {\bf gs} & & & & {\bf 99.50 $\pm$ 0.10} & \\[-4mm]

\footnotetext[1]{Result also incorporates data from Table~\ref{BDG}}
\footnotetext[2]{designates total Gamow-Teller transitions to levels not explicitly listed; values were derived
with the help of calculations in [Ha02].}
\end{tabular}
\end{ruledtabular}
\end{table*}
\endgroup

\begingroup
\squeezetable
\begin{table*}
\caption{Relative intensities of $\beta$-delayed $\gamma$-rays in the superallowed $\beta$-decay daughters.  These data
are used to determine some of the branching ratios presented in Table~\ref{R}.  (See Table~\ref{ref}
for the correlation between the alphabetical reference code used in this table and the actual reference numbers.)
\label{BDG}}
\begin{ruledtabular}
\begin{tabular}{llllllll}
\multicolumn{2}{c}{Parent/Daughter}
 & \multicolumn{1}{c}{daughter}
 & \multicolumn{2}{c}{Measured $\gamma$-ray Ratio}
 & \multicolumn{1}{c}{}
 & \multicolumn{2}{c}{Average value} \\[1mm]
\cline{4-6} 
\cline{7-8} 
   \multicolumn{2}{c}{nuclei}
 & \multicolumn{1}{c}{ratios\footnotemark[1]}
 & \multicolumn{1}{c}{1}
 & \multicolumn{1}{c}{2} 
 & \multicolumn{1}{c}{} 
 & \multicolumn{1}{c}{Ratio} 
 & \multicolumn{1}{c}{scale} \\

\hline
& & & & & & & \\[-4mm]
$^{18}$Ne & $^{18}$F & $\gamma_{660}/\gamma_{1042}$ & ~~~0.021 $\pm$ 0.003 ~~~[Ha75] & 0.0169 $\pm$ 0.0004 [He82] & & & \\
 & & & ~$\,$0.0172 $\pm$ 0.0005 ~$\,$[Ad83]& & & 0.0171 $\pm$ 0.0003 & 1.0 \\
$^{26}$Si & $^{26}$Al & $\gamma_{1622}/\gamma_{829}$ & ~~~0.149 $\pm$ 0.016 ~~~[Mo71] & ~$\,$0.134 $\pm$ 0.005 ~$\,$[Ha75]& & & \\
 & & & ~$\,$0.1245 $\pm$ 0.0023 ~$\,$[Wi80]& & & 0.1265 $\pm$ 0.0036 & 1.7 \\
 & & $\gamma_{1655}/\gamma_{829}$ & 0.00145 $\pm$ 0.00032 [Wi80] & & & 0.0015 $\pm$ 0.0003 & \\
 & & $\gamma_{1843}/\gamma_{829}$ & ~~~0.013 $\pm$ 0.003 ~~~[Mo71] & ~$\,$0.016 $\pm$ 0.003 ~$\,$[Ha75] & & & \\
 & & & 0.01179 $\pm$ 0.00027 [Wi80] & & & 0.0118 $\pm$ 0.0003 & 1.0 \\
 & & $\gamma_{2512}/\gamma_{829}$ &0.00282 $\pm$ 0.00010 [Wi80] & & & 0.0028 $\pm$ 0.0001 & \\
 & & $\gamma_{\rm total}/\gamma_{829}$ & & & & 0.1426 $\pm$ 0.0036 & \\
$^{30}$S & $^{30}$P & $\gamma_{709}/\gamma_{677}$ & ~~~0.006 $\pm$ 0.003 ~~~[Mo71] & 0.0037 $\pm$ 0.0009 [Wi80] & & 0.0039 $\pm$ 0.0009 & 1.0 \\
 & & $\gamma_{2341}/\gamma_{677}$ & ~~~0.033 $\pm$ 0.002 ~~~[Mo71] & 0.0290 $\pm$ 0.0006 [Wi80] & & 0.0293 $\pm$ 0.0011 & 1.9 \\
 & & $\gamma_{3019}/\gamma_{677}$ & 0.00013 $\pm$ 0.00006 [Wi80] & & & 0.0001 $\pm$ 0.0001 & \\
 & & $\gamma_{\rm total}/\gamma_{677}$ & & & & 0.0334 $\pm$ 0.0014 & \\
$^{34}$Ar & $^{34}$S & $\gamma_{461}/\gamma_{666}$ & ~~~~$\,$0.28 $\pm$ 0.16 ~~~~$\:$[Mo71] & ~$\,$0.365 $\pm$ 0.036 ~$\,$[Ha74a] & & ~$\,$0.361 $\pm$ 0.035 & 1.0 \\
 & & $\gamma_{2580}/\gamma_{666}$ & ~~~~$\,$0.38 $\pm$ 0.09 ~~~~$\:$[Mo71] & ~$\,$0.345 $\pm$ 0.01 ~~~[Ha74a] & & ~$\,$0.345 $\pm$ 0.010 & 1.0 \\
 & & $\gamma_{3129}/\gamma_{666}$ & ~~~~$\,$0.67 $\pm$ 0.08 ~~~~$\:$[Mo71] & ~$\,$0.521 $\pm$ 0.012 ~$\,$[Ha74a] & & ~$\,$0.524 $\pm$ 0.022 & 1.8 \\
 & & $\gamma_{\rm total}/\gamma_{666}$ & & & & ~$\,$1.231 $\pm$ 0.043 & \\
$^{42}$Ti & $^{42}$Sc & $\gamma_{2223}/\gamma_{611}$ & ~~~0.012 $\pm$ 0.004 ~~~[Ga69] & & & ~$\,$0.012 $\pm$ 0.004 & \\
 & & $\gamma_{\rm total}/\gamma_{611}$ & ~~~~~~~~2 $\times$ 0.012 ~~~$\,$[En90] & & & ~$\,$0.024 $\pm$ 0.008 & \\

\footnotetext[1]{$\gamma$-ray intensities are denoted by $\gamma_{E}$, where $E$ is the $\gamma$-ray energy in keV.}
\end{tabular}
\end{ruledtabular}
\end{table*}
\endgroup

\begingroup
\squeezetable
\begin{table*}
\caption{References for which the original decay-energy results have been updated to incorporate the most recent calibration standards.  (See Table~\ref{ref} for the correlation between the alphabetical reference code used in this table and the actual reference numbers.)
\label{update}}
\vskip 1mm
\begin{ruledtabular}
\begin{tabular}{lll}
  \multicolumn{1}{l}{References (parent nucleus)\footnotemark[1]}
& \multicolumn{1}{l}{}
& \multicolumn{1}{l}{Update procedure} \\
\hline
\\[-3mm]
\textbullet~Bo64$\,$($^{18}$Ne), Ba84$\,$($^{10}$C), Br94$\,$($^{26}$Al$^m$) &~~~~ & \textbullet~We have converted all original (p,n) threshold measurements to $Q$-values \\ 
 Ba98$\,$($^{10}$C), Ha98$\,$($^{38}$K$^m$), To03$\,$($^{14}$O) & &  using the most recent mass excesses [Au03]. \\[1mm]
\textbullet~Ry73a$\,$($^{34}$Cl), Ho74$\,$($^{54}$Co), Sq76$\,$($^{46}$V) & & \textbullet~These (p,n) threshold measurements have been adjusted to reflect recent \\
Ba77c$\,$($^{34}$Cl), Wh77$\,$($^{14}$O) & & calibration $\alpha$-energies [Ry91] before being converted to $Q$-values. \\[1mm]
\textbullet~Pr67$\,$($^{18}$Ne) & & \textbullet~Before conversion to a $Q$-value, this (p,n) threshold was adjusted to reflect a  \\
 & & new value for the $^7$Li(p,n) threshold [Wh85], which was used as calibration. \\[1mm]
\textbullet~Ja78$\,$($^{38}$K$^m$) & & \textbullet~This (p,n) threshold was measured relative to those for $^{10}$C and $^{14}$O; we have \\
 & & adjusted it based on average $Q$-values obtained for those decays in this work. \\[1mm]
\textbullet~Bu79$\,$($^{38}$K$^m$) & & \textbullet~Before conversion to a $Q$-value, this (p,n) threshold was adjusted to reflect the  \\
 & & modern value for the $^{35}$Cl(p,n) threshold [Au03], which was used as calibration. \\[1mm]
\textbullet~Bu61$\,$($^{14}$O), Ba62$\,$($^{14}$O) & & \textbullet~These $^{12}$C($^3$He,n) threshold measurements have been adjusted for updated \\
 & & calibration reactions based on current mass excesses [Au03]. \\[1mm]
\textbullet~Ha74d$\,$($^{34}$Cl, $^{50}$Mn, $^{54}$Co) & & \textbullet~These ($^3$He,t) reaction $Q$-values were calibrated by the $^{27}$Al($^3$He,t) reaction \\
 & & to excited states in $^{27}$Si; they have been revised according to modern mass \\
 & & excesses [Au03] and excited-state energies [En98]. \\[1mm]
\textbullet~Ki89$\,$($^{42}$Sc) & & \textbullet~This $^{41}$Ca(p,$\gamma$) reaction $Q$-value was measured relative to that for $^{40}$Ca(p,$\gamma$); \\
 & & we have slightly revised the result based on modern mass excesses [Au03]. \\[1mm]
\textbullet~Ha74c$\,$($^{22}$Mg, $^{26}$Si, $^{30}$S, $^{34}$Ar), & & \textbullet~These (p,t) reaction $Q$-values have been adjusted to reflect the current $Q$- \\
Se74$\,$($^{38}$Ca) & & value for the $^{16}$O(p,t) reaction [Au03], against which they were calibrated. \\[-4mm]    

\footnotetext[1]{These references all appear in Table~\ref{QEC} under the appropriate parent nucleus.}
\end{tabular}
\end{ruledtabular}
\end{table*}
\endgroup

\begingroup
\squeezetable
\begin{table*}
\caption{References from which some or all results have been rejected.  (See Table~\ref{ref} for the correlation between the alphabetical reference code used in this table and the actual reference numbers.)
\label{reject}}
\vskip 1mm
\begin{ruledtabular}
\begin{tabular}{llll}
  \multicolumn{2}{l}{References (parent nucleus)}
& \multicolumn{1}{l}{}
& \multicolumn{1}{l}{Reason for rejection} \\
\hline
1. & Decay-energies: &~~~~ & \\[1mm]
 & \textbullet~Pa72$\,$($^{30}$S, $^{38}$Ca) & & \textbullet~No calibration is given for the measured (p,t) reaction $Q$-values; update \\
 & & & is clearly required but none is possible. \\[1mm]
 & \textbullet~No74$\,$($^{22}$Mg) & & \textbullet~Calibration reaction $Q$-values have changed but calibration process is too \\
 & & & complex to update. \\[1mm]
 & \textbullet~Ro74$\,$($^{10}$C) & & \textbullet~P.H. Barker (co-author) later considered that inadequate attention had \\
 & & & been paid to target surface purity [Ba84]. \\[1mm]
 & \textbullet~Ba77b$\,$($^{10}$C) & & \textbullet~P.H. Barker (co-author) later stated [Ba84] that the (p,t) reaction $Q$-value \\
 & & & could not be updated to incorporate modern calibration standards. \\[1mm]
 & \textbullet~Wh81 and Ba98$\,$($^{14}$O) & & \textbullet~The result in [Wh81] was updated in [Ba98] but then eventually withdrawn \\
 & & & by P.H. Barker (co-author) in [To03]. \\[2mm]  
2. & Half-lives: & & \\[1mm]
 & \textbullet~Ja60$\,$($^{26}$Al$^m$), He61$\,$($^{14}$O), Ba62$\,$($^{14}$O),  & & \textbullet~Quoted uncertainties are too small, and results likely biased, in light of  \\
 & Ea62$\,$($^{10}$C), Ba63$\,$($^{10}$C), Fr63$\,$($^{14}$O, $^{26}$Si), & &  statistical difficulties more recently understood (see [Fr69a]).  In particular, \\
 & Fr65b$\,$($^{42}$Sc, $^{46}$V, $^{50}$Mn), Si72$\,$($^{14}$O) & & ``maximum-likelihood" analysis was not used. \\[1mm]
 & \textbullet~Ha72a$\,$($^{26}$Al$^m$, $^{34}$Cl, $^{38}$K$^m$, $^{42}$Sc) & & \textbullet~All four quoted half-lives are systematically higher than more recent and \\
 & & & accurate measurements. \\[1mm]
 & \textbullet~Ro74$\,$($^{10}$C) & & \textbullet~P.H. Barker (co-author) later considered that pile-up had been \\
 & & & inadequately accounted for [Ba90]. \\[1mm]
 & \textbullet~Ch84$\,$($^{38}$K$^m$) & & \textbullet~``Maximum-likelihood" analysis was not used.  \\[2mm]
3. & Branching-ratios: & & \\[1mm]
 & \textbullet~Fr63$\,$($^{26}$S)& & \textbullet~Numerous impurities present; result is obviously wrong. \\

\end{tabular}
\end{ruledtabular}
\end{table*}
\endgroup

\begingroup
\squeezetable
\begin{table*}
\caption{Reference key, relating alphabetical reference codes used in Tables~\ref{QEC}-\ref{t1/2} to the actual reference numbers.
\label{ref}}
\vskip 1mm
\begin{ruledtabular}
\begin{tabular}{llllllllllll}
  Table & Reference & Table & Reference & Table & Reference & Table & Reference & Table & Reference & Table & Reference \\ 
  code & number & code & number & code & number & code & number & code & number & code & number \\
\hline
  Ad83  & \cite{Ad83}  & Ba98  & \cite{Ba98}  & Dr75  & \cite{Dr75}  & He61  & \cite{He61}  & Mc67  & \cite{Mc67}  & Se74  & \cite{Se74} \\
  Aj88  & \cite{Aj88}  & Ba00  & \cite{Ba00}  & Ea62  & \cite{Ea62}  & He81  & \cite{He81}  & Mi67  & \cite{Mi67}  & Se04  & \cite{Se04} \\
  Aj91  & \cite{Aj91}  & Ba01  & \cite{Ba01}  & En90  & \cite{En90}  & He82  & \cite{He82}  & Mo71  & \cite{Mo71}  & Sh55  & \cite{Sh55} \\
  Al69  & \cite{Al69}  & Ba04  & \cite{Ba04}  & En98  & \cite{En98}  & He02  & \cite{He02}  & Mu04  & \cite{Mu04}  & Si66  & \cite{Si66} \\ 
  Al70  & \cite{Al70}  & Ba05  & \cite{Ba05}  & Fr63  & \cite{Fr63}  & Ho64  & \cite{Ho64}  & Na91  & \cite{Na91}  & Si72  & \cite{Si72} \\ 
  Al72  & \cite{Al72}  & Be68  & \cite{Be68}  & Fr65b & \cite{Fr65b} & Ho74  & \cite{Ho74}  & Ni69  & \cite{Ni69}  & Sq75  & \cite{Sq75} \\ 
  Al75  & \cite{Al75}  & Be78  & \cite{Be78}  & Fr69a & \cite{Fr69a} & Hu82  & \cite{Hu82}  & No74  & \cite{No74}  & Sq76  & \cite{Sq76} \\ 
  Al77  & \cite{Al77}  & Be85  & \cite{Be85}  & Fr75  & \cite{Fr75}  & Hy03  & \cite{Hy03}  & Oi01  & \cite{Oi01}  & Ti95  & \cite{Ti95} \\ 
  Al78  & \cite{Al78}  & Bi03  & \cite{Bi03}  & Fu99  & \cite{Fu99}  & Ia03  & \cite{Ia03}  & Pa72  & \cite{Pa72}  & To03  & \cite{To03} \\ 
  Al82  & \cite{Al82}  & Bl02  & \cite{Bl02}  & Ga69  & \cite{Ga69}  & In77  & \cite{In77}  & Pi03  & \cite{Pi03}  & Vo77  & \cite{Vo77} \\
  An70  & \cite{An70}  & Bl04a & \cite{Bl04a} & Ga01  & \cite{Ga01}  & Is80  & \cite{Is80}  & Pr67  & \cite{Pr67}  & Wa83  & \cite{Wa83} \\
  As70  & \cite{As70}  & Bl04b & \cite{Bl04b} & Go72  & \cite{Go72}  & Ja60  & \cite{Ja60}  & Pr90  & \cite{Pr90}  & Wa92  & \cite{Wa92} \\
  Au03  & \cite{Au03}  & Bo64  & \cite{Bo64}  & Gi72  & \cite{Gi72}  & Ja78  & \cite{Ja78}  & Ra83  & \cite{Ra83}  & We68  & \cite{We68} \\ 
  Az74  & \cite{Az74}  & Br94  & \cite{Br94}  & Ha67  & \cite{Ha67}  & Ka68  & \cite{Ka68}  & Re85  & \cite{Re85}  & Wh77  & \cite{Wh77} \\ 
  Az75  & \cite{Az75}  & Bu61  & \cite{Bu61}  & Ha68  & \cite{Ha68}  & Ka69  & \cite{Ka69}  & Ro70  & \cite{Ro70}  & Wh81  & \cite{Wh81} \\ 
  Ba62  & \cite{Ba62}  & Bu79  & \cite{Bu79}  & Ha72a & \cite{Ha72a} & Ke04  & \cite{Ke04}  & Ro72  & \cite{Ro72}  & Wh85  & \cite{Wh85} \\
  Ba63  & \cite{Ba63}  & Bu88  & \cite{Bu88}  & Ha74a & \cite{Ha74a} & Ki89  & \cite{Ki89}  & Ro74  & \cite{Ro74}  & Wi76  & \cite{Wi76} \\
  Ba67  & \cite{Ba67}  & Ca05  & \cite{Ca05}  & Ha74b & \cite{Ha74b} & Ki91  & \cite{Ki91}  & Ro75  & \cite{Ro75}  & Wi78  & \cite{Wi78} \\
  Ba77a & \cite{Ba77a} & Ch84  & \cite{Ch84}  & Ha74c & \cite{Ha74c} & Ko83  & \cite{Ko83}  & Ry73a & \cite{Ry73a} & Wi80  & \cite{Wi80} \\
  Ba77b & \cite{Ba77b} & Cl73  & \cite{Cl73}  & Ha74d & \cite{Ha74d} & Ko87  & \cite{Ko87}  & Ry91  & \cite{Ry91}  & Zi72  & \cite{Zi72} \\
  Ba77c & \cite{Ba77c} & Da79  & \cite{Da79}  & Ha75  & \cite{Ha75}  & Ko97a & \cite{Ko97a} & Sa80  & \cite{Sa80}  & Zi87  & \cite{Zi87} \\
  Ba84  & \cite{Ba84}  & Da80  & \cite{Da80}  & Ha94  & \cite{Ha94}  & Ko97b & \cite{Ko97b} & Sa95  & \cite{Sa95}  & & \\
  Ba88  & \cite{Ba88}  & Da85  & \cite{Da85}  & Ha98  & \cite{Ha98}  & Kr91  & \cite{Kr91}  & Sa04  & \cite{Sa04}  & & \\
  Ba89  & \cite{Ba89}  & De69  & \cite{De69}  & Ha02  & \cite{Ha02}  & Li94  & \cite{Li94}  & Sc05  & \cite{Sc05}  & & \\
  Ba90  & \cite{Ba90}  & De78  & \cite{De78}  & Ha03  & \cite{Ha03}  & Ma94  & \cite{Ma94}  & Se73  & \cite{Se73}  & & \\

\end{tabular}
\end{ruledtabular}
\end{table*}
\endgroup

There are two cases, $^{26}$Al$^m$ and $^{38}$K$^m$, in which the superallowed decay originates from an isomeric state.  For both,
there are $Q_{EC}$-value measurements that correspond to the ground state as well as to the isomer. Obviously, these two sets of
measurements are simply related 
to one another by the excitation energy of the isomeric state in the parent.  In Table~\ref{QEC},
the set of measurements for the ground-state $Q_{EC}$-value and for the excitation energy of the isomeric state appear in separate rows, each
with its identifing property given in column 3 and its weighted average appearing in column 7.  In the row below, the average value
given in column 7 for the superallowed transition is the weighted average not only of the direct superallowed $Q_{EC}$-value
measurements in that row, but also of the result derived from the two preceeding rows.  
Note that in all cases the $Q_{EC}$-value
for the superallowed transition appears in bold-face type.

For those 11 superallowed decays that lead to radioactive daughter nuclei, there are very few direct measurements of the
$Q_{EC}$-value for the superallowed transition.  In general, that $Q_{EC}$-value must be deduced from the measured mass excesses
of the parent and daughter nuclei, together with the excitation energy of the analog 0$^+$ state in the daughter.  Each of these
properties is identified in column 3 of Table~\ref{QEC}, with the individual measurements of that property, their weighted average
and a scale factor appearing in columns to the right.  The average $Q_{EC}$-value listed for the corresponding superallowed
transition is obtained from these separate averages.  If a direct measurement of the superallowed $Q_{EC}$-value exists, then it is
also included in the final average.

Especially in these latter 11 cases, it might be imagined that it would have been sufficient for us to use the 2003 Mass
tables \cite{Au03} to derive the $Q_{EC}$-values of interest.  There are, however, significant differences in our approach.  We have
included all pertinent measurements for each property as described in section~\ref{eval}; typically, only a subset of the
available data is included as input to the mass tables.  
Furthermore, we have examined each reference in detail and either accepted the result, updated it
to modern calibration standards or rejected it for cause.  The updating procedures are outlined, reference by reference, in 
Table~\ref{update} and the rejected results are similarly documented in Table~\ref{reject}.  With a comparatively small data set,
we could afford to pay the kind of individual attention that is impossible when one is considering all nuclear masses.

The half-life data appear in Table~\ref{t1/2} in similar format to Table~\ref{QEC}.  For obvious reasons, half-life measurements do not lend themselves
to being updated.  Consequently, a number of mostly pre-1970 measurements have been rejected because they were not analyzed with the
``maximum-likelihood" method.  The importance of using this technique for precision measurements was not recognized until that
time \cite{Fr69a} and, without access to the primary data, there is no way a new analysis can be applied retroactively.  All
rejected half-life measurements are also documented in Table~\ref{reject}.

Finally, the branching-ratio measurements are presented in Table~\ref{R}.  The decays of the $T_z = 0$ parents are the most
straightforward since, in all these cases, the superallowed branch accounts for $>$99.5\% of the total decay strength.  Thus,
even imprecise measurements of the weak non-superallowed branches can be subtracted from 100\% to yield the superallowed
branching ratio with good precision.  For the higher-Z parents of this type, particularly $^{62}$Ga and heavier, it has
been shown theoretically \cite{Ha02} and experimentally \cite{Pi03, Hy03} that numerous very-weak Gamow-Teller transitions occur,
which, in total, can carry significant decay strength.  Where such unobserved transitions are expected to exist, we have used
a combination of experiment and theory to account for the unobserved strength, with uncertainties being adjusted accordingly.

The branching ratios for decays from $T_z = -1$ parents are much more challenging to determine, since the superallowed branch
is usually one of several strong branches -- with the notable exception of $^{14}$O -- and, in two of the measured cases, it
actually has a branching ratio of less than 10\%.  The decays of $^{18}$F, $^{26}$Si, $^{30}$S, $^{34}$Ar and $^{42}$Ti required
special treatment in our presentation.  In each of these five cases, the absolute branching ratio for a single $\beta$-transition
has been measured. The branching ratios for other $\beta$-transitions must then be determined from the relative intensities of
$\beta$-delayed $\gamma$-rays in the daughter.  The relevant $\gamma$-ray intensity measurements appear in Table~\ref{BDG}, with
their averages then being used to determine the superallowed branching-ratio averages shown in bold type in Table~\ref{R}.  These
cases are also flagged with a table footnote. 

\section{The $\F t$ values}
\label{s:Ft}

Having surveyed the primary experimental data, we now turn to
producing a set of $ft$-values for the twenty superallowed
transitions being considered.  The statistical rate function, $f$, 
for each transition depends primarily on the charge of the
daughter nucleus, $Z$, and on the $Q_{EC}$-value to the fifth power.
Consequently the uncertainty in the value of $f$ due
to the experimental uncertainty in $Q_{EC}$ is given by
$(\Delta f/f) \approx 5 (\Delta Q_{EC}/Q_{EC})$.  Our goal in
computing $f$ therefore is to ensure that the computation itself
yields percentage errors much less than those due to the uncertainty in the
$Q_{EC}$-value, which can be $<$0.02\% in the best cases.  To this end we have
written a new code, the details of which are given in Appendix \ref{s:srf}.  Our
final $f$-values and their uncertainties are recorded in column
two of Table~\ref{t:tab5}.

\begingroup
\squeezetable
\begin{table*}[t]
\begin{center}
\caption{Derived results for superallowed Fermi beta decays.
\label{t:tab5}}
\vskip 1mm
\begin{ruledtabular}
\begin{tabular}{rrrrrrrr}
& & & & & & & \\[-3mm]
Parent & & $P_{EC}$ &  Partial half-life & &
&  &  \\
nucleus & \multicolumn{1}{c}{$f$} & (\%) &  \multicolumn{1}{c}{$t(ms)$} &  
\multicolumn{1}{c}{$ft(s)$} &
\multicolumn{1}{c}{$\delta_R^{\prime}$ (\%)} & $\delta_C - \delta_{NS}$ (\%) &
 \multicolumn{1}{c}{$\F t (s)$}  \\[1mm] 
\hline
& & & & & & & \\[-3mm]
\multicolumn{2}{l}{$T_z = -1$:} & & & & & & \\
$^{10}$C & $2.3009 \pm 0.0012$ & 0.297 & $1321000 \pm 1900$ & $3039.5 \pm 4.7$~~&
$1.652 \pm 0.004$ & $0.540 \pm 0.039$ & $3073.0 \pm 4.9$~  \\
$^{14}$O & $42.772 \pm 0.024$~ & 0.088 & $71151 \pm 16$~~~ & $3043.3 \pm 1.9$~ &
$1.520 \pm 0.008$ & $0.570 \pm 0.056$ & $3071.9 \pm 2.6$~ \\
$^{18}$Ne & $134.48 \pm 0.15$~~ & 0.081 & $21730 \pm 590$~ & $2922 \pm 80$~~ &
$1.484 \pm 0.012$ & $0.910 \pm 0.047$ & $2938 \pm 80$~~ \\
$^{22}$Mg & $418.44 \pm 0.18$~~ & 0.069 & $7295 \pm 17$~~~ & $3052.4 \pm 7.2$~ &
$1.446 \pm 0.017$ & $0.505 \pm 0.024$ & $3080.9 \pm 7.4$~ \\
$^{26}$Si & $1023.3 \pm 3.7$~~~  & 0.064 & $2978 \pm 40$~~~ & $3047 \pm 42$~~ &
$1.420 \pm 0.023$ & $0.600 \pm 0.024$ & $3072 \pm 42$~~ \\
$^{30}$S  & $1967.1 \pm 7.8$~~~  & 0.066 & $1524 \pm 21$~~~ & $2998 \pm 44$~~ &
$1.405 \pm 0.029$ & $1.125 \pm 0.039$ & $3006 \pm 44$~~ \\
$^{34}$Ar & $3414.2 \pm 1.5$~~~  & 0.069 & $896.0 \pm 3.5$~~ & $3059 \pm 12$~~ &
$1.394 \pm 0.035$ & $0.825 \pm 0.044$ & $3076 \pm 12$~~ \\
$^{38}$Ca & $5338 \pm 22$~~~~ & 0.075 &     &     & 
$1.397 \pm 0.042$ & $0.910 \pm 0.053$ &  \\
$^{42}$Ti & $7043 \pm 30$~~~~ & 0.088 & $470 \pm 160$~ & $3300 \pm 1100$ &
$1.412 \pm 0.050$ & $1.015 \pm 0.110$ & $3300 \pm 1100$ \\[5mm]
\multicolumn{2}{l}{$T_z = 0 $:} & & & & & & \\
$^{26}$Al$^m$ & $478.20 \pm 0.11$~~ & 0.082 & $6350.2 \pm 1.9$~~ & $3036.7 \pm 1.2$~ &
$1.458 \pm 0.020$ & $0.261 \pm 0.024$ & $3072.9 \pm 1.5$~ \\
$^{34}$Cl & $1996.39 \pm 0.41$~~ & 0.080 & $1527.99 ^{+0.44}_{-0.47}$~~ & $3050.5 \pm 1.1$~ & 
$1.425 \pm 0.032$ & $0.720 \pm 0.039$  & $3071.7 \pm 1.9$~ \\
$^{38}$K$^m$ & $3298.10 \pm 0.33$~~ & 0.085 & $925.11 \pm 0.27$~ & $3051.1 \pm 1.0$~~&
$1.423 \pm 0.039$ & $0.720 \pm 0.047$ & $3072.2 \pm 2.1$~ \\
$^{42}$Sc & $4470.03 \pm 1.46$~~ & 0.099 & $681.43 \pm 0.26$~ & $3046.0 \pm 1.5$~ &
$1.437 \pm 0.047$ & $0.460 \pm 0.047$ & $3075.6 \pm 2.5$~ \\
$^{46}$V & $7200.0 \pm 5.0$~~~ & 0.101 & $422.99 \pm 0.11$~ & $3045.5 \pm 2.2$~ &
$1.429 \pm 0.054$ & $0.465 \pm 0.033$ & $3074.7 \pm 3.0$~ \\
$^{50}$Mn & $10731.2 \pm 1.8$~~~ & 0.107 & $283.71 \pm 0.13$~ & $3044.5 \pm 1.5$~ &
$1.429 \pm 0.062$ & $0.547 \pm 0.037$ & $3071.1 \pm 2.7$~ \\
$^{54}$Co & $15749.3 \pm 3.0$~~~ & 0.111 & $193.495 ^{+0.063}_{-0.086}$~ &$3047.4 ^{+1.2}_{-1.5}~~\,$ &
$1.428 \pm 0.071$ & $0.639 \pm 0.043$ & $3071.2 \pm 2.8$~ \\
$^{62}$Ga & $26250 \pm 400$~~ & 0.137 & $116.509 ^{+0.070}_{-0.179}$~ & $3058 \pm 47$~~ & 
$1.445 \pm 0.087$ & $1.42 \pm 0.16$~ & $3058 \pm 47$~~ \\
$^{66}$As & $31610 \pm 890$~~ & 0.156 &       &     & 
$1.457 \pm 0.095$ & $1.45 \pm 0.16$~ &       \\
$^{70}$Br & $38600 \pm 3600$~ & 0.175 &       &    &
$1.47 \pm 0.11$~ & $1.41 \pm 0.20$~ &       \\
$^{74}$Rb & $47280 \pm 100$~~ & 0.194 & $65.227 \pm 0.078$ & $3083.8 \pm 7.5$~ &
$1.49 \pm 0.12$~ & $1.50 \pm 0.41$~ & $3083 \pm 15$~~ \\[5mm]
& & & & & \multicolumn{2}{r}{Average (best 12), $\overline{\F t}$} & $3072.7 \pm 0.8$~ \\
& & & & & \multicolumn{2}{r}{$\chi^2/\nu$} & \multicolumn{1}{c}{0.42} \\
\end{tabular}
\end{ruledtabular}
\end{center}
\end{table*}
\endgroup

The partial half-life, $t$, for each transition is obtained from its total
half-life, $t_{1/2}$, and branching ratio, $R$, according to
the formula
\be
t = \frac{t_{1/2}}{R} \left ( 1 + P_{EC} \right )
\label{partialt}
\ee
where $P_{EC}$ is the calculated electron-capture fraction.  The
evaluation of $P_{EC}$ is discussed by Bambynek \etal \cite{Ba77}
and is based on the equation
\be
P_{EC} = \frac{1}{2} \pi \left [ \sum_x \beta_x^2
\left ( W_{EC} - |W_x| \right )^2 B_x \right ] / f
\label{pec}
\ee
The sum extends over all atomic subshells from which an electron can be
captured.  The factor $\beta_x$ is the Coulomb amplitude of the 
appropriate bound-state electron radial wave function; $W_{EC}$ is the
$Q_{EC}$-value expressed in electron rest-mass units; $W_x$ is
the $x$-subshell binding energy also in electron rest-mass units; and
$B_x$ takes account of the effects of electron exchange and overlap.
We have computed $P_{EC}$ for the cases of interest here using our
$Q_{EC}$-values from Table~\ref{QEC} and the values
of $\beta_x^2 B_x$ and $W_x$ from, respectively, Tables 1 and 2 of
Appendix F in ref.~\cite{Fi96}.
The $P_{EC}$ results are shown (as percentages) in column
three of Table~\ref{t:tab5}.  Based on experimental tests of such
$P_{EC}$ calculations \cite{Ba77}, we expect these results to be
accurate to a few parts in 100; thus they do not contribute perceptibly
to the overall uncertainties.  Partial half-lives derived from 
Eq.~(\ref{partialt}), and corresponding $ft$ values appear in
columns four and five. 

To obtain $\F t$-values according to Eq.~(\ref{Ftfactor}) we must now
deal with the small correction terms. The term $\delta_R^{\prime}$
has been calculated from standard QED, and is currently evaluated to
order $Z \alpha^2$ and estimated in order $Z^2 \alpha^3$ \cite{Si87,JR87};
its values, listed in column six of Table~\ref{t:tab5}, are around 1.4\%
and can be considered to be very reliable.  The
structure-dependent terms $\delta_{NS}$ and $\delta_{C}$, have also
been calculated in the past but at various times over three decades and
with a variety of different models.  Their uncertainties are
larger.  This topic has been reviewed recently by Towner and Hardy \cite{TH02},
who presented new calculations of these corrections in which consistent model
spaces and approximations have been used for both correction terms.  The results
of these new calculations are recorded in column seven of Table~\ref{t:tab5}.
Finally, the resulting $\F t$-values are listed in column eight.

\subsection{CVC test}
\label{ss:cv}

\begin{figure*}[t]
\epsfig{file=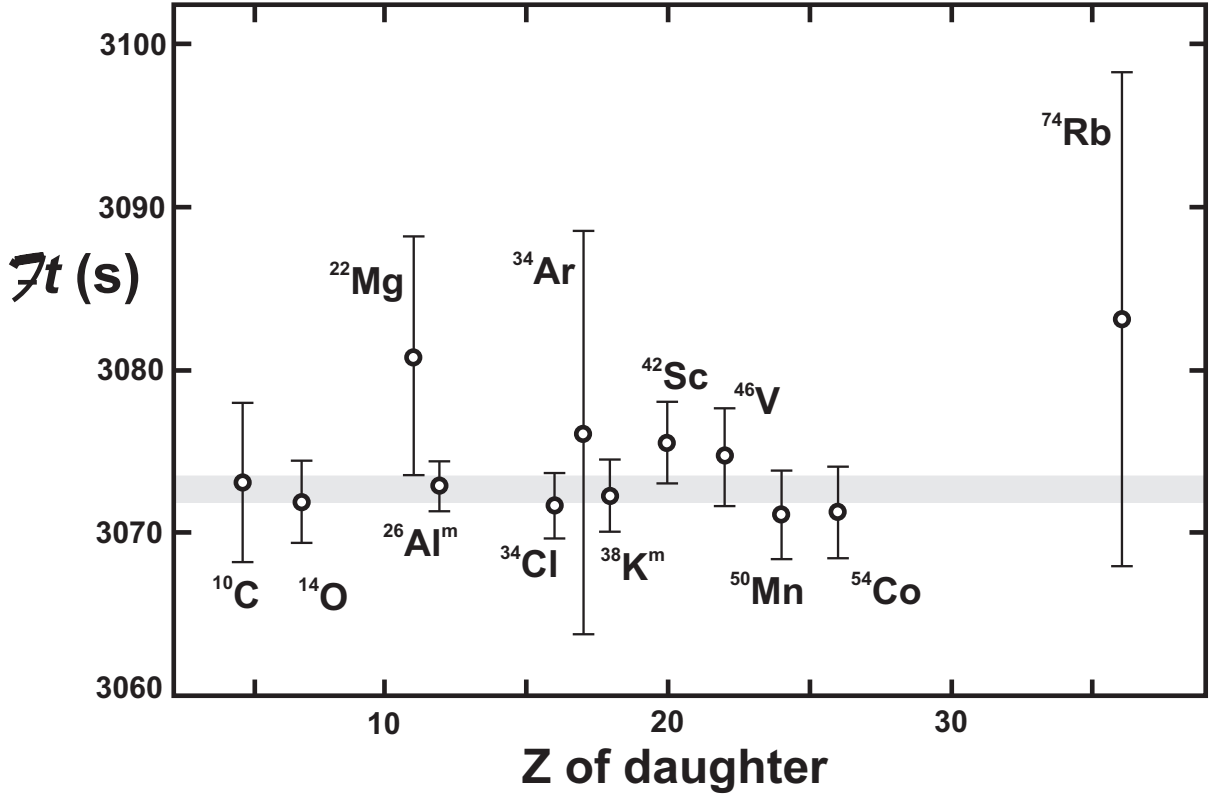,width=16cm}
\caption{$\F t$ values plotted as a function of the charge on the
daughter nucleus, $Z$.  The shaded horizontal band gives one standard
deviation around the
average $\overline{\F t}$ value, Eq.~(\protect\ref{Ftavgstat}).}
\label{f:Ftvalues}
\end{figure*}

We are now ready to test the CVC assertion that the $\F t$ values
should be constant for all nuclear superallowed transitions of this type.
The data in Table~\ref{t:tab5}
clearly satisfy the test; the weighted average of the 12 most-precise results
(with ``statistical" uncertainty only) is
\be
\overline{\F t} = 3072.7 \pm 0.8 s
\label{Ftavgstat}
\ee
with a corresponding chi-square per degree of freedom of $\chi^2/\nu = 0.42$.  In 
Fig.~\ref{f:Ftvalues} we plot the same 12 values, all of
whose statistical accuracy is better than 0.5\%.  It is evident
from both the figure and the table that the data form a consistent set, thus verifying
the expectation of the CVC hypothesis at the level of $3 \times 10^{-4}$,
which is the fractional uncertainty quoted in Eq.~(\ref{Ftavgstat}).
This is a 30\% improvement over the results from our last survey
in 1990 \cite{HT90} and is principally due to improvements
in the experimental data themselves.

\begin{figure*}[t]
\epsfig{file=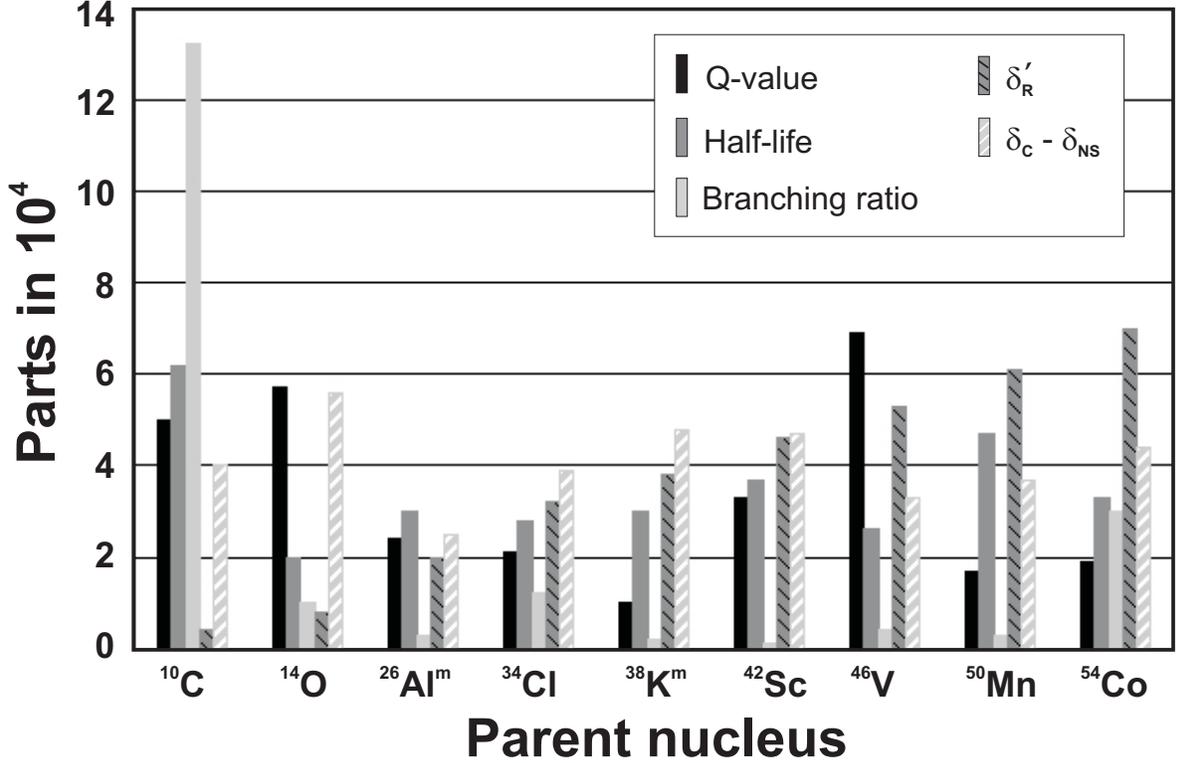,width=16cm}
\caption{Summary histogram of the fractional uncertainties attributable to each 
experimental and theoretical input factor that contributes to the final
$\protect\F t$ values for the nine most precise superallowed decay data.}
\label{f:hist9}
\end{figure*}

\begin{figure*}[t]
\epsfig{file=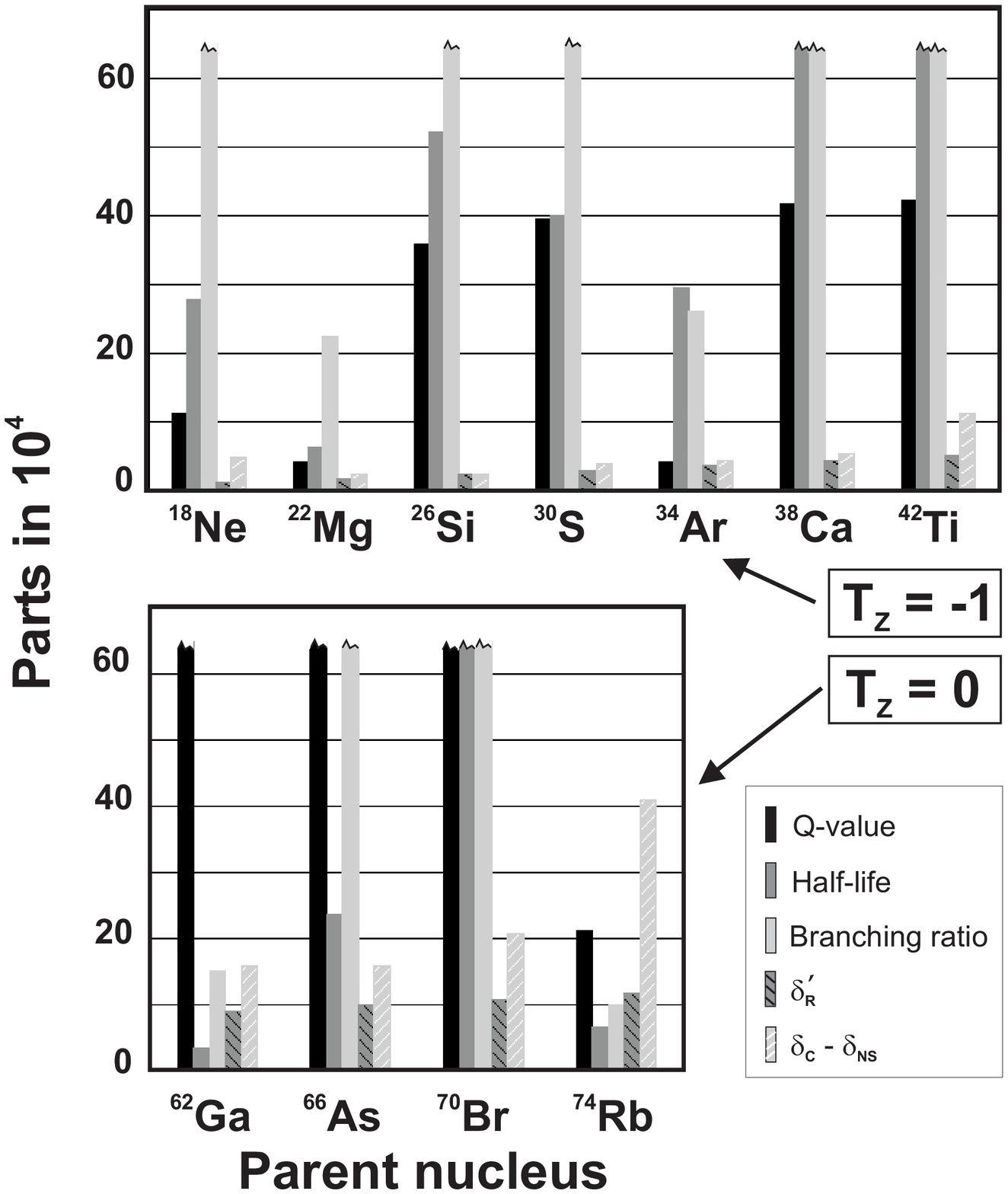,width=10cm}
\caption{Summary histogram of the fractional uncertainties attributable to each 
experimental and theoretical input factor that contributes to the final
$\protect\F t$ values for the twelve other superallowed decay data.  Where
the error is shown as exceeding 60 parts in $10^4$, no useful experimental
measurement has been made.}
\label{f:hist12}
\end{figure*}

\subsection{$\F t$-value error budgets}
\label{ss:su}

In Figs.~\ref{f:hist9} and \ref{f:hist12} we show the contributing 
factors to the individual $\F t$-value uncertainties.  For the
most precise data, $^{14}$O to $^{54}$Co, which appear in Fig.~\ref{f:hist9},
the theoretical uncertainties are greater than,
or comparable to, the experimental ones.  The nuclear-structure-dependent
correction, $\delta_C - \delta_{NS}$, contributes an almost constant
uncertainty of 4 parts in $10^4$ across these nuclei, while the
nucleus-dependent radiative correction, $\delta_R^{\prime}$, has an
uncertainty that grows as $Z^2$.  This is because the contribution
to $\delta_R^{\prime}$ from order $Z^2 \alpha^3$ has only been
estimated from its leading logarithm \cite{Si87} and the magnitude
of this estimate has been taken as the uncertainty in
$\delta_R^{\prime}$.  In fact, for $^{50}$Mn and $^{54}$Co this becomes
the leading uncertainty, indicating that a closer look at the order 
$Z^2 \alpha^3$ contribution would now be worthwhile.  For the
eight precise data, the experimental branching ratios are
$> 99 \%$ and have very small associated uncertainties with the exception
of $^{54}$Co, which has a $3\times10^4$ fractional
uncertainty attributed to its branching ratio.  This is because $^{54}$Co
is predicted to have several weak Gamow-Teller branches that have not yet
been observed.  We have used an estimate of the strength of the missing
branches, taken from a shell-model calculation \cite{Ha02}, to assign an
uncertainty to the superallowed branching ratio.  Missing weak
branches become a larger issue for the heavier-mass nuclei with $A \geq 62$,
where they contribute significantly to the branching-ratio uncertainty.

For the less precisely known decay of $^{10}$C, and for the twelve decays
depicted in Fig.~\ref{f:hist12}, the predominant uncertainties
are all experimental in origin with the single exception of
$^{74}$Rb, for which the nuclear-structure calculation is quite
difficult \cite{TH02}, resulting in a larger uncertainty on
$\delta_C - \delta_{NS}$.  Many of the experimental $Q$-values,
half-lives and branching ratios have yet to be measured 
for the cases in Fig.~\ref{f:hist12}, but recent advances
in experimental techniques are likely to change this situation dramatically within
the next few years.

\subsection{Accounting for systematic uncertainties}
\label{ss:se}

So far, we have dealt with the inter-nuclear behavior of $\F t$-values,
examining their constancy as a test of CVC.  With that test passed
at high precision, we are now in a position to use the average
$\F t$-value obtained from these concordant nuclear data to go
beyond nuclei, obtaining first the vector coupling constant (see
Eq.~(\ref{Ftconst})) and then the $V_{ud}$ matrix element.  Before
doing so, however, we must address one more possible source of
uncertainty.  Though the average $\F t$ value given in Eq.~(\ref{Ftavgstat})
includes a full assessment of the uncertainties attributable to
experiment and to the particular calculations used to obtain the
correction terms, it does not incorporate any provision for a
common systematic error that could arise from the type of
calculation chosen to model the nuclear-structure effects.  In
this section we look more critically at the
nuclear-structure-dependent corrections, and in particular at
the isospin-symmetry-breaking correction\footnote{The reason
we do not consider further the nuclear-structure-dependent
radiative correction, $\delta_{NS}$, is that it is very small for
the series of transitions that have $T_z=0$ parent states \cite{TH02}.
Of the nine precisely known transitions we are concentrating on, seven
are of this type.}, $\delta_C$.  

There have been a
number of previous calculations of $\delta_C$ besides those of
ours \cite{TH02}:  Hartree-Fock calculations
of Ormand and Brown \cite{OB95}, RPA calculations of Sagawa, van
Giai and Suzuki \cite{SVS96}, R-matrix calculations of Barker \cite{Ba92},
and Woods-Saxon calculations of Wilkinson \cite{Wi02}, to name 
some of the more recent publications.  Of these, we will only retain
the Ormand-Brown (OB) calculations since they, like ours (TH),
are constrained to reproduce other isospin properties of the nuclei
involved:  They reproduce the measured coefficients
of the relevant isobaric multiplet mass equation, the known 
proton and neutron separation energies, and the measured $ft$ values
of weak non-analog $0^+ \rightarrow 0^+$ transitions \cite{Ha94},
where they are known.  The other calculations are not constrained
by experiment in any way and thus offer no independent means to assess
their efficacy.

Unfortunately, calculations of $\delta_C$ by OB
are not available for all the cases listed in Table~\ref{t:tab5}, so we 
must concentrate on the nine most precise data:  $^{10}$C, $^{14}$O,
$^{26}$Al$^m$, $^{34}$Cl, $^{38}$K$^m$, $^{42}$Sc, $^{46}$V,
$^{50}$Mn, and $^{54}$Co.  When the TH values of $\delta_C$ are used, the
average $\F t$-value for these nine cases alone is $\overline{\F t} = 3072.6 \pm 0.8$ s
with $\chi^2/\nu = 0.35$.  When 
OB values are used for $\delta_C$ instead, the weighted average is $\overline{\F t} =
3074.5 \pm 0.8$ s with $\chi^2/\nu = 0.92$.  Although the chi-square
with the OB values is nearly a factor of three worse,
we do not argue that this is sufficient reason to reject the OB calculation.
Rather, we observe that the OB values of $\delta_C$ are systematically
smaller and hence the $\F t$ values systematically larger than ours.
Evidently there is a systematic difference between our Woods-Saxon
and OB's Hartree-Fock calculations of $\delta_C$ and that difference
should be accounted for in the final result.  Thus, we adopt the average
of these two results for our recommended $\overline{\F t}$, and assign
a systematic uncertainty equal to half the spread between them: {\it viz}
\bea
\overline{\F t} & = & 3073.5 \pm 0.8_{\rm stat} \pm 0.9_{\rm syst}~s
\nonumber \\
& = & 3073.5 \pm 1.2~s,
\label{Ftavg}
\eea 
where the two errors have been combined in quadrature.

\section{Impact on weak-interaction physics}
\label{s:iwip}

\subsection{The Value of $V_{ud}$}
\label{ss:vVud}

With a mutually consistent set of $\F t$ values, we can now use their
average value in Eq.~(\ref{Ftavg}) to determine the vector coupling constant,
$\GV$, from Eq.~(\ref{Ftconst}).  The value of $\GV$ itself
is of little interest, but it can be related to the weak
interaction constant for the purely leptonic muon decay, $\GF$, to yield
the much more interesting 
up-down matrix element of the Cabibbo-Kobayashi-Maskawa (CKM) 
quark-mixing matrix\footnote{More completely we could write
$\GV = \GF V_{ud} \gV (q^2 \rightarrow 0)$, where $\gV$ is the
vector form factor given in Eq.~(\protect\ref{VuAu}), or as 
$\GV = \GF V_{ud} C_V$, where $C_V$ is the vector coupling constant
in the Jackson, Treiman and Wyld \protect\cite{JTW57} Hamiltonian in
Eq.~(\protect\ref{JTWHam}), with $\gV (q^2 \rightarrow 0) = C_V =1$.}:
$\GV = \GF V_{ud}$.  The relation we use is
\be
V_{ud}^2 = \frac{K}{2 \GF^2 (1 + \DRV ) \overline{\F t}},
\label{Vudeq}
\ee
where $\DRV$ is the nucleus-independent radiative correction.  The
currently accepted value for this correction is derived from
the expression
\cite{MS86,Si94}
\be
\DRV = \frac{\alpha}{2 \pi} \left [ 4 \ln (\mZ / m_p) + \ln (m_p / \mA)
+ 2C_{\rm Born} \right ] + \cdots ,
\label{DRVeq}
\ee
where the ellipses represent further small terms of order 0.1\%.
Here $\mZ$ is the $Z$-boson mass, $m_p$ the proton mass, $\mA$
the mass parameter in the dipole form of the axial-vector form
factor, and $C_{\rm Born}$ is the universal order-$\alpha$ 
axial-vector contribution.  The various terms have the values
\be
\DRV = 2.12 - 0.03 + 0.20 + 0.10 \%,
\label{DRVest}
\ee
with the first term, the leading logarithm, being essentially
unambiguous in value.  The final value recommended by Sirlin \cite{Si94} is
\be
\DRV = (2.40 \pm 0.08 ) \%.
\label{DRV}
\ee
The uncertainty is almost entirely due to the value selected for the
axial-vector form factor mass, which Sirlin argues should lie
in the range $(m_{a_1} /2) \leq \mA \leq 2 m_{a_1}$, where $m_{a_1}$
is the physical $a_1$ meson mass.

Using the Particle Data Group (PDG) \cite{PDG04} value for the
weak interaction coupling constant from muon decay of
$\GF /(\hbar c )^3 = (1.16639 \pm 0.00001) \times 10^{-5}$ GeV$^{-2}$,
we obtain from Eq.~(\ref{Vudeq}) the result
\be
|V_{ud}|^2 = 0.9482 \pm 0.0008.
\label{Vudsq}
\ee
Note that the total uncertainty here -- 0.00083, if the next significant
figure is included --  is almost entirely due to the uncertainties
contributed by the theoretical corrections.  By far the largest
contribution, 0.00074, arises from the uncertainty in $\DRV$; 0.00031
comes from the nuclear-structure-dependent corrections $\delta_C - \delta_{NS}$ 
(principally from the systematic difference between the OB and TH 
calculations discussed in Sect.~\ref{ss:se}) and 0.00012 is attributable to
$\delta_R^{\prime}$.  Only 0.00016 can be considered to be experimental in origin.

The corresponding value of $V_{ud}$ is
\be
|V_{ud}| = 0.9738 \pm 0.0004,
\label{Vudvalu}
\ee
a result that differs by two units in the last quoted digit from
our previously recommended result \cite{TH03}.  This shift, well within the quoted
one standard deviation, is due to the improvements in the experimental
data and to our re-computing of the statistical rate function
(see Appendix \ref{s:srf}), in which a number of different parameter 
choices were made for the charge-density distribution, the
oscillator length parameter for nuclear radial functions, and
for the screening correction.  Coincidentally, the value of $V_{ud}$ quoted
in Eq.~(\ref{Vudvalu}) is identical to the currently recommended
PDG value \cite{PDG04}, although our uncertainty
is one digit smaller.
  
\subsection{Unitarity of the CKM matrix}
\label{ss:uCKM}

The CKM matrix yields the transformation equations for a change of
basis from quark weak-interaction
eigenstates to quark mass eigenstates.  As such, the CKM matrix must be unitary
in order that the bases remain orthonormal.  
With the CKM matrix elements determined from experimental data,
one important test they should satisfy is that they yield a 
unitary matrix.  Currently, the sum of the squares of the top-row elements, which
should equal one, constitutes the most demanding available test.
With our experimental value for $|V_{ud}|^2$ given in Eq.~(\ref{Vudsq})
and the PDG's recommended values \cite{PDG04} of
$|V_{us}| = 0.2200 \pm 0.0026$ and $|V_{ub}| = 0.00367 \pm 0.00047$,
this unitarity test yields:
\be
|V_{ud}|^2 + |V_{us}|^2 + |V_{ub}|^2 = 0.9966 \pm 0.0014
\label{sumfail}
\ee
The test fails by 2.4 standard deviations.  Two-thirds of the
assigned error is attributed to the uncertainty in $|V_{us}|^2$, \viz
0.0011, and one-third to the error in $|V_{ud}|^2$, \viz  0.0008.
The latter, as we have already demonstrated,
is {\em not} predominantly experimental in origin, but is dominated
by the uncertainty in the nucleus-independent radiative
correction, $\DRV$.

A recent measurement of the $K^+ \rightarrow \pi^0 e^+ \nu_e ~ (
K_{e3}^+)$ branching ratio from the Brookhaven E865 experiment \cite{Sh03}
obtains $V_{us} = 0.2272 \pm 0.0030$.  Although this result is included
in the PDG average value, it is considerably higher than the older experimental
results from $K_{e3}^+$ and $K_{e3}^0$ decays, with which it is inconsistent.
Experiments now in progress should help clarify the situation.  If, for the
moment, we adopt the E865 value for $V_{us}$ rather than the PDG average, then the
result in Eq.~(\ref{sumfail}) is modified to
\be
|V_{ud}|^2 + |V_{us}|^2 + |V_{ub}|^2 = 0.9999 \pm 0.0016
\label{sumsucceed}
\ee
and unitarity is fully satisfied.

Another, at present, less demanding test is to examine the first column
of the CKM matrix.  The PDG value for $V_{cd}$ is $0.224 \pm 0.012$, but
little is known about $V_{td}$ other than it is expected to lie in the
range $0.0048 \leq V_{td} \leq 0.014$.  In this range it has negligible impact on the
unitarity sum.  With our value of $|V_{ud}|^2$, this unitarity sum
becomes
\be
|V_{ud}|^2 + |V_{cd}|^2 + |V_{td}|^2 = 0.9985 \pm 0.0054
\label{sum2succeed}
\ee
Here the error is given entirely by the uncertainty in the value of
$V_{cd}$ and unitarity is evidently satisfied at this level of
accuracy.

\subsection{Fundamental Scalar Interaction}
\label{ss:si}

For the past 40 years, the weak interaction has been described by an equal mix 
of vector and axial-vector interactions that maximizes parity violation.
The theory is known colloquially as the `$V-A$' theory.  Despite the 
ever increasing precision possible in weak-interaction experiments, no defect has
been found in the $V-A$ theory.  Prior to the establishment of the
$V-A$ theory, other forms of fundamental couplings, notably scalar and tensor
interactions, were popular.  Today there is still interest in
searching for scalar and tensor interactions, not because we expect
them to contribute importantly, but rather because we wish to
establish limits to their possible contribution.

A general form of the weak-interaction Hamiltonian was written
down by Jackson, Treiman and Wyld \cite{JTW57}.  In examining superallowed
Fermi transitions, we are only interested in scalar and vector
couplings, for which that Hamiltonian becomes  
\bea
H_{S+V} &  =  & (\overline{\psi}_p \psi_n)
        (C_S \overline{\phi}_e \phi_{\overline{\nu}_e}
        + C_S^{\prime} \overline{\phi}_e \gamma_5 \phi_{\overline{\nu}_e})
\nonumber \\
& &
 + (\overline{\psi}_p \gamma_{\mu} \psi_n)
        (C_V \overline{\phi}_e \gamma_{\mu} \phi_{\overline{\nu}_e}
+ C_V^{\prime} \overline{\phi}_e \gamma_{\mu}\gamma_5 \phi_{\overline{\nu}_e}) .
\label{JTWHam}
\eea
If we assume that the Hamiltonian is invariant under time reversal, then
all the coupling constants must be real.
Those coupling constants carrying a prime represent parity-nonconserving
interactions.  If we further assume that parity violation is
maximal, then
$C_i^{\prime} = C_i$. 
In this limit, the scalar and vector terms can be written
\bea
H_{S+V} & = &  \left ( \overline{\psi}_p C_S \psi_n \right )
\left ( \overline{\phi}_e (1 + \gamma_5 ) \phi_{\overline{\nu}_e}
\right )
\nonumber \\
& &
+ \left ( \overline{\psi}_p
C_V \gamma_{\mu}  \psi_n \right )
\left ( \overline{\phi}_e \gamma_{\mu} 
( 1 + \gamma_5 ) \phi_{\overline{\nu}_e} \right )  .
\label{JTWSV}
\eea
A nonrelativistic reduction of the hadron matrix element for the
scalar and the time part of the vector interaction shows that
they both reduce simply to constants in leading order.  However,
under charge conjugation the matrix element
$ ( \overline{\psi}_p C_S \psi_n ) $
changes sign relative to
$ ( \overline{\psi}_p
C_V \gamma_{4}  \psi_n ) $.
Thus we write $\pm C_S$ in the ensuing formulae with the upper sign 
being for electron emission and the lower sign for positron emission.
The lepton matrix elements are different in the two terms in Eq.~(\ref{JTWSV}) so
the contribution to the shape-correction function from the scalar
interaction will involve a different combination of electron
and neutrino radial functions than that from the vector interaction.
The final formula for $C(Z,W)$ is
\bea
C(Z,W) & = &
\sum_{k_e k_{\nu} K} \lambda_{k_e}
\Bigm{\{} \bigm{(} M_K(k_e,k_{\nu}) + \overline{m}_K(k_e,k_{\nu}) \bigm{)}^2
\nonumber \\
& &  
~~~~~~~~~~+ \bigm{(} m_K(k_e,k_{\nu}) + \overline{M}_K(k_e,k_{\nu}) \bigm{)}^2 
\nonumber \\
& &  
- \frac{2 \mu_{k_e} \gamma_{k_e}}{k_e W}  
\bigm{(} M_K(k_e,k_{\nu}) + \overline{m}_K(k_e,k_{\nu}) \bigm{)}  
\nonumber \\
& &  
~~~~~~~~~~\bigm{(} m_K(k_e,k_{\nu}) + \overline{M}_K(k_e,k_{\nu}) \bigm{)} 
\Bigm{\}} , 
\label{CWnew}
\eea
where $M_K(k_e,k_{\nu})$ and
$m_K(k_e,k_{\nu})$ are the reduced matrix elements given in
Eq.~(\ref{MmKkk}), which incorporate the radial functions, $F(r)$ and $f(r)$,
defined in Eq.~(\ref{Ffr3}).  The reduced matrix elements $\overline{M}_K(k_e,k_{\nu})$
and $\overline{m}_K(k_e,k_{\nu})$ are the same as $M_K(k_e,k_{\nu})$ and
$m_K(k_e,k_{\nu})$ except that the radial functions, $F(r)$ and $f(r)$, are
replaced by $\overline{F}(r)$ and $\overline{f}(r)$, where
\bea
\overline{F}(r) & = & H(r)
\left \{ - G_{--}~
j_{k_{\nu}-1}(p_{\nu}r) + G_{-+}~
j_{k_{\nu}}(p_{\nu}r) \right \}
\nonumber \\
& & + D(r)
\left \{ G_{+-}~
j_{k_{\nu}-1}(p_{\nu}r) - G_{++}~
j_{k_{\nu}}(p_{\nu}r) \right \}
\nonumber \\
\overline{f}(r) & = & h(r)
\left \{ - G_{--}~
j_{k_{\nu}-1}(p_{\nu}r) + G_{-+}~
j_{k_{\nu}}(p_{\nu}r) \right \}
\nonumber \\
& & + d(r)
\left \{ G_{+-}~
j_{k_{\nu}-1}(p_{\nu}r) - G_{++}
j_{k_{\nu}}(p_{\nu}r) \right \} .
\label{Ffr}
\eea
The functions $H$, $D$, $h$ and $d$ are
linear combinations of the electron functions, $f_{\kappa}(r)$ and
$g_{\kappa}(r)$, as given in Eq.~(\ref{HDhd}); and the angular
momentum factors $G_{\pm \pm }$ are defined in Eq.~(\ref{GKLs}).

\subsubsection{Order of magnitude estimates}
\label{sss:estimates}

For a pure Fermi transition, the multipolarity of the transition operators
is $K = 0$.  Keeping only the lowest lepton partial waves, $k_e = 1$ and
$k_{\nu} =1$, we expand the lepton radial functions in a power series in $r$.  The
order of magnitude of the lepton wave functions at small $r$ are
\bea
f_1(r) & = & 1 - \ldots
\nonumber \\
g_{-1}(r) & = & 1 - \ldots
\nonumber \\
f_{-1}(r) & = & small
\nonumber \\
g_{1}(r) & = & small
\nonumber \\
j_0(p_{\nu} r) & = & 1 - \ldots
\nonumber \\
j_1(p_{\nu} r) & = & small ,
\label{fgj}
\eea
We retain only $f_1(r)$, $g_{-1}(r)$ and $j_0(p_{\nu} r)$, setting their values
to unity, and drop the other small terms.  The angular momentum factors for $K=L=s=0$ are
$G_{++} = G_{--} = 1$, and
$G_{+-} = G_{-+} = 0$.
Then the amplitudes become
\bea
M_0(1,1) & = & C_V M_F + \ldots
\nonumber \\
m_0(1,1) & = & {\rm small} ,
\nonumber \\
\overline{M}_0(1,1) & = & \mp C_S M_F + \ldots
\nonumber \\
\overline{m}_0(1,1) & = & {\rm small},
\label{Mmest}
\eea
where $M_F$ is the Fermi matrix element.  The shape-correction function is then
\bea
C(Z,W) & = & |M_F|^2 \left ( C_V^2 + C_S^2 \pm \frac{2 \mu_1 \gamma_1}{W}
C_S C_V \right )
\nonumber \\
& \simeq & |M_F|^2 C_V^2 \left ( 1 + b_F \gamma_1/W + \ldots \right ) ,
\label{CZWapp}
\eea
where it is assumed that $C_S \ll C_V$.  The term in $b_F \gamma_1/W$ is called
the Fierz interference term, with $b_F = \pm 2 \mu_1 C_S/C_V$.
This is the well-known expression given by Jackson, Treiman
and Wyld \cite{JTW57}.  Here $\mu_1$ is one of the beta-decay Coulomb
functions, Eq.~(\ref{bdcf}), 
and is of order unity, and $\gamma_1 = (1 - (\alpha Z)^2)^{1/2}$.

\subsubsection{Determining a limit on $C_S/C_V$}
\label{sss:dlcs}

With the results of our data survey, we can now search for any evidence
of a $1/W$ term in the shape-correction
function, and hence set a limit on $C_S$.  The test is based on
the corrected $\F t$ values being a constant for all
superallowed transitions between isospin $T = 1$ analogue states.  For
optimum sensitivity, we do not use Eq.~(\ref{CZWapp}) for $C(Z,W)$ in the
evaluation of the statistical rate function, $f$, because of
the extreme nature of some of the approximations made in deriving that equation.
Instead we use the exact numerically computed expression.
Since this calculated value of $f$ depends on the value of $C_S$ we simply treat
$C_S$ as an adjustable parameter and seek a
value that minimizes $\chi^2$, in a least-squares fit
to the expression $\F t = {\rm constant}$.  The result is
\be
C_S/C_V = - ( 0.00005 \pm 0.00130 ) .
\label{Cssign}
\ee
The sign of $C_S/C_V$ is determined from the fit, since
the calculated $f$ depends on the interference 
between vector and scalar interactions.  The interpretation
of the sign is a little more delicate.  We define $C_S$ to be
the strength of the scalar interaction in electron-emission beta
decay, and this is the value quoted in Eq.~(\ref{Cssign}). 
Since all the superallowed data involve positron emitters
there is a sign change mentioned earlier due to charge conjugation
that operationally is included in the computations.  The
corresponding Fierz interference constant, $b_F$,
is just $-2$ times this quantity\footnote{In 
our previous work described in
ref.~\protect\cite{HT75} and adopted in our subsequent publications, we
explicitly included a minus sign in the formulae in recognition that
all the superallowed Fermi transitions involved positron emitters.
Thus the shape-correction function $C(Z,W)$ was modified to
$C(Z,W)(1 - \gamma_1 b_F / W)$ and a fit of 
$\F t (1 - \gamma_1 b_F / \langle W \rangle )$
to a constant yielded a value of $b_F$ that was negative.  Currently in
Eq.~(\protect\ref{CZWapp}) we have defined $b_F$ such that
$C(Z,W)$ is modified to 
$C(Z,W)(1 + \gamma_1 b_F / W)$ and hence we are now quoting $b_F$
with a positive sign.}:
$b_F = 0.0001 \pm 0.0026$.  
Had we not assumed that parity was violated maximally
then the outcome would be
\be
\frac{C_S C_V + C_S^{\prime} C_V^{\prime} }
{|C_V|^2 + |C_V^{\prime}|^2 + |C_S|^2 + |C_S^{\prime}|^2 }
= - (0.00005 \pm 0.00130) .
\label{CsCsplimit}
\ee
This result shows a factor of 30 reduction in the central value compared
to our previously published
result \cite{TH03}, with the standard
deviation being essentially unchanged.  This is by far
the most stringent limit on $C_S/C_V$ ever obtained from nuclear beta decay.

\subsection{Induced Scalar Interaction}
\label{ss:fsi}

If we consider only the vector part of the weak interaction, for composite
spin-1/2 nucleons the
most general form of that interaction is written \cite{BB82} as 
\be
H_V = \overline{\psi}_p ( \gV \gamma_{\mu} - \fM \sigma_{\mu \nu} q_{\nu}
+ i \fS q_{\mu} ) \psi_n ~ \overline{\phi}_e \gamma_{\mu} ( 1 +
\gamma_5 ) \phi_{\overline{\nu}_e},
\label{HVnucl}
\ee
with $q_{\mu}$ being the four-momentum transfer,
$q_{\mu} = (p_p - p_n)_{\mu}$.  The values of the coupling constants
$\gV$ (vector), $\fM$ (weak magnetic) and $\fS$ (induced scalar) are
prescribed if the CVC hypothesis -- that the weak vector current
is just an isospin rotation of the electromagnetic vector current --
is correct.  In particular, since CVC implies that
the vector current is divergenceless, we anticipate that $\fS = 0$. 
An independent argument \cite{We58}, that there be no second-class currents
in the hadronic weak interaction, also requires $\fS$ to vanish.  Our goal
in this section is to use the data from superallowed beta decay
to set limits on the possible value of the induced scalar coupling
constant, $\fS$.  This will provide a test of the CVC hypothesis and
simultaneously set limits on the presence of second-class currents
in the hadronic vector weak interaction.

\subsubsection{Relation between $f_S$ and $C_S$}
\label{sss:orderest}

Considering, then, just the induced scalar term in the vector part of
the weak interaction,
\be
H_V(S) = \overline{\psi}_p (i \fS q_{\mu} ) \psi_n
~ \overline{\phi}_e \gamma_{\mu} ( 1 + \gamma_5 ) \phi_{\overline{\nu}_e} ,
\label{HVS1}
\ee
we see that this term can be reorganised to match closely the Hamiltonian from
the fundamental scalar interaction shown in Eq.~(\ref{JTWHam}).  The momentum
transfer, $q_{\mu} = (p_p-p_n)_{\mu} = - (p_e + p_{\overline{\nu}_e} )_{\mu}$,
can be moved into the lepton matrix element where, in combination
with $\gamma_{\mu}$, it can be replaced with the free-particle Dirac
equation: $\gamma_{\mu} (p_e)_{\mu} \phi_e = i m_e \phi_e$,
$\gamma_{\mu} (p_{\overline{\nu}_e})_{\mu} \phi_{\overline{\nu}_e} = 
i m_{\overline{\nu}_e} 
\phi_{\overline{\nu}_e}$,  with $m_e$ and $m_{\overline{\nu}_e}$ being the
electron and neutrino masses, respectively.  On setting the neutrino
mass to zero, we find that $H_V(S)$ becomes
\be
H_V(S) = \overline{\psi}_p m_e \fS \psi_n
~ \overline{\phi}_e ( 1 + \gamma_5 ) \phi_{\overline{\nu}_e}.
\label{HVS2}
\ee
This expression is equivalent to the 
fundamental scalar interaction in Eq.~(\ref{JTWHam}) with $C_S$
simply replaced by $m_e \fS $.  Thus, its effect
on the shape-correction function can be described by the same
replacement in Eq.~(\ref{CZWapp}).  An equivalent result was
obtained by Holstein \cite{Ho84}.

\subsubsection{Determining a limit on $\fS$}
\label{sss:dlfs}

We have now established the mathematical equivalence of the effects that
$\fS$ and $C_S$ have on the shape-correction function, $C(Z,W)$.  As a
result, we can use Eq.~(\ref{Cssign}) to conclude that
\be
m_e \fS / \gV = - ( 0.00005 \pm 0.00130) .
\label{fsgv}
\ee
The sign of $\fS/\gV$ follows the same convention as that described
after Eq.~(\ref{Cssign}). This result is a vindication for the CVC hypothesis, which predicts
$\gV = 1$ and $\fS = 0$.  Our result confirms this prediction at the
level of 13 parts in $10^4$.  As already mentioned, this result can also
be interpreted as setting a limit on vector second-class currents in
the semi-leptonic weak interaction, which therefore have not been
observed here at the same level of precision.

\subsection{Right-hand Currents}
\label{ss:rhc}

Let us no longer consider parity violation to be maximal.  The
general form of the weak interaction Hamiltonian \cite{JTW57} for just the
vector couplings of relevance for superallowed beta decay is
\be
H_{V}  =  
  (\overline{\psi}_p \gamma_{\mu} \psi_n)
        (C_V \overline{\phi}_e \gamma_{\mu} \phi_{\overline{\nu}_e}
+ C_V^{\prime} \overline{\phi}_e \gamma_{\mu}\gamma_5 \phi_{\overline{\nu}_e})
\label{JTWHV}
\ee
With $C_V^{\prime} \neq C_V$ we cannot associate the coupling
constants with the hadron matrix elements as we did in Eq.~(\ref{JTWSV}).  Instead,
the lepton and neutrino radial functions remain combined with
$C_V$ or $C_V^{\prime}$.  
The final formula for the shape-correction function then becomes
\begin{widetext}
\bea
C(Z,W)  & = & 
\sum_{k_e k_{\nu} K} \lambda_{k_e}
\biggm{\{} \frac{1}{2} \Bigm{(} M_K^2(k_e,k_{\nu}) + m_K^2(k_e,k_{\nu})
+ N_K^2(k_e,k_{\nu}) + n_K^2(k_e,k_{\nu})  \Bigm{)} 
\nonumber \\
& & ~~~~~~~~~   
- \frac{2 \mu_{k_e} \gamma_{k_e}}{k_e W}  
~\frac{1}{2} \Bigm{(} M_K(k_e,k_{\nu}) m_K(k_e,k_{\nu})    
+ N_K(k_e,k_{\nu}) n_K(k_e,k_{\nu}) \Bigm{)} \biggm{\}} 
\label{CWrhc}
\eea
where $M_K(k_e,k_{\nu})$,
$m_K(k_e,k_{\nu})$,
$N_K(k_e,k_{\nu})$,
and $n_K(k_e,k_{\nu})$ are reduced matrix elements as defined
in Eq.~(\ref{MmKkk}), with their 
respective  radial functions
being $F(r)$, $f(r)$, $G(r)$, and $g(r)$.
These radial functions are
\bea
F(r) & = & H(r)
\left \{ C_V G_{--}~
j_{k_{\nu}-1}(p_{\nu}r) - C_V^{\prime} G_{-+}~
j_{k_{\nu}}(p_{\nu}r) \right \}
\nonumber \\
& &
+ D(r)
\left \{ C_V^{\prime} G_{-+}~
j_{k_{\nu}-1}(p_{\nu}r) - C_V G_{++}~
j_{k_{\nu}}(p_{\nu}r) \right \}
\nonumber \\
f(r) & = & h(r)
\left \{ C_V G_{--}~
j_{k_{\nu}-1}(p_{\nu}r) - C_V^{\prime} G_{-+}~
j_{k_{\nu}}(p_{\nu}r) \right \}
\nonumber \\
& &
+ d(r)
\left \{ C_V^{\prime} G_{-+}~
j_{k_{\nu}-1}(p_{\nu}r) - C_V G_{++}~
j_{k_{\nu}}(p_{\nu}r) \right \} 
\nonumber \\
G(r) & = & H(r) 
\left \{ - C_V^{\prime} G_{--}~
j_{k_{\nu}-1}(p_{\nu}r) + C_V G_{-+}
j_{k_{\nu}}(p_{\nu}r) \right \}
\nonumber \\
& &
+ D(r)
\left \{ - C_V G_{-+}~
j_{k_{\nu}-1}(p_{\nu}r) + C_V^{\prime} G_{++}~
j_{k_{\nu}}(p_{\nu}r) \right \}
\nonumber \\
g(r) & = & h(r)
\left \{ - C_V^{\prime} G_{--}~
j_{k_{\nu}-1}(p_{\nu}r) + C_V G_{-+}~
j_{k_{\nu}}(p_{\nu}r) \right \}
\nonumber \\
& &
+ d(r) 
\left \{ - C_V G_{+-}~
j_{k_{\nu}-1}(p_{\nu}r) + C_V^{\prime} G_{++}~
j_{k_{\nu}}(p_{\nu}r) \right \} 
\label{Ffr2}
\eea
\end{widetext}
where the functions $H$, $D$, $h$ and $d$ are
linear combinations of the electron functions, $f_{\kappa}(r)$ and
$g_{\kappa}(r)$ as given in Eq.~(\ref{HDhd}).  The angular
momentum factors $G_{\pm ,\pm }$ are defined in Eq.~(\ref{GKLs}).

\subsubsection{Order of magnitude estimates}
\label{sss:estimates2}

Consider a pure Fermi transition for which the multipolarity is
$K=0$ and only the lowest lepton partial waves, $k_e = 1$ and
$k_{\nu} = 1$, are kept.  
Then, as in Sect.~\ref{sss:estimates}, the amplitudes become
\bea
M_0(1,1) & = & C_V M_F + \ldots
\nonumber \\
m_0(1,1) & = & {\rm small}
\nonumber \\
N_0(1,1) & = & - C_V^{\prime} M_F + \ldots
\nonumber \\
n_0(1,1) & = & {\rm small} ,
\label{Mmest2}
\eea
The shape-correction function is then
\be
C(Z,W) = |M_F|^2 \sfrac{1}{2}\left ( C_V^2 + C_V^{\prime \, 2} 
\right ) .
\label{CZWapp2}
\ee
We see that the dominant impact of
the right-hand current is simply to scale the statistical
rate function by $\sfrac{1}{2} (1 + C_V^{\prime \, 2}/C_V^2)$.  
This has no impact on the CVC
test that demonstrates that $\F t = ~{\rm constant}$, but it does
shift the value of the vector coupling
constant and thus the deduced value of $V_{ud}^2$.  However, $V_{ud}^2$
is obtained from the ratio of beta-decay to muon-decay rates, so 
before we can make any definitive statement on the effect of a right-hand
current on $V_{ud}^2$, we must first examine
the impact of that current on muon decay.  We will show next
that the correction due to a right-hand current is second order
in small quantities in muon decay, but first order in vector beta decay.
To this end we
examine a more general Hamiltonian presented by
Herczeg \cite{He86}.

\subsubsection{The effect on $V_{ud}^2$ }
\label{sss:dvud2}

In the $SU(2)_L \times U(1)$ Standard Model, the semi-leptonic weak
interaction Hamiltonian can be written schematically as
\be
H_{SM} = \frac{\GF}{\sqrt{2}} V_{ud} (V-A)(V-A),
\label{HSM}
\ee
where the first factor of $V-A$ represents the lepton currents:
$V = \overline{\phi}_e \gamma_{\mu} \phi_{\overline{\nu}_e}$ and
$-A = \overline{\phi}_e \gamma_{\mu} \gamma_5 \phi_{\overline{\nu}_e}$,
while the second $V-A$ represents the hadron currents:
$V = \overline{\psi}_p \gamma_{\mu} \psi_n$ and 
$-A = \overline{\psi}_p \gamma_{\mu} \gamma_5 \psi_n$.  The weak
interaction coupling $\GF/\sqrt{2} = g^2/8 \MW^2$, where
$g$ is the basic coupling constant of the Weinberg-Salam Standard Model
and $\MW$ is the mass of the exchanged $W$-boson.

Herczeg \cite{He86,He01} considers an extension that is the most general form
for non-derivative local four-fermion couplings
\bea
H & = & a_{LL} (V-A)(V-A)
+ a_{LR} (V-A)(V+A)
\nonumber \\
& &
+ a_{RL} (V+A)(V-A)
+ a_{RR} (V+A)(V+A),
\label{Haa}
\eea
where again the first factor represents the lepton currents, the second
the hadron currents.  The lepton fields are now written as
$V = \overline{\phi}_e \gamma_{\mu} \phi^L_{\overline{\nu}_e}$ or
$\overline{\phi}_e \gamma_{\mu} \phi^R_{\overline{\nu}_e}$ depending
whether the chirality of the neutrino is left-handed for $V-A$
coupling or right-handed for $V+A$ coupling.  The neutrino states
are in general linear combinations of mass eigenstates,
\be
\phi^L_{\overline{\nu}_e} = {\textstyle \sum_i} U_{e i} \phi^L_i
\qquad
\phi^R_{\overline{\nu}_e} = {\textstyle \sum_i} V_{e i} \phi^R_i ,
\label{Numatrix}
\ee
where $U_{ei}$ and $V_{ei}$ are first-row elements of the neutrino mixing
matrix.  The observed beta decay probability is the sum of the 
probabilities of decays into the energetically allowed neutrino
mass eigenstates.  We follow Herczeg \cite{He86,He01} in assuming
that the neutrinos produced in beta decay are light
enough that the effects of their masses can be neglected.  In
particular, the terms that arise from the interference between
amplitudes involving neutrinos of different chirality are dropped.
Then the effect of neutrino mass mixing can be taken into 
account by our multiplying the coupling constants $a_{LL}$ and
$a_{LR}$ by $\sqrt{u_e}$, and $a_{RL}$ and $a_{RR}$ by
$\sqrt{v_e}$ where
\be 
u_e = {\textstyle \sum^{\prime}_i} | U_{ei} |^2   \qquad
v_e = {\textstyle \sum^{\prime}_i} | V_{ei} |^2 .  
\label{ueve}
\ee
The prime on the summation indicates that the sum extends only over
the neutrinos that are light enough to be produced in beta
decay.  Note that if all the neutrinos are light  for both left-handed
and right-handed chiralities, then $u_e = v_e =1$ as a consequence
of the unitarity of the neutrino mixing matrix.

Herczeg's Hamiltonian, Eq.~(\ref{Haa}), can be rewritten
\bea
H & = & ( a_{LL}+a_{LR} + a_{RL} + a_{RR} ) VV 
\nonumber \\
& &
+ (-a_{LL}-a_{LR} + a_{RL} + a_{RR} ) AV 
\nonumber \\
& &
+ (-a_{LL}+a_{LR} - a_{RL} + a_{RR} ) VA 
\nonumber \\
& &
+ ( a_{LL}-a_{LR} - a_{RL} + a_{RR} ) AA 
\label{altH}
\eea
We can compare this with the Jackson, Treiman and Wyld (JTW) Hamiltonian \cite{JTW57}, which
in the current notation becomes
\bea
H_{JTW} & = & (C_V V - C_V^{\prime} A ) V + (-C_A A + C_A^{\prime} V ) A 
\nonumber \\
& = & C_V VV - C_V^{\prime} AV + C_A^{\prime} VA - C_A AA
\label{HJTW}
\eea
Thus we identify the correspondences as\footnote{Herczeg \cite{He86,He01}
employs a metric that leads to a different sign on the $\gamma_5$
matrix, so his correspondences yield a different overall sign from ours
for $C_V^{\prime}$ and $C_A$.}
\bea
C_V & = & a_{LL} + a_{LR} + a_{RL} + a_{RR}
\nonumber \\
C_V^{\prime} & = & a_{LL} + a_{LR} - a_{RL} - a_{RR}
\nonumber \\
C_A & = & -a_{LL} + a_{LR} + a_{RL} - a_{RR}
\nonumber \\
C_A^{\prime} & = & -a_{LL} + a_{LR} - a_{RL} + a_{RR}
\label{CVaLL}
\eea
For Fermi beta decay, only the vector part of the weak hadron current
contributes, so the decay rate, $\Gamma_{\beta}$, 
as shown earlier in Eq.~(\ref{CZWapp2}), is proportional to
\bea
\Gamma_{\beta} & \propto & \sfrac{1}{2} \left ( |C_V|^2 +
|C_V^{\prime}|^2 \right )
\nonumber \\
& = & |a_{LL} + a_{LR} |^2 + |a_{RL} + a_{RR} |^2
\nonumber \\
& = & |a_{LL}|^2 \left ( |1 + \overline{a}_{LR} |^2
+ |\overline{a}_{RL} + \overline{a}_{RR} |^2 \right )
\nonumber \\
& \simeq & |a_{LL}|^2 \left ( 1 + 2 Re \overline{a}_{LR} + \ldots \right )
\label{twoalr}
\eea
where $\overline{a}_{ij} = a_{ij}/a_{LL}$.

To continue our determination of $V_{ud}$ we need to consider the purely
leptonic muon decay.  Herczeg \cite{He86} writes the effective
Hamiltonian for muon decay in analogy to Eq.~(\ref{Haa}) as
\bea
H & = & c_{LL}(V-A)(V-A) + c_{LR} (V-A)(V+A) 
\nonumber \\
& &
+ c_{RL} (V+A)(V-A)
+ c_{RR} (V+A)(V+A)
\label{Hcc}
\eea
The coupling constants in Eqs.~(\ref{Hcc}) and (\ref{Haa}) are 
related by the CKM matrix elements by
\bea
a_{LL} & = & c_{LL} V_{ud}^L
\nonumber \\
a_{LR} & = & c_{LR} e^{i \alpha } V_{ud}^R
\nonumber \\
a_{RL} & = & c_{RL} V_{ud}^L
\nonumber \\
a_{RR} & = & c_{RR} e^{i \alpha } V_{ud}^R .
\label{aLLcLL}
\eea
Here $V_{ud}^L$ is the $ud$-matrix element of the CKM matrix for
left-handed chirality quarks, and $V_{ud}^R$ is for right-handed
chirality quarks.  The phase $\alpha$ is a CP-violating phase 
in the right-handed CKM matrix.  The decay rate, $\Gamma_{\mu}$,
for muon decay is proportional to
\bea
\Gamma_{\mu} & \propto & |c_{LL}|^2 + |c_{LR}|^2 
+ |c_{RL}|^2 + |c_{RR}|^2 
\nonumber \\
& = & |c_{LL}|^2 \left ( 1 + |\overline{c}_{LR}|^2
+ |\overline{c}_{RL}|^2 + |\overline{c}_{RR}|^2 \right ) 
\label{ratemu}
\eea
where $\overline{c}_{ij} = c_{ij}/c_{LL}$.

Combining Eqs.~(\ref{twoalr}) and (\ref{ratemu}), we obtain an expression
that connects the ratio of beta-decay to muon-decay rates with the value
of $|V_{ud}^L|^2$, \viz
\be
\frac{\Gamma_{\beta}}{\Gamma_{\mu}} = |V_{ud}^L|^2
\frac{|1 + \overline{a}_{LR} |^2 + |\overline{a}_{RL} + \overline{a}_{RR} |^2}
{1 + |\overline{c}_{LR}|^2 + |\overline{c}_{RL}|^2 + |\overline{c}_{RR} |^2} .
\label{rrates}
\ee
In the Standard Model, only $a_{LL}$ and $c_{LL}$ are non-zero; in any case, the
quantities $\overline{a}_{ij}$ and
$\overline{c}_{ij}$ with $ij = LR$, $RL$, or $RR$ can certainly be considered
small.  The correction to the muon decay rate from right-handed
interactions is therefore seen to be second order in small quantities,
while the correction to Fermi beta decay rate is first order.  Keeping only
first-order small quantities, Eq.~(\ref{rrates}) reduces to\footnote{
Herczeg \protect\cite{He01} also considers the possibility that
the relation between purely leptonic and semi-leptonic Hamiltonians,
Eq.~(\protect\ref{aLLcLL}), is not sufficiently general.  He writes
$a_{LL} = (a_{LL})_{SM} + a_{LL}^{\prime}$, with
$(a_{LL})_{SM} = c_{LL} V_{ud}^L$.  Then Eq.~(\protect\ref{rrates1})
becomes $\Gamma_{\beta}/\Gamma_{\mu} = |V_{ud}^L|^2 ( 1 +
2 Re ( \overline{a}_{LL}^{\prime} + \overline{a}_{LR} ))$,
where $\overline{a}_{LL}^{\prime} = a_{LL}^{\prime} / (a_{LL})_{SM}$.
We will not pursue this further, but it is obvious the formulae above
can accommodate this extension with a simple replacement of
$\overline{a}_{LR}$ with $\overline{a}_{LL}^{\prime} +
\overline{a}_{LR}$.}
\be
\frac{\Gamma_{\beta}}{\Gamma_{\mu}} = |V_{ud}^L|^2
\left ( 1 + 2 Re \overline{a}_{LR} \right ) .
\label{rrates1}
\ee
If the neutrino masses are such that $u_e \neq 1$ and $v_e \neq 1$ then
this equation is modified to
\be
\frac{\Gamma_{\beta}}{\Gamma_{\mu}} = |V_{ud}^L|^2
\frac{(u_e)_{\beta}}{[(u_e)_{\mu} (u_{\mu})_{\mu} ]^{1/2} }
\left ( 1 + 2 Re \overline{a}_{LR} \right ) ,
\label{rrates2}
\ee
where $(u_e)_{\beta}$ in the numerator is given by the $u_e$ in Eq.~(\ref{ueve}),
with the sum extended over neutrinos light enough to be produced in
beta decay, while in the denominator $(u_e)_{\mu}$ is given by the
same expression but with the sum extended over neutrinos light
enough to be produced in muon decay.  Note that the $Q$-value for muon decay 
is 105 MeV, a factor of ten times larger than the $Q$-value
for any Fermi beta decay we are considering.  Also $u_{\mu}$ in
Eq.~(\ref{rrates2}) is defined as $\sum_i^{\prime} | U_{\mu i}|^2$,
where $U_{\mu i}$ are second-row elements of the neutrino mixing
matrix.  In what follows, we will assume
$(u_e)_{\beta} = (u_e)_{\mu} = (u_{\mu})_{\mu} = 1$.

Before proceeding to numeric limits, it is worth showing how the current
formulae relate to the simpler and more restrictive manifest 
left-right symmetric model \cite{BBMS77}.  In this model the departure
from maximal parity violation is entirely due to the
presence of a second $W$-boson whose mass is much heavier than the
usual $W$-boson.  If left-hand couplings are mediated by the boson, $W_L$,
and right-hand couplings by $W_R$, then $W_L$ and $W_R$ will be linear
combinations of the mass eigenstates $W_1$ and $W_2$, \viz
\bea
W_L & = & W_1 \cos \zeta + W_2 \sin \zeta
\nonumber \\
W_R & = & e^{i \omega } \left ( - W_1 \sin \zeta + W_2 \cos \zeta \right )
\label{WLWR}
\eea
and $\omega$ is a CP violating phase.  If it is further assumed that,
apart from the different masses of the $W_1$ and $W_2$ bosons, the
coupling constants and CKM matrix elements are identical for left-hand
and right-hand couplings, then there are only two parameters in this model.  
These parameters are: $\delta = ( m_1 / m_2 )^2$ and $\zeta$, where $m_1$
and $m_2$ are the masses of $W_1$ and $W_2$ respectively.  Both parameters
are small and, of course, are zero in the Standard Model.  The parameters of
Herczeg's Hamiltonian, Eq.~(\ref{Haa}), can be expressed in terms of
$\delta$ and $\zeta$ \cite{He01}:
\bea
a_{LL} & = & \frac{g^2}{8 m_1^2} V_{ud}^L ~~~~~~~~~~ \overline{a}_{RR} = \delta
\nonumber \\
\overline{a}_{LR} & = & \overline{a}_{RL} = - e^{i \omega } \zeta
\rightarrow - \zeta
\label{azeta}
\eea
for negligible CP-violating effects.  In this limit, the expression
for the ratio of Fermi beta to muon decay rates, Eq.~(\ref{rrates1}),
reduces to
\be
\frac{\Gamma_{\beta}}{\Gamma_{\mu}} = |V_{ud}^L|^2
\left ( 1 - 2 \zeta \right ) .
\label{rrates3}
\ee
This is the expression we used in our earlier work \cite{TH03} to set
limits on the extent of right-hand currents.

\subsubsection{Numeric Limit}
\label{sss:nl}

Let us now insert the experimental values from our survey data for the beta-decay
and muon-decay rates to determine an experimental value for
$|V_{ud}|^2$, which we will write as $|V_{ud}|^2_{\rm{expt}}$.
This is the value we recorded earlier in Eq.~(\ref{Vudsq}).
Then Eq.~(\ref{rrates1}) can be written as
\bea
|V_{ud}|^2_{\rm{expt}} & = & |V_{ud}^L|^2 
( 1 + 2 Re \overline{a}_{LR} )
\nonumber \\
& = & \left ( 1 - |V_{us}^L|^2 - |V_{ub}^L|^2 \right )
( 1 + 2 Re \overline{a}_{LR} ),
\label{rrates4}
\eea
where in the second line we have inserted the condition for unitarity of the CKM
matrix.  Adopting the PDG's recommended values \cite{PDG04} for $|V_{us}^L|$ and
$|V_{ub}^L|$ (see text preceding Eq.~(\ref{sumfail})) we obtain the following
result from Eq.~(\ref{rrates4}):
\bea
(0.9482 \pm 0.0008) & = & (0.9516 \pm 0.0011) 
( 1 + 2 Re \overline{a}_{LR} )
\nonumber \\
Re \overline{a}_{LR} & = & -0.00176 \pm 0.00074.
\label{aLRPDG}
\eea
Within the context of the manifest left-right symmetric model (see Eq.~\ref{rrates3}),
this result corresponds to $\zeta = 0.00176 \pm 0.00074$, a similar value to
the one we reported previously \cite{TH03}.
The result of a non-zero $a_{LR}$ or $\zeta$ simply reflects the fact
that the experimental values of the first-row CKM matrix elements
do not satisfy the unitarity requirement.

If, instead, we adopt the
E865 value \cite{Sh03} for $V_{us}$ rather than the PDG average, then the
result in Eq.~(\ref{aLRPDG}) is modified to
\bea
(0.9482 \pm 0.0008) & = & (0.9484 \pm 0.0014) 
( 1 + 2 Re \overline{a}_{LR} )
\nonumber \\
Re \overline{a}_{LR} & = & - 0.00007 \pm 0.00084.
\label{aLR865}
\eea
This result is consistent with no right-hand currents and unitarity
being satisfied in the experimental CKM matrix elements.

\section{Conclusions}
\label{s:concl}

Previous surveys of superallowed Fermi beta decay have at times noted 
disagreement \cite{TH73,Ko84} among the derived $\F t$ values, and at other
times agreement \cite{HT75,HT90}.  When disagreement was evident, subsequent
attention paid to the problem led to both theoretical and experimental advances.
As presented here, in Sect.~\ref{s:Ft}, the status now is of excellent 
agreement among all $\F t$ values -- to better than 3 parts in $10^4$
over a wide range of nuclei from $A = 10$ to $A = 74$.  Such agreement 
confirms the expectations of CVC, allows very restrictive limits to be set on the
possible presence of scalar currents and makes it possible to go forward
with confidence to the next steps -- the determination of $V_{ud}$ and
the unitarity test of the CKM matrix.

The outstanding challenge at this time is that the value obtained for $V_{ud}$,
when combined with the current PDG-recommended values of $V_{us}$ and $V_{ub}$,
leads to a unitarity test that fails by more than two standard deviations.  There
are no evident defects in the calculated radiative and isospin-symmetry-breaking
corrections that could remove this problem and, indeed, a shift in any one of these
corrections large enough to restore unitarity would be almost impossible to
justify \cite{TH03}.  Moreover, the derived value of $V_{ud}$ from nuclear
decays has been remarkably stable for three decades despite a vast increase
in the quantity of high quality data and many theoretical refinements in the
calculations of the correction terms.

So if any progress is to be made in firmly establishing (or eliminating) the
discrepancy with unitarity, both theory and experiment must be brought to bear
afresh on the principal sources of uncertainty.  Although we will focus here on
improving the nuclear contribution to the unitarity test, additional experiments
are also required for neutron, pion and kaon decays.  The first two provide
independent, though so far much less precise, values for $V_{ud}$; the third
establishes the value of $V_{us}$, which may ultimately turn out to be solely
responsible for restoring the CKM matrix to unitarity.  Whatever the outcome
for unitarity, however, the results of all these studies will provide crucial
information, either in characterizing new physics beyond the standard model or
in setting a tight limit on its possible existence.

We have taken pains throughout this work to pay careful attention to all
uncertainties, theoretical and experimental.  In Section \ref{ss:vVud}
we detailed the various contributions to the uncertainty in $|V_{ud}|^2$.  
Of these, by far the largest is due to the nucleus-independent radiative
correction, $\DRV$.  Its uncertainty arises primarily from a box diagram
involving the exchange of one $W$ boson and one photon between the hadron
and the lepton. To make the loop integration tractable, it is divided by a scale
parameter into high- and low-energy portions.  The high-energy contribution
can be computed reliably \cite{ER04} but the low-energy one, as calculated
originally by Sirlin \cite{Si78}, depends on the choice of scale parameter.
Sirlin chose \cite{MS86,Si94} a reasonable range for this parameter, which
has been retained by subsequent authors \cite{To92,To94}.  It is this choice
of range that drives the overall uncertainty on $\DRV$.  Recent work \cite{An04}
with effective field theories based on chiral perturbation theory has been unable
to improve the situation: although this approach replaces the low-energy contributions
to the loop diagrams by well-defined low-energy constants, the values of these
constants are not known {\it a priori}.  How to obtain a more refined,
first-principles computation of the low-energy contribution remains an open
theoretical problem \cite{ER04}, but one of considerable importance and
urgency.  Not only is this uncertainty the principal limitation on the
precision with which $V_{ud}$ can be determined from nuclear superallowed
$\beta$ decay, but it will have a similar limiting effect on its determination
from neutron and pion decays as well.

The next largest contributor to the error budget on $|V_{ud}|^2$ is the
isospin-symmetry-breaking correction, $\delta_C$.  Although uncertainties have
been individually determined for the most recent calculations \cite{TH02} of
$\delta_C-\delta_{NS}$ (see Table~\ref{t:tab5}), the dominant source of
$|V_{ud}|^2$ uncertainty attributable to $\delta_C$ arises from the small
systematic difference between the results from different theoretical techniques
used to calculate $\delta_C$. Our approach \cite{TH02}, using Woods-Saxon
functions, yields larger $\delta_C$ values than the Ormand-Brown one \cite{OB95},
using Hartree-Fock functions.  Here we have taken the democratic approach, considering
that these two sets of calculations are equally likely to be correct and letting
the difference between their results determine a systematic uncertainty that we
apply to the final result (see Eq.~(\ref{Ftavg})).  

\begin{figure*}[t]
\epsfig{file=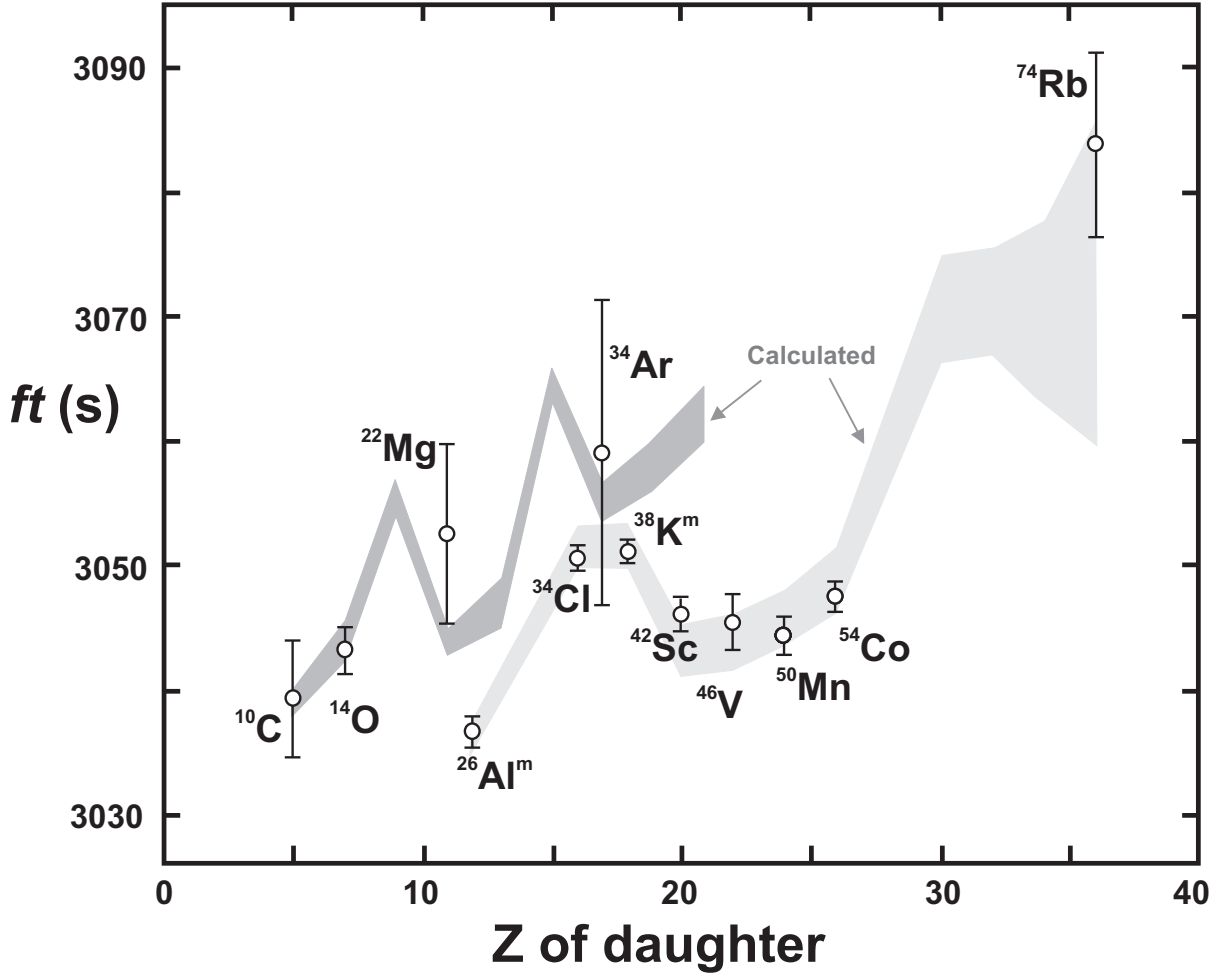,width=16cm}
\caption{Experimental $ft$ values plotted as a function of the charge
on the daughter nucleus, $Z$.  The bands represent the theoretical
quantity $\overline{\F t}/((1 + \delta_R^{\prime})(1 - \delta_C + \delta_{NS}))$.
The two groups distinguish those beta emitters whose parent nuclei
have isospin $T_z = -1$ (darker shading) from those with $T_z =0$
(lighter shading).}
\label{f:dcvar}
\end{figure*}

If reducing the uncertainty on $\DRV$ must rank as the first priority for future
theoretical work, then improving our confidence in $\delta_C$ can be taken as
the top priority challenge for experiment.  Although there is no way to check the
correctness of the absolute values of $\delta_C$ from experiment, it is possible
to check the nucleus-to-nucleus variations in the calculated values.  The method,
which is illustrated in Fig.~\ref{f:dcvar}, is based on the validity of the CVC
hypothesis that the corrected $\F t$ values for the superallowed $0^+ \rightarrow
0^+$ decays should be constant.  In the figure we compare the uncorrected measured
$ft$ values (points and error bars) with the theoretical quantity $\overline{\F t}/
((1 + \delta_R^{\prime})(1 - \delta_C + \delta_{NS}))$ shown as a band, the width
of which represents its estimated error.  With the average $\F t$ value, 
$\overline{\F t}$, taken from Table~\ref{t:tab5}, this comparison specifically
tests the collective ability of all three calculated correction terms to reproduce the
variations in $ft$ from one transition to another.  However, since $\delta_R^{\prime}$ is
almost independent of Z when $Z>10$, this test really probes directly the effectiveness
of the calculated values of $\delta_C-\delta_{NS}$.   

It can be seen that there is remarkable agreement between theory and experiment.  
In assessing the significance of this agreement, it is important to recognize
that the calculations of $\delta_C$ and $\delta_{NS}$ for $Z\leq 26$ are based on
well-established shell-model wave functions and were further tuned to reproduce measured
binding energies, charge radii and coefficients of the isobaric multiplet mass
equation \cite{TH02}.  The origins of the calculated correction terms for all
cases are completely independent of the superallowed
decay data.  Thus, the agreement in the figure between the measured superallowed data
points and the theoretical band is already a powerful validation of the calculated
corrections used in determining that band.  The validation becomes even more convincing
when we consider that it would require a pathological fault indeed in the theory to
allow the observed nucleus-to-nucleus variations in $\delta_C$ to be reproduced in
such detail while failing to obtain the {\em absolute} values to comparable precision.  
Pleasing as the agreement in Fig.~\ref{f:dcvar} is, though, new experiments can
still improve the test, making it even more demanding, and can ultimately reduce
the uncertainty in $\delta_C$ further.

These new experiments can follow different paths.  One is to improve the
precision on the nine superallowed transitions whose $ft$ values are already known
to within $0.15 \%$ or better.  On the one hand, these are the easiest cases to study,
all having stable daughters and all, except $^{10}$C, decaying predominantly ($>$ 99\%)
through a single superallowed transition.  On the other hand, they have been the subject of
intense scrutiny for at least the past four decades and, given the number of careful
measurements already published, the prospects for really significant improvements in
these cases, at least in the near future, do not seem bright.  Nevertheless, a glance at
Fig.~\ref{f:hist9} shows that some modest improvements are certainly possible.  If we accept
as a goal that experiment should be a factor of two more precise than theory, then we
see that the $Q_{EC}$-values for $^{10}$C, $^{14}$O, $^{26}$Al$^m$, $^{42}$Sc and
$^{46}$V, the half-lives of $^{10}$C, $^{26}$Al$^m$, $^{42}$Sc and $^{50}$Mn, and the
branching ratio for $^{10}$C can all bear improvement.  

A second path is to expand the number of precisely measured superallowed emitters to
include cases for which the calculated nuclear-structure-dependent corrections are larger,
or show larger variations from nuclide to nuclide, than the values applied to the nine
currently best-known transitions.  We argue that if the calculations reproduce the
experimentally observed variations where they are large, then that must surely verify their
reliability for the original nine transitions whose corrections are considerably smaller.  
The recent results for $^{22}$Mg, $^{34}$Ar and $^{74}$Rb are the first cases of this type to reach
sufficient precision that they can contribute to the test (see Fig.~\ref{f:dcvar}) but
more are sure to follow.  We have included in our survey all cases that we believe are
potential candidates within the next few years.

Without doubt, these new cases present serious experimental challenges.  In general, the
parent nuclei are more exotic and thus more difficult to produce in pure and statistically
useful quantities.  They also exhibit more complex branching patterns, which for the cases
with $A \geq 62$ include Gamow-Teller transitions that may be unobservable individually but
collectively can play a non-negligible role \cite{Ha02}.  These heavier nuclei also have very short
half-lives, which currently limit the precision with which their $Q_{EC}$-values can be
measured.  Even so, all these obstacles are obviously now being overcome and we may reasonably
hope that before long there will not only be more cases with precisely measured parameters, but
there will be more than one measurement of each parameter, an essential prerequisite for
reliable results at the level of precision needed to constrain the correction terms.

Although it would not impact significantly on the unitarity question, there is an additional
reason to improve the precision with which the $ft$ values are known, particularly for the
cases with $A \leq 26$.  A scalar current, if it exists, would manifest itself as a $1/W$-dependence
in the shape-correction function used in the $f$-value calculations (see Eq.~(\ref{CZWapp})).  
Since the superallowed transition energies decrease with $A$, this effect would be strongest for
the lightest nuclei, $^{10}$C and $^{14}$O.  Improved precision in the $\F t$ values for those
two nuclei would act directly to reduce the limits set on a possible scalar current.   

In conclusion, we can assert that world data for superallowed $0^+ \rightarrow 0^+$ $\beta$
decays strongly support the CVC expectation of an unrenormalized vector coupling constant, and
also set a new limit on the possible existence of scalar currents.  The nuclear-structure-dependent
corrections used in the analyses of these data have so far stood up very favorably to experimental
tests, and the value currently obtained with them for $V_{ud}$ is deemed to be very robust, even
though it is an important component of the top-row test of CKM unitarity that fails by more than
two standard deviations.  We have indicated the improvements required from both theory and experiment
to increase the precision in future so as to produce a more definitive result for the unitarity test.

\acknowledgments

The work of JCH was supported by the U. S. Dept. of Energy under Grant
DE-FG03-93ER40773 and by the Robert A. Welch Foundation.
IST would like to thank the Cyclotron Institute of Texas A \& M University
for its hospitality during several two-month visits.

\appendix

\section{Statistical Rate Function}
\label{s:srf}

The statistical rate function is an integral over phase space,
\be
f = \int_{1}^{W_0} p W (W_0 - W)^2 F(Z, W) S(Z, W) dW ,
\label{fdef}
\ee
where
$W$ is the electron total energy in electron rest-mass  
units,   
$W_0$ is the maximum value of $W$,        
$p = (W^2 -1)^{1/2}$ is the electron momentum,           
$Z$ is the charge number of the daughter nucleus 
(positive for  
electron emission, 
negative for positron emission),  
 $F(Z, W)$ is 
the Fermi function, and
 $S(Z, W)$ is the shape-correction function.
If the shape-correction function is put to unity, the integral becomes the
customarily defined one for beta decay, which we will denote
as $f_{\rm stat}$:
\be
f_{\rm stat} = \int_1^{W_0} p W (W_0 - W)^2 F(Z, W) dW .
\label{fstat}
\ee 
The exact evaluation of $f$ differs from $f_{\rm stat}$ by 0.2\% at
$A = 10$ up to 5.7\% at $A = 74$.  Thus, to maintain 0.1\% accuracy for 
$f$ over that range, we must determine the shape-correction function itself
with 2\% accuracy.  Obtaining this accuracy requires consideration
of the following issues:

\bi

\item The electron wave functions can no longer be simply those of the
lowest partial wave ($j=1/2$) generated by a point nuclear charge
and evaluated at radius R, the nuclear surface, 
but must be the exact functions for some
chosen nuclear charge-density distribution; 

\item  The lepton wavefunctions exhibit some $r^2$ dependence
over the nuclear volume, leading to what are called second-forbidden
corrections.  Furthermore, a more accurate treatment of the weak interaction
leads to relativistic and induced-current corrections.  All these
effects must be incorporated since they impact on the nuclear matrix elements and inject a
mild nuclear-structure dependence into the evaluation of $f$.

\item  The atomic electrons cannot be ignored, but must be accommodated
approximately in a screening correction;

\ei

In what follows we describe the ingredients of a code we have
written that incorporates these effects.  It is based on the formalism
of Behrens and B\"{u}hring \cite{BB82}.  Note that they define 
\be 
F(Z, W) S(Z, W) = F_0 L_0 C(Z, W) ,
\label{F0L0C}
\ee
where $F_0 = 2 F(Z, W)/(1 + \gamma_1)$.  The purpose of their redefining
the shape correction factor in this way was to remove the
historic requirement to evaluate the electron wave functions at
the nuclear surface.  The product $F_0 L_0$ is given entirely
in terms of the amplitudes of the electron wave function at
the origin (see Eq.~(\ref{bdcf}) below), and $C(Z,W)$ is the shape-correction function
defined with respect to this choice.

\subsection{Electron Radial Wave Functions}
\label{ss:erwf}

The wave function for the electron emitted in beta decay is given by
the solution to the Dirac equation with an external electromagnetic
field present, but restricted to the special case where the vector
potential vanishes identically and the scalar potential is static
and spherically symmetric.  We solve in spherical coordinates and
introduce a partial wave expansion such that the basis states are
written:
\be 
\psi_{\kappa}^{\mu} = \left ( \begin{array}{c}
{\rm sign}(\kappa ) f_{\kappa} (r) \chi_{\hhphen \kappa}^{\mu} \\
g_{\kappa} (r) \chi_{\kappa}^{\mu} \end{array} \right ) ,
\label{psiKmu}
\ee
where $f_{\kappa}$, $g_{\kappa}$ are radial functions, and $\chi_{\kappa
}^{\mu}$ are the usual spin-angular-momentum wave functions describing the
coupling of the orbital angular momentum $l$ and spin $\sfrac{1}{2}$
to give a total angular momentum $j$ with $z$-component $\mu$:
\be
\chi_{\kappa}^{\mu} = i^l \sum_{m_l m_s}
\langle l m_l \, \sfrac{1}{2} m_s | j \mu \rangle
Y_{l \, m_l}(\hat{{\bf r}}) \chi_{m_s} .
\label{chiKmu}
\ee
The eigenvalue $\kappa$ is
\be
\kappa = -j (j+1) + l (l+1) - \sfrac{1}{4}
\label{Kappadef}
\ee
and has values
\bea
\kappa & = & - (l+1)  = - (j + \sfrac{1}{2} ) ~~~~ {\rm if} ~
j = l + \sfrac{1}{2}
\nonumber \\[1mm]
\kappa & = &  l   =  (j + \sfrac{1}{2} ) ~~~~ {\rm if} ~
j = l - \sfrac{1}{2}.
\label{Kappavalu}
\eea
The radial functions are solutions to a pair of coupled equations
\bea
\frac{d g_{\kappa}}{d r} + \frac{(\kappa +1)}{r} g_{\kappa}
- ( W + 1 - V(r) ) f_{\kappa} & = & 0
\nonumber \\[1mm]
\frac{d f_{\kappa}}{d r} - \frac{(\kappa -1)}{r} f_{\kappa}
+ ( W - 1 - V(r) ) g_{\kappa} & = & 0 .
\label{radialeq}
\eea
Here $V(r)$ is a spherically symmetric static potential
that represents the interaction of the electron with the charge
distribution of the nucleus.  

Our task is to solve the pair of coupled radial equations, 
Eq.~(\ref{radialeq}), in three regions, $0 \leq r \leq R_1$,
$R_1 \leq r \leq R_2$, $R_2 \leq r \leq \infty$, matching the solutions at each
region boundary.  The first region is the one over which the nuclear
charge density is non-zero.  
Here, we establish a power-series solution, regular at the origin,
as the starting solution and integrate numerically to $R_1$.
In the second region, we have a pure Coulomb potential, for which
an analytic solution can be found in terms of confluent hypergeometric
functions of complex argument.  
The asymptotic solution in the third region is expressed
in terms of the desired outgoing waves and a phase shift.  The
unknowns in the calculation are the phase shift, $\Delta_{\kappa}$,
 and the normalization
of the interior solution, $\alpha_{\kappa}$: they are determined from
the matching conditions.

In the derivation of expressions for beta-decay observables, 
certain combinations of amplitudes
and phase shifts characterising the electron
wave functions appear.  These combinations are called
the beta-decay Coulomb functions, and are generally organized so that
they are of order unity, with
corrections of order $(\alpha Z)^2$.  Since we are only interested
here in the shape-correction
function, the only such functions we need are
$F_0L_0$, $\lambda_k$ and $\mu_k$, where $k = | \kappa |$, which
are defined as
\bea
F_0L_0 & = & \frac{\alpha_{\hhphen 1}^2 + \alpha_{+1}^2}{2p^2}
\nonumber \\
\lambda_k & = &
\frac{\alpha_{\hhphen k}^2 + \alpha_{+k}^2}
{\alpha_{\hhphen 1}^2 + \alpha_{+1}^2}
\nonumber \\
\mu_k & = &
\frac{\alpha_{\hhphen k}^2 - \alpha_{+k}^2}
{\alpha_{\hhphen k}^2 + \alpha_{+k}^2}
\frac{k W}{\gamma_k} ,
\label{bdcf}
\eea
where $\gamma_k = \{ k^2 -
(\alpha Z)^2 \}^{1/2}$.  Although these functions
have actually been tabulated
by Behrens and J\"{a}necke \cite{BJ69}, we  
have not used the tables but, in the interests of precision, have
explicitly computed the
functions from the calculated values of $\alpha_{\kappa}$.

\subsection{Shape-correction function}
\label{ss:scf}

Behrens and B\"{u}hring \cite{BB82} give the following expression
for the shape-correction function:
\bea
C(Z,W) & = &
\sum_{k_e k_{\nu} K} \lambda_{k_e}
\left \{ M_K^2(k_e,k_{\nu}) + m_K^2(k_e,k_{\nu}) \right.
\nonumber \\ 
& & \left. - \frac{2 \mu_{k_e} \gamma_{k_e}}{k_e W}  
M_K(k_e,k_{\nu})  m_K(k_e,k_{\nu}) \right \} ,
\label{CWfinal}
\eea
where the sums over $k_e$ and $k_{\nu}$ are partial-wave expansions
of the electron and neutrino wave functions with
$k_e = j_e + \sfrac{1}{2}$ and
$k_{\nu} = j_{\nu} + \sfrac{1}{2}$, $j_e$ and $j_{\nu}$
being the electron and neutrino total
angular momenta.  The integer $K$ represents the multipolarity
of the transition operators and is limited to the range
$|j_e - j_{\nu}| \leq K \leq j_e + j_{\nu}$.
The functions $M_K(k_e,k_{\nu})$ and
$m_K(k_e,k_{\nu})$ are given in terms of form factors \cite{BB82}
and we evaluate these form
factors in the ``impulse approximation".  In this
approximation, the nucleus is treated as a collection of non-interacting
nucleons so it is only necessary to consider the weak interaction
as acting upon a single nucleon.  All the many-body aspects of the
nucleus can thus be handled in the standard shell-model way.  Let $O$ be
a one-body operator, which can be written 
\be
O = \sum_{\alpha \beta} \langle \alpha | O | \beta \rangle
a_{\alpha}^{\dag} a_{\beta}  ,
\label{defO}
\ee
where $a_{\alpha}^{\dag}$ is the creation operator for a nucleon
in quantum state $\alpha$, and $a_{\beta}$ is an annihilation operator
destroying a nucleon in state $\beta$.  The matrix element
of $O$ for many-body states becomes
\be
\langle f | O | i \rangle = \sum_{\alpha \beta}
\langle \alpha | O | \beta \rangle
\langle f | a_{\alpha}^{\dag} a_{\beta} | i \rangle ;
\label{defOme}
\ee
that is, the matrix element is a linear combination of single-particle
matrix elements,
$\langle \alpha | O | \beta \rangle $.
The coefficients in the expansion,
$\langle f | a_{\alpha}^{\dag} a_{\beta} | i \rangle $,
are called one-body density matrix elements (OBDME).  We leave it,
then, as the job of the shell model to provide the OBDMEs and only deal
here with the single-particle matrix elements.
The functions $M_K(k_e,k_{\nu})$ and
$m_K(k_e,k_{\nu})$ are now given in terms of reduced nuclear
matrix elements for a single-particle transition
$j_{\beta} \rightarrow j_{\alpha}$ by
\bea
M_K(k_e,k_{\nu}) & = & \frac{\sqrt{4 \pi }}{\hat{K} \hat{J}_i}
\sum_{L s} (-)^{K-L}
\langle j_{\alpha} \dbar F(r) \hat{T}_{K L s} \dbar j_{\beta} \rangle 
\nonumber \\
m_K(k_e,k_{\nu}) & = & \frac{\sqrt{4 \pi }}{\hat{K} \hat{J}_i}
\sum_{L s} (-)^{K-L}
\langle j_{\alpha} \dbar f(r) \hat{T}_{K L s} \dbar j_{\beta} \rangle ,
\label{MmKkk}
\eea
where $\hat{\jmath}$ is a short-hand notation for
$(2 j + 1 )^{1/2}$ and $J_i$ is the spin of the decaying
nucleus.  The radial functions are
\bea
F(r) & = & 
H(r)
\left \{ G_{--} ~
j_{k_{\nu}-1}(p_{\nu}r) - G_{-+} ~
j_{k_{\nu}}(p_{\nu}r) \right \}
\nonumber \\
&  & + 
D(r)
\left \{ G_{+-} ~
j_{k_{\nu}-1}(p_{\nu}r) - G_{++} ~
j_{k_{\nu}}(p_{\nu}r) \right \}
\nonumber \\
f(r) & = & 
h(r)
\left \{ G_{--} ~
j_{k_{\nu}-1}(p_{\nu}r) - G_{-+} ~
j_{k_{\nu}}(p_{\nu}r) \right \}
\nonumber \\
&  & + 
d(r)
\left \{ G_{+-} ~
j_{k_{\nu}-1}(p_{\nu}r) - G_{++} ~
j_{k_{\nu}}(p_{\nu}r) \right \} ,
\label{Ffr3}
\eea
with
\bea
H(r) & = &
\sfrac{1}{2} \left ( f_{k_e}(r) + g_{-k_e}(r) \right )
\nonumber \\
D(r) & = &
\sfrac{1}{2} \left ( g_{k_e}(r) - f_{-k_e}(r) \right )
\nonumber \\
h(r) & = &
\sfrac{1}{2} \left ( f_{k_e}(r) - g_{-k_e}(r) \right )
\nonumber \\
d(r) & = &
\sfrac{1}{2} \left ( g_{k_e}(r) + f_{-k_e}(r) \right ) .
\label{HDhd}
\eea
Here $f_{\kappa}(r)$ and
$g_{\kappa}(r)$ are radial electron functions\footnote{The internal
normalisations at the origin are here set to unity.  Recall that
these normalisations, $\alpha_{\kappa}$, have been separated out
into the beta-decay Coulomb functions, $F_0L_0$, (see Eq.~(\ref{bdcf})).}, 
while the spherical
Bessel functions represent radial neutrino wave functions.
The functions $G_{++}$, $G_{+-}$, $G_{-+}$ and $G_{--}$ are
short-hand notations for $G_{KLs}(k_e,k_{\nu})$,
$G_{KLs}(k_e,\hhphen k_{\nu})$,
$G_{KLs}(\hhphen k_e,k_{\nu})$, and
$G_{KLs}(\hhphen k_e,\hhphen k_{\nu})$ respectively, where
the functions $G_{KLs}(\kappa_e,\kappa_{\nu})$ contain all
the angular momentum factors for the leptons: 
\bea
\lefteqn{G_{K L s}(\kappa_e , \kappa_{\nu})  = 
i^{l_{\nu} + L + l_e }
(-)^{j_e - j_{\nu}} }
\nonumber \\
& &
\hat{s} \hat{K} \hat{\jmath}_e \hat{\jmath}_{\nu}
\hat{l}_e \hat{l}_{\nu} ~
\langle l_e 0 \, l_{\nu} 0 | L 0 \rangle
\left \{ \begin{array}{ccc} K & s & L \\[1mm] j_e & \sfrac{1}{2} & l_e \\[1mm]
j_{\nu} & \sfrac{1}{2} & l_{\nu} \end{array} \right \} . 
\label{GKLs}
\eea
Lastly, the operators $\hat{T}_{KLs}$ depend on the angle and
spin coordinates and are defined as
\bea
\hat{T}_{K L s}^M(\hat{{\bf r}}) & = & (V_0 + A_0) i^L Y_{L M}(\hat{{\bf r}})
\delta_{K,L} ~~~~~~ {\rm if} ~ s=0
\nonumber \\
& = & ({\bf V} + {\bf A}) .  i^L
{\bf Y}_{K L M} (\hat{{\bf r}}) ~~~~~~ {\rm if} ~ s=1 ,
\label{TKLSM}
\eea
where $V_0$, $A_0$ are the time parts of the vector and axial-vector
hadronic currents and ${\bf V}$, ${\bf A}$ are the space parts.
Further
${\bf Y}_{K L M} (\hat{{\bf r}})$ 
is a vector spherical harmonic \cite{Ed64}, which in Eq.~(\ref{TKLSM})
forms a scalar product with vectors ${\bf V}$ and ${\bf A}$.

\subsection{Hadronic matrix element}
\label{ss:hadme}

For nucleons, the
vector and axial-vector interactions are written
\bea
V_{\mu} & = & \gV \gamma_{\mu} - \fM \sigma_{\mu \nu} q_{\nu}
+ i \fS q_{\mu}
\nonumber \\
A_{\mu} & = & \gA \gamma_{\mu} \gamma_5 - \fT \sigma_{\mu \nu} q_{\nu}
\gamma_5 + i \fP q_{\mu} \gamma_5 .
\label{VuAu}
\eea
Were we discussing the weak interaction between point-like spin-1/2 fermions,
then we would set $\gV = \gA = 1$, and
$\fM = \fS = \fT = \fP = 0$.  However, in considering nucleons, we
recognize that they are not point-like and furthermore they are influenced
by the presence of the strong interaction.  Thus Eq.~(\ref{VuAu})
presents the most general form of a vector and axial-vector
interaction that is consistent with Lorentz invariance and excludes
momentum operators beyond the first power.  Here $q_{\mu} =
( p_f - p_i )_{\mu}$ is the momentum transfer.  The coefficients,
in principle, could be functions of $q^2$ but, because of the low
four-momentum transfer in beta decay, this $q^2$ dependence can be
neglected and the coefficients are referred to as coupling constants
with individual titles: $\gV$ being vector; $\gA$, axial-vector; $\fM$,
weak magnetic; $\fS$, induced scalar; $\fT$, induced tensor
and $\fP$, induced pseudoscalar.   

Our aim is to reduce the matrix element $i\overline{u}_f ( V_{\mu} +
A_{\mu} ) u_i$, where $\overline{u}_f$ and $u_i$ are Dirac
spinors characterising nucleons of momentum ${\bf p}_f$ and ${\bf p}_i$,
to the nonrelativistic form involving Pauli two-component spinors,
\be
\chi_{m_f}^{\dag} \left ( V_0 + {\bf V} + A_0 + {\bf A} \right )
\chi_{m_i} ,
\label{Paulime}
\ee
by simply multiplying out the Dirac matrices involved and keeping terms
to first order in $|{\bf p}|/\MN$, and dropping terms in
$|{\bf p}^2|/\MN^2$ and higher order.  
This multiplication yields
\bea
V_0 & = & \gV + \fS ( W_0 - V(r))
\label{V0} \\[2mm]
{\bf V} & = & \frac{\gV}{2 \MN} \left [ {\bf p} + i \mbox{\boldmath
$\sigma$} \times {\bf q} \right ] +
\fM i \mbox{\boldmath $\sigma$} \times {\bf q} - \fS {\bf q}
\label{vectV} \\
A_0 & = & - \frac{\gA}{2 \MN} \mbox{\boldmath $\sigma$}.{\bf p}
- \fT  \mbox{\boldmath $\sigma$}.{\bf q}
\label{A0} \\[2mm]
{\bf A} & = & -\gA \mbox{\boldmath $\sigma$}
+  \fT ( W_0 - V(r)) \mbox{\boldmath $\sigma$} ,
\label{vectA}
\eea
where ${\bf p} = {\bf p}_f + {\bf p}_i$ and
${\bf q} = {\bf p}_f - {\bf p}_i$.  
Note that the large terms of order unity occur in $V_0$ and ${\bf A}$.
Each of the four coupling constants denoted by an $f$ are small
and of order
$1/\MN$ and, as a consequence, terms in $f/\MN$ have been dropped.  
Eqs.~(\ref{V0}) to (\ref{vectA}) are the quantities needed
in the operators $\hat{T}_{KLs}^M(\hat{\bf r})$ in Eq.~(\ref{TKLSM}).

\subsection{Reduced matrix elements}
\label{ss:hdme}

All the beta-decay observables \cite{BB82} can be expressed in terms
of the functions $M_K(k_e,k_{\nu})$ and
$m_K(k_e,k_{\nu})$ defined in Eq.~(\ref{MmKkk}).  Here we are only
interested in the shape-correction function, Eq.~(\ref{CWfinal}), which
is a particularly simple combination of these quantities.  Before
proceding to evaluate $M_K(k_e,k_{\nu})$ and $m_K(k_e,k_{\nu})$, we
note that the expressions for both differ only in the presence of  
$F(r)$ in one case and $f(r)$ in the other.  For simplicity in what
follows, we will only explicitly display formulae incorporating $f(r)$;
obviously an equivalent set can be written with $F(r)$.  

The operators $\hat{T}_{KLs}^M(\hat{{\bf r}})$ are shown in Eq.~(\ref{TKLSM}),
but it is tidier if we incorporate the phase,
$(-)^{K-L}$ (see Eq.~(\ref{MmKkk})), 
into the operator.  All the
operators (some after rearrangement) can then be expressed as a 
product of spherical tensors, one in orbital space and one in
spin space.  So, generically the operators take the form
\be
T_{KLS}^M({\bf r}) = f(r) (-)^{K-L} T_{KM}(\Lambda_L \otimes \Sigma_S ) ,
\label{tproduct}
\ee
where $\Lambda_L$ is a spherical tensor in orbital space of multipolarity
$L$, and $\Sigma_S$ a spherical tensor in spin space\footnote{
Note, the upper case $S$ in Eq.~(\ref{tproduct}) referring
to the multipolarity of the spin operator is not the same as the 
lower case $s$ in Eq.~(\ref{TKLSM}).}.  
We have introduced notation for a composite spherical tensor
obtained from the combination of two other spherical tensors: \viz
\be
T_{K M}(\Lambda_L \otimes \Sigma_S) =
\sum_{M_L M_S}
\langle L M_L \, S M_S | K M \rangle
\Lambda_{L \, M_L} \Sigma_{S \, M_S} .
\label{compst}
\ee
The single-particle wave functions from the shell model can also be
expressed as products of orbital and spin space functions:
\be
| j m \rangle 
=  \sum_{m_l m_s} \langle l m_l \, \sfrac{1}{2} m_s | j m \rangle
R_{n l}(r) i^l Y_{l m_l}(\hat{{\bf r}}) \chi_{m_s} ,
\label{spwf}
\ee
where $n$ is the principal quantum number designating the number of nodes
in the radial function.  Notice the presence of $i^l$ with the spherical
harmonics\footnote{ If one-body density matrix elements (OBDME)
are imported from a shell-model calculation into this 
beta-decay environment, then it is important that these OBDME
be computed with similar $i^l$ phases included in the definition
of single-particle wave functions.}.

The first step in evaluating the reduced matrix element is to 
factorize it into orbital and spin reduced matrix elements:
\bea
\lefteqn{
\langle (l_{\alpha} \sfrac{1}{2} ) j_{\alpha} \dbar T_{KLS} \dbar
(l_{\beta} \sfrac{1}{2} ) j_{\beta} \rangle    = }
\nonumber \\
& &
(-)^{K-L}
A_{(LS)K} ~ \langle l_{\alpha} \dbar f(r) \Lambda_L \dbar l_{\beta} \rangle
\langle \sfrac{1}{2} \dbar \Sigma_S \dbar \sfrac{1}{2}\rangle , 
\label{LSfactor}
\eea
where
\be    
A_{(LS)K} = \hat{\jmath}_{\alpha} \hat{K} \hat{\jmath}_{\beta} ~
\left \{ \begin{array}{ccc} l_{\alpha} & \sfrac{1}{2} & j_{\alpha} \\[1mm]
l_{\beta} & \sfrac{1}{2} & j_{\beta} \\[1mm] L & S & K \end{array} \right \} .
\label{AKLS}
\ee
Our
conventions on reduced matrix elements are those of Edmonds \cite{Ed64}.
Next, we need to define two spin matrix elements, denoted $S_0$ and $S_1$ and given by
\bea
S_0 & \equiv & \langle \sfrac{1}{2} \dbar 1 \dbar 
\sfrac{1}{2} \rangle  ~~ = ~~ \sqrt{2} ~ \delta_{S,0}
\nonumber \\
S_1 & \equiv & \langle \sfrac{1}{2} \dbar \mbox{\boldmath $\sigma$} \dbar 
\sfrac{1}{2} \rangle  ~~ = ~~ \sqrt{6} ~ \delta_{S,1} ;
\label{spinme}
\eea
and two two orbital matrix elements denoted
$L_L$ and $L_{(J 1)L}({\bf Q})$.  The first is
\bea
L_L & \equiv & \langle l_{\alpha} \dbar f(r) i^L Y_L(\hat{{\bf r}})
\dbar l_{\beta} \rangle
\nonumber \\[2mm]
& = & i^{l_{\alpha} + L + l_{\beta}} ~ \frac{\hat{l}_{\alpha} \hat{L} \hat{l}_{\beta}}
{\sqrt{4 \pi }} 
\left ( \begin{array}{ccc} l_{\alpha} & L & l_{\beta} \\ 0 & 0 & 0 \end{array}
\right )
\langle R_{\alpha} | f(r) | R_{\beta} \rangle ,
\label{YLme}
\eea
where the last factor is the radial integral:
\be
\langle R_{\alpha} | f(r) | R_{\beta} \rangle \equiv \int_0^{\infty}
R_{\alpha}(r) f(r) R_{\beta}(r) r^2 dr .
\label{rradint}
\ee
The second involves derivative operators and requires a little
more care.  The matrix element is
\be 
L_{(J1)L}({\bf Q}) \equiv \langle l_{\alpha} \dbar f(r) i^J
T_L(Y_J \otimes {\bf Q} ) \dbar l_{\beta} \rangle ,
\label{L2def}
\ee
where ${\bf Q}$ is either ${\bf p} = {\bf p}_f + {\bf p}_i$ or 
${\bf q} = {\bf p}_f - {\bf p}_i$.  Thus we write ${\bf Q}$ as
${\bf p}_f \pm {\bf p}_i$ with the upper sign appropriate for ${\bf p}$
and the lower sign for ${\bf q}$.  Now ${\bf Q}$ is also
$-i ( \mbox{\boldmath $\nabla$}_f \pm
\mbox{\boldmath $\nabla$}_i )$, where the gradient operator acts on 
either the initial or final nuclear wave function but not on the
integrand, $f(r)$.  Thus the interpretation is as follows:
$\langle \phi_f | f {\bf Q} | \phi_i \rangle = -i \{ \pm
\langle \phi_f | f | \mbox{\boldmath $\nabla$} \phi_i \rangle -
\langle \mbox{\boldmath $\nabla$} \phi_f | f | \phi_i \rangle \}$.
The result for $L_{(J1)L}({\bf Q})$ is
\begin{widetext}
\bea
\lefteqn{
L_{(J1)L}({\bf Q}) = i^{l_{\alpha}+l_{\beta}+J+1} (-)^{J-L} \frac{\hat{J}}{\sqrt{4 \pi }}
\times }
\nonumber \\
& &
\biggm{\{}
\pm U(l_{\beta} 1 l_{\alpha} J ; l_{\beta}+1 L)
\left ( \begin{array}{ccc} l_{\alpha} & J & l_{\beta}+1 \\ 0 & 0 & 0 \end{array} \right )
\hat{l}_{\alpha} (l_{\beta} + 1 )^{1/2} 
\bigm{\langle} R_{\alpha} | f | \bigm{(} \sfrac{d}{dr} - \sfrac{l_{\beta}}{r} 
\bigm{)} R_{\beta} \bigm{\rangle}
\nonumber \\[2mm]
& &
~~ \mp U(l_{\beta} 1 l_{\alpha} J ; l_{\beta}-1 L)
\left ( \begin{array}{ccc} l_{\alpha} & J & l_{\beta}-1 \\ 0 & 0 & 0 \end{array} \right )
\hat{l}_{\alpha} (l_{\beta} )^{1/2} 
\bigm{\langle} R_{\alpha} | f | \bigm{(} \sfrac{d}{dr} + \sfrac{l_{\beta}+1}{r}
\bigm{)} R_{\beta} \bigm{\rangle}
\nonumber \\[2mm]
& &
~~ - (-)^{J+1-L}  U(l_{\beta} J l_{\alpha} 1 ; l_{\alpha}+1 L)
\left ( \begin{array}{ccc} l_{\alpha}+1 & J & l_{\beta} \\ 0 & 0 & 0 \end{array} \right )
\hat{l}_{\beta} (l_{\alpha}+1)^{1/2} 
\bigm{\langle} \bigm{(} \sfrac{d}{dr} - \sfrac{l_{\alpha}}{r} 
\bigm{)} R_{\alpha} | f | R_{\beta} \bigm{\rangle}
\nonumber \\[2mm]
& &
~~ + (-)^{J+1-L}  U(l_{\beta} J l_{\alpha} 1 ; l_{\alpha}-1 L)
\left ( \begin{array}{ccc} l_{\alpha}-1 & J & l_{\beta} \\ 0 & 0 & 0 \end{array} \right )
\hat{l}_{\beta} (l_{\alpha} )^{1/2}  
\bigm{\langle} \bigm{(} \sfrac{d}{dr} + \sfrac{l_{\alpha}+1}{r} 
\bigm{)} R_{\alpha} | f | R_{\beta} \bigm{\rangle}
\biggm{\}} ,
\nonumber \\
& &
\label{L2value}
\eea
where the upper sign is used for ${\bf Q} = {\bf p}$ and the lower sign for
${\bf Q} = {\bf q}$. 
The $U$-coefficient is a recoupling of
three angular momenta and is related to a $6j$-symbol:
\be
U(a b c d; e f) = (-)^{a+b+c+d} ~ \hat{e} \hat{f} ~
\left \{ \begin{array}{ccc} a & b & e \\ d & c & f \end{array} \right \} .
\label{Ucoef}
\ee

Finally, we are ready to write down the specific reduced matrix elements for all
the different hadronic components of the weak interaction,
Eqs.~(\ref{V0}) to (\ref{vectA}):  

\vspace{2mm}

\noindent{\it Time-like Vector Current}
\be
\langle (l_{\alpha} \sfrac{1}{2} ) j_{\alpha} \dbar f(r) V_0 i^K Y_K \dbar
(l_{\beta} \sfrac{1}{2} ) j_{\beta} \rangle
= 
\left ( \gV + \fS \left ( W_0 + \sfrac{6}{5} \sfrac{\alpha Z}{R} \right )
\right ) A_{(K0)K} ~ L_K ~ S_0 . 
\label{V0me}
\ee

\noindent{\it Space-like Axial Current}
\be
\langle (l_{\alpha} \sfrac{1}{2} ) j_{\alpha} \dbar f(r) {\bf A}. i^L {\bf Y}_{K L} \dbar
(l_{\beta} \sfrac{1}{2} ) j_{\beta} \rangle
= (-)^{K-L}
\left ( -\gA + \fT \left ( W_0 + \sfrac{6}{5} \sfrac{\alpha Z}{R} \right )
\right ) A_{(L1)K} ~ L_L ~ S_1 . 
\label{vectAme}
\ee
In Eqs.~(\ref{V0me}) and (\ref{vectAme}), for simplicity we have replaced
the function, $V(r)$ (see Eqs.~(\ref{V0}) and (\ref{vectA})) by the potential
due to a uniform charge distribution for small $r$, with $r^2$ replaced by its
expectation value.  In our computations we actually included the function $V(r)$
in the integrand of the appropriate radial integral.

\noindent{\it Space-like Vector Current}
\bea
\lefteqn{
\langle (l_{\alpha} \sfrac{1}{2} ) j_{\alpha} \dbar f(r) {\bf V}. i^L {\bf Y}_{K L} \dbar
(l_{\beta} \sfrac{1}{2} ) j_{\beta} \rangle
 = 
}
\nonumber \\
& & ~~ (-)^{K-L} \left \{
\frac{\gV}{2 \MN } A_{(K0)K} ~ L_{(L1)K}({\bf p}) ~ S_0
- \fS  A_{(K0)K} ~ L_{(L1)K}({\bf q}) ~ S_0 \right .
\nonumber \\
& & ~~ \left .
- \sqrt{2} \sum_J U(1 1 K L ; 1 J) 
\left ( \frac{\gV}{2 \MN } + \fM \right ) 
A_{(J1)K} ~ L_{(L1)J}({\bf q}) ~ S_1  \right \} .
\label{vectVme}
\eea

\noindent{\it Time-like Axial Current}
\bea
\lefteqn{
\langle (l_{\alpha} \sfrac{1}{2} ) j_{\alpha} \dbar f(r) A_0 i^K Y_K \dbar
(l_{\beta} \sfrac{1}{2} ) j_{\beta} \rangle
 = }
\nonumber \\
& &
\sum_J (-)^{J-K} ~ \frac{\hat{J}}{\hat{K}} ~ \left \{
- \frac{\gA}{2 \MN} A_{(J1)K} ~ L_{(K1)J}({\bf p}) ~ S_1 
- \fT 
A_{(J1)K} ~ L_{(K1)J}({\bf q}) ~ S_1 \right \} .
\label{A0me}
\eea
\end{widetext}

\subsection{Numerical Results}
\label{ss:numresult}

The key ingredient for the computation of exact electron wave functions
in beta decay is the charge-density distribution of the daughter
nucleus.  
There are various parameterizations available in the literature,
of which the following are the most common:

\bi

\item {\it Two-parameter Fermi distribution (2pF)}.  
This charge density distribution,
\be
\rho (r) = \frac{\rho_0}{1 + \exp \{ (r - c)/a \} } ,
\label{Fermirho}
\ee
has two parameters, $c$ and $a$, other than its normalization.

\item {\it Three-parameter Fermi distribution (3pF)}.  
This is an extension of the two-parameter model, which introduces a dimensionless
``wine-bottle" parameter, $w$, that impacts on the small-$r$
behaviour of the density distribution.  The functional form is
\be
\rho (r) = \frac{\rho_0 ( 1 + w r^2/c^2 )}{1 + \exp \{ (r - c)/a \} } .
\label{3Fermirho}
\ee

\item {\it Three-parameter Gaussian distribution (3pG)}.  
This is an alternative three-parameter model with a Gaussian rather than a Fermi
distribution:
\be
\rho (r) = \frac{\rho_0 ( 1 + w r^2/c^2 )}{1 + \exp \{ (r^2 - c^2)/a^2 \} } .
\label{3Gaussrho}
\ee

\item {\it Harmonic-oscillator distribution (HO)}.  
In light $p$-shell nuclei, where only $s$ and $p$ orbitals are occupied,
a density distribution can be constructed from the
harmonic oscillator radial wavefunctions.  Its form is
\be
\rho (r) = \rho_0 ( 1 + \alpha r^2/b^2 ) \exp (- r^2/b^2) ,
\label{HOdensity}
\ee
where $b$ is the harmonic oscillator length parameter, and
$\alpha$ is related to the number of $p$-shell protons,
$\alpha = (Z-2)/3$.  However, in practise both $b$ and $\alpha$ are treated
as free parameters and adjusted to fit the elastic-electron scattering data.

\ei

\begingroup
\squeezetable
\begin{table}[t]
\begin{center}
\caption{Charge-density distributions from elastic electron-scattering
data \protect\cite{De87}.  The radius parameter, in some cases, has
been adjusted to reproduce $\langle r^2 \rangle^{1/2}$.
Parameters, $c$ and $a$, are in fm units, parameter $w$ is dimensionless.
\label{t:tab2}}
\vskip 1mm
\begin{ruledtabular}
\begin{tabular}{rrrrrrr}
Daughter & $\langle r^2 \rangle^{1/2}$ & &  &
&  &  $\Delta f$\footnotemark[1]\\
Nucleus & fm~~~ & Model\footnotemark[2] &  $c$~~ &
$a$~~ & $w$~~ & \% ~ \\
\hline
& & & & & & \\[-3mm]
$T_z = -1$: & & & & & & \\
$^{10}$B & 2.45(10) & HO & 1.709\footnotemark[3] & 0.837\footnotemark[4] &
& 0.001 \\
$^{14}$N & 2.52(2) & 3pF & 2.572 & 0.5052 & $-0.180$ & 0.000 \\
$^{18}$F & 2.90(3) & 2pF & 2.574 & 0.567 &  & 0.001 \\
$^{22}$Na & 2.95(5) & 2pF & 2.750 & 0.549 & & 0.001 \\
$^{26}$Al & 3.03(2) & 2pF & 2.791 & 0.569 & & 0.001 \\
$^{30}$P & 3.18(3) & 3pF & 3.350 & 0.582 & $-0.173$ & 0.002 \\
$^{34}$Cl & 3.39(2) & 3pF & 3.479 & 0.599 & $-0.100$ & 0.001 \\
$^{38}$K & 3.41(4) & 3pF & 3.738 & 0.585 & $-0.201$ & 0.004 \\
$^{42}$Sc & 3.50(5) & 3pF & 3.794 & 0.586 & $-0.161$ & 0.004 \\[5mm]
$T_z = 0 $: & & & & & & \\
$^{26}$Mg & 3.06(5) & 2pF & 3.049 & 0.523 &  & 0.002 \\
$^{34}$S & 3.29(1) & 3pG & 2.810 & 2.191 &  0.160 & 0.001 \\
$^{38}$Ar & 3.36(5) & 2pF & 3.590 & 0.507 & & 0.004 \\
$^{42}$Ca & 3.48(3) & 3pF & 3.765 & 0.586 & $-0.161$ & 0.002 \\
$^{46}$Ti & 3.61(3) & 2pF & 3.711 & 0.588 & & 0.003 \\
$^{50}$Cr & 3.66(4) & 2pF & 3.868 & 0.566 & & 0.004 \\
$^{54}$Fe & 3.69(2) & 3pG & 3.541 & 2.270 & 0.403 & 0.003 \\
$^{62}$Zn & 3.90(2) & 3pG & 3.570 & 2.465 & 0.342 & 0.005 \\
$^{66}$Ge & 4.04(4) & 2pF & 4.398 & 0.585 &  & 0.011 \\
$^{70}$Se & 4.07(5) & 2pF & 4.442 & 0.585 &  & 0.011 \\
$^{74}$Kr & 4.10(5) & 2pF & 4.489 & 0.585 &  & 0.013 \\
\end{tabular}
\end{ruledtabular}
\footnotetext[1]{Percentage uncertainty in $f$ due to the uncertainty in        
$\langle r^2 \rangle^{1/2}$.}
\footnotetext[2]{See Eqs.(\protect\ref{Fermirho}) to 
(\protect\ref{HOdensity}).}
\footnotetext[3]{This is parameter, $b$, of Eq.~(\protect\ref{HOdensity})
in fm units.}
\footnotetext[4]{This is the dimensionless parameter, $\alpha$, of 
Eq.~(\protect\ref{HOdensity}).}
\end{center}
\end{table}
\endgroup

These model distributions typically contain
two or three parameters.  Where possible, the parameters
are determined from experimental data on elastic electron
scattering, since the measured electron-scattering form factors
are just the Fourier transforms of the charge-density
distributions.  A compilation of charge-density distributions
determined from electron scattering is given by De Vries \etal \cite{De87}.
We have assessed these data and selected for each daughter nucleus what we believe to be the
`best' value of the rms radius, $\langle r^2 \rangle^{1/2}$, and
its probable error.  In cases where data are not available on the
isotope of interest, we have examined the nearest isotope that is available
and applied a modest isotope shift to its value of
$\langle r^2 \rangle^{1/2}$. 
Our final selected values are listed in Table~\ref{t:tab2}.
We also list the percentage uncertainty in the
exact value of $f$ due solely to the uncertainty in
$\langle r^2 \rangle^{1/2}$. 
Clearly, the uncertainty in the charge-density distribution is
not a factor in the determination of $f$ to $0.1 \%$ accuracy.

Before the final evaluation of the statistical rate function,
$f$, there are two further corrections to consider:  for screening 
and recoil.  To accommodate these corrections and to
remove -- as is customary -- the leading matrix element
from the definition of $f$, we rewrite $f$ as follows:
\be
f = \xi R(W_0) \int_1^{W_0}  p W (W_0-W)^2
F_0 L_0 C(Z,W) Q(Z,W) dW .
\label{fcorrect}
\ee
Comparison with Eqs.~(\ref{fdef}) and (\ref{F0L0C}) reveals three new
factors, $Q(Z,W)$, $R(W_0)$ and $\xi$.  We will deal with them in that order.

The calculation of the Fermi function
presented so far 
makes no allowance for the screening of the atomic
electrons.  Rose \cite{Ro36} was the first to find a simple
analytic prescription to obtain the Fermi function for
a screened field from the Fermi function for the corresponding
unscreened field.  That prescription is to incorporate 
a correction factor into the integrand for $f$: {\it viz}
\be
Q(Z, W) = \frac{\tilde{p} \tilde{W}}{p W}
\frac{ F(Z, \tilde{W})}{F(Z, W)} ,
\label{Qscreen}
\ee
where $\tilde{W} = W - V_0$, $\tilde{p} = (\tilde{W}^2 -1 )^{1/2}$
and $V_0 = N(\tilde{Z}) \alpha^2 \tilde{Z}^{4/3}$, with
$\tilde{Z}$ being the electronic charge of the parent atom
and $N(\tilde{Z})$ being a weak function of
$\tilde{Z}$, which varies from $N = 1.42$ at $\tilde{Z} = 8$ to
$N = 1.56$ at $\tilde{Z} = 29$ (see Matese and Johnson \cite{MJ66}).
Since the factor $Q(Z,W)$ yields a correction to $f$
of order $\sim 0.2 \%$, we only need Rose's screening correction to
be accurate to within 50\% of its central value in order to assure
us an accuracy in $f$ of 0.1\%. Matese and Johnson \cite{MJ66} have tested the
Rose formula by comparing it with numerical solutions of the Dirac
equation for a self-consistent Hartree-Fock-Slater potential.
They conclude that $Q(Z,W)$ has an accuracy
of four significant figures or better
for all energies except the very lowest in
positron emitters.  Since we integrate $Q(Z,W)$
over the whole beta spectrum, which actually de-emphasizes the lowest
positron energies, we conclude that the Rose formula has far more than sufficient
accuracy for our purpose.

\begin{table}[t]
\begin{center}
\caption{Comparison of statistical rate functions, $f_{{\rm stat}}$,
$f_{{\rm approx}}$ and the exact value, $f$.
\label{t:tab1}}
\vskip 1mm
\begin{ruledtabular}
\begin{tabular}{rrrrrr}
Parent & $Q_{EC}$ & $W_0$~~ & $f_{{\rm stat}}$~~ & 
$f$~~~~ & $\Delta f_{{\rm stat}}$\footnotemark[1]
\\
& keV & & & & \%~~~ \\
\hline
& & & & & \\[-3mm]
$T_z = -1$: & & & & & \\
$^{10}$C  &  1907.9 &  2.7336 & 2.29484 & 2.30089 & $-0.26$ \\
$^{14}$O  &  2831.0 &  4.5400 & 42.6147 & 42.7485 & $-0.31$ \\
$^{18}$Ne &  3402.0 &  5.6575 & 133.867 & 134.484 & $-0.46$ \\
$^{22}$Mg &  4124.6 &  7.0716 & 415.826 & 418.440 & $-0.62$ \\
$^{26}$Si &  4836.9 &  8.4656 & 1014.75 & 1023.28 & $-0.83$ \\
$^{30}$S  &  5459.5 &  9.6840 & 1945.49 & 1967.05 & $-1.10$ \\
$^{34}$Ar &  6062.8 & 10.8647 & 3366.22 & 3414.21 & $-1.41$ \\
$^{38}$Ca &  6614.2 & 11.9437 & 5247.54 & 5338.46 & $-1.70$ \\
$^{42}$Ti &  7000.9 & 12.7004 & 6904.47 & 7042.83 & $-1.96$ \\[5mm]
$T_z = 0 $: & & & & & \\
$^{26m}$Al&  4232.5 &  7.2828 & 474.691 & 478.176 & $-0.73$ \\
$^{34}$Cl &  5491.8 &  9.7472 & 1971.95 & 1996.39 & $-1.22$ \\
$^{38m}$K &  6044.4 & 10.8286 & 3248.45 & 3298.10 & $-1.51$ \\
$^{42}$Sc &  6425.7 & 11.5748 & 4391.71 & 4470.41 & $-1.76$ \\
$^{46}$V  &  7050.7 & 12.7979 & 7044.04 & 7199.96 & $-2.17$ \\
$^{50}$Mn &  7632.5 & 13.9364 & 10456.3 & 10731.6 & $-2.57$ \\
$^{54}$Co &  8242.7 & 15.1305 & 15281.4 & 15750.1 & $-2.98$ \\
$^{62}$Ga &  9171.0 & 16.9472 & 25187.9 & 26247.6 & $-4.04$ \\
$^{66}$As &  9550.0 & 17.6889 & 30146.4 & 31613.7 & $-4.64$ \\
$^{70}$Br &  9970.0 & 18.5108 & 36605.4 & 38602.2 & $-5.17$ \\
$^{74}$Rb & 10416.5 & 19.3846 & 44606.1 & 47277.8 & $-5.65$ \\
\end{tabular}
\end{ruledtabular}
\footnotetext[1]{$\Delta f_{\rm stat} = 100*(f_{\rm stat} - f)/f$}
\end{center}
\end{table}

The second new factor in Eq.~(\ref{fcorrect}) is $R(W_0)$, which is the
correction for recoil: it recognizes that the daughter nucleus is not
at rest but has a small amount of recoiling kinetic energy.  As a result,
the leptons' maximum energy is actually slightly less than $W_0$.  The
recoil correction \cite{BB82} is 
\be
R(W_0) = 1 - \frac{3 W_0}{2 M_A} ,
\label{recoil}
\ee
where $M_A$ is the average of the initial and final nuclear masses.  For use in
eq.~(\ref{recoil}), $M_A$ must, like $W_0$, be expressed in electron
rest-mass units.  The resulting correction is very small, 
being of order $0.02 \%$ for the superallowed beta decays from
$A = 10$ to $A = 74$.

Lastly, for allowed transitions it is customary to remove the leading
matrix element from the definition of $f$.  Thus, we have introduced $\xi$
in Eq.~(\ref{fcorrect}), where $\xi = 1/|M_F|^2$ for superallowed Fermi
transitions, $M_F$ being the Fermi matrix element.
For pure Gamow-Teller transitions $\xi = 1/|M_{GT}|^2$, with
$M_{GT}$ being the Gamow-Teller matrix element.

In Table \ref{t:tab1} we list both the values of $f_{\rm stat}$, Eq.~(\ref{fstat}),
and the exact values of $f$ for cases of interest in
superallowed beta decay.  The relevant $Q_{EC}$ value is listed as
well.  For the exact calculations, we imported one-body density
matrix elements, OBDME, from a shell-model code.  For each case
we performed several shell-model calculations for various
sets of effective interactions and model spaces.  We used, in
fact, the same wave functions that we used \cite{TH02} to compute the
nuclear-structure corrections $\delta_C$ and $\delta_{NS}$.
Thus our $f$ calculations can be considered to be entirely consistent
with the calculation of the nuclear-structure-dependent
corrections.  The $f$ calculation, however, is not very
sensitive to the shell-model inputs.  In light nuclei,
different shell-model OBDME gave changes in $f$ at the
0.01\% level, increasing to around 0.1\% in $A = 74$,
our heaviest-mass case.  Where we have more than one
shell-model calculation for a given nucleus, we have
averaged the $f$ values for the entry in Table \ref{t:tab1}.

\end{document}